\documentclass[aps,prd,amsmath,amssymb,showpacs]{revtex4}

\usepackage{epsfig,float}
\usepackage{graphicx}
\usepackage{dcolumn}% Align table columns on decimal point
\usepackage{morefloats}% bold math
\usepackage{color}
\usepackage{slashed}
\usepackage{bbm}

\begin{document}
\title{Internal particle width effects on the the triangle singularity mechanism in the study of the $\eta(1405)$ and $\eta(1475)$ puzzle}

\author{Meng-Chuan Du$^{1,2}$\footnote{{\it Email address:} dumc@ihep.ac.cn} }

\author{Qiang Zhao$^{1,2,3}$\footnote{{\it Email address:} zhaoq@ihep.ac.cn} }

\affiliation{$^1$ Institute of High Energy Physics and Theoretical Physics Center for Science Facilities,
        Chinese Academy of Sciences, Beijing 100049, China}
\affiliation{$^2$ School of Physical Sciences, University of Chinese Academy of Sciences, Beijing 100049, China}
\affiliation{$^3$ Synergetic Innovation Center for Quantum Effects and Applications (SICQEA),
Hunan Normal University, Changsha 410081, China}

\begin{abstract}
In this article, the analyticity of triangle loop integral with complex masses of internal particles is discussed in a new perspective, base on which we obtain the explicit width dependence of the absorptive part of the triangle amplitude. We reanalyze the decay pattern of $\eta(1405/1475)$ with the width effects included in the triangle singularity (TS) mechanism. Based on the present experimental information, we provide a self-consistent description of the $K\bar{K}\pi$, $\eta\pi\pi$, and $3\pi$ decay channels for $\eta(1405/1475)$. Our results confirm the claim that the TS mechanism plays a decisive role in the understanding of the $\eta(1405)$ and $\eta(1475)$ puzzle. Namely, the observed differences of $\eta$ resonances within the mass region of $1.40\sim 1.48$ GeV are originated from the same state. For the isospin violated process $J/\psi\to\gamma\eta(1405/1475)\to f_0(980)\pi\to 3\pi$, we identify an additional contribution to the $a_0(980)-f_0(980)$ mixing via the TS mechanism.
\end{abstract}
\maketitle

\section{Introduction}

It has been a long-standing question on the existence of glueball in hadron spectroscopy. This exotic object as the bound state of gluons predicted by QCD has been a crucial piece of information for our understanding of strong interaction theory in the non-perturbative regime. The corresponding theoretical study and experimental search for glueball states have been the topical subjects in hadron physics. However, although tremendous efforts have been made, the indisputable evidence for their existence is still lacking. In the glueball spectrum the low-lying states include scalar ($J^{PC}=0^{++}$), tensor ($2^{++}$) and pseudoscalar ($0^{-+}$). So far, the lattice QCD (LQCD) simulations~\cite{Morningstar:1999rf,Bali:1993fb,Chen:2005mg,Chowdhury:2014mra,Richards:2010ck,Sun:2017ipk} suggest that their typical masses are about $1.5\sim 1.7$, $2.2\sim 2.3$, and $2.4\sim 2.6$ GeV, respectively. The mass hierarchy $M_{0^{++}}<M_{2^{++}} < M_{0^{-+}}$ has been a stable feature from LQCD. While there have been topical reviews on the glueball spectrum in the literature for the scalar and tensor, our focus in this work is on the pseudoscalar glueball. We will discuss the long-standing controversial issues involved in the identification of the pseudoscalar glueball candidate, and stress that a self-consistent picture can only be obtained with a special kinematic effect, i.e. the triangle singularity (TS) or Landau singularity mechanism~\cite{Landau:1959fi,Cutkosky:1960sp,bonnevay:1961aa,Peierls:1961zz}, implemented.

In the literature the most promising candidate for the pseudoscalar glueball has been assigned to the $\eta(1405)$ since it was introduced as an additional state to the nearby $\eta(1475)$ and $\eta(1295)$ around early 1990s~\cite{Bai:1990hs,Bolton:1992kb,Augustin:1989zf,Augustin:1990ki,Bertin:1995fx,Bertin:1997zu,Cicalo:1999sn,Bai:2004qj}. These three states of isospin 0 and similar masses cannot fit the pattern arising from the SU(3) flavor symmetry of quark model in the light quark sector. A broadly accepted classification has been that the $\eta(1295)$ and $\eta(1475)$ belong to the isospin-0 radial excitation states in the SU(3) flavor multiplet due to the mixing between the flavor singlet and octet. The $\eta(1405)$ as an out-numbered state was then proposed to be the ground state pseudoscalar glueball candidate. Such an assignment was based on phenomenological studies which predicted the mass of the pseudoscalar glueball around 1.4 GeV~\cite{Faddeev:2003aw}. This proposal seemed to be able to accommodate the experimental observations with the theoretical prediction, and had attracted a lot of efforts to further explore the structure and production mechanism of the $\eta(1405)$ as the pseudoscalar glueball candidate~\cite{Donoghue:1980hw,Close:1980rv,Barnes:1981kp,Close:1987er,Amsler:2004ps,Masoni:2006rz,Klempt:2007cp}.

Notice that the mass of $\eta(1405)$ is far below the expected value from LQCD~\cite{Morningstar:1999rf,Bali:1993fb,Chen:2005mg,Chowdhury:2014mra,Richards:2010ck,Sun:2017ipk}. In the literature a lot of theoretical studies have focused on the consequence of the quark and glueball mixings by assuming the $\eta(1405)$ to be the pseudoscalar glueball candidate. Investigations of the pseudoscalar glueball mixings with the light $q\bar{q}$ and its mass positions can be categorized into the following classes: (i) Quantify mixings among the ground state pseudoscalar mesons $\eta$ and $\eta'$, and the pseudoscalar glueball which is assigned to $\eta(1405)$~\cite{Rosenzweig:1981cu,Cheng:2008ss,Close:1996yc,Li:2007ky,Gutsche:2009jh,Li:2009rk,Eshraim:2012jv};  (ii) Identify mechanisms that cause the low mass of pseudoscalar glueball~\cite{Faddeev:2003aw,Cheng:2008ss} compared with the lattice QCD (LQCD) calculations~\cite{Morningstar:1999rf,Bali:1993fb,Chen:2005mg,Chowdhury:2014mra,Richards:2010ck,Sun:2017ipk}. However, because of model-dependence it has been very difficult to make progress on establishing unambiguously the glueball nature of $\eta(1405)$.

It is a challenge to bring down the pure gauge glueball mass from $\sim 2.5$ GeV to $\sim 1.4$ GeV. The very relevant issue is that whatever the mechanism could be it requires an abnormally strong coupling between the light quark states and pure gauge glueball. It has been quoted broadly that the QCD sum rules for pseudoscalar glueball led to relatively low masses. However, it should be noted that the pseudoscalar glueball sum rules are very sensitive to assumptions made in the calculations. As noted explicitly in Ref.~\cite{Senba:1981iy} the pseudoscalar glueball mass in QCD sum rules has large uncertainties and is very sensitive to the gluon condensation. The question about the pseudoscalar glueball mass is inevitably correlated with the $\eta$-$\eta'$ mixing because of the axial anomaly~\cite{Witten:1979vv,Veneziano:1979ec,Novikov:1979uy}. The deviation of the $\eta$-$\eta'$ mixing angle from the ideal one between the flavor octet and singlet indicates the crucial role played by the anomaly. Meanwhile, the $\eta$-$\eta'$ mixing angle does not determine the $\eta'$ mass as a usual mixing scheme would suggest. On the contrary, its dependence of the topological charge density $\langle 0|G_{\mu\nu}^a\tilde{G}^{a\mu\nu}|\eta'\rangle$ has to be taken into account.

In Ref.~\cite{Cheng:2008ss} a dynamical approach for the $\eta$-$\eta'$-glueball mixing was explored by implementing the mixing into the equations of motion for the anomalous Ward identity and a low mass about 1.4 GeV for the physical pseudoscalar glueball. This approach was extended to accommodate the $\eta_c$ in Ref.~\cite{Tsai:2011dp} and a similar result was extracted. However, an analysis of Ref.~\cite{Mathieu:2009sg} based on the same dynamics yields a lower bound of about 2 GeV for the pseudoscalar glueball mass. In Ref.~\cite{Qin:2017qes} a revisit of the mixing scheme of Refs.~\cite{Cheng:2008ss,Tsai:2011dp} was carried out, and the numerical results of Refs.~\cite{Cheng:2008ss,Tsai:2011dp} were confirmed except that the approximation for extracting the pseudoscalar glueball mass appeared to be problematic~\footnote{A detailed deduction can be found in Ref.~\cite{Qin:2017qes}.}. After curing this problem, it shows that the physical glueball mass will favor to be higher than 2 GeV which is remarkably consistent with the conclusion of Ref.~\cite{Mathieu:2009sg} and matches the LQCD simulations. It is interesting to note that the analysis of Ref.~\cite{Gabadadze:1997zc} also suggests a pseudoscalar glueball mass above 2 GeV, although the physical state should be lighter than the quenched state from the pure Yang-Mills gauge theory~\cite{Gabadadze:1997zc}. The mass difference between the quenched pure gauge state and the QCD state is of the order of $1/N_c$, which means that a low mass state around 1.4 GeV is unfavored. In fact, a lot of puzzling questions arised from not only conflicts between the experimental observation and LQCD results, but also between the early phenomenological studies and the first principle LQCD simulations.

The change of situation was triggered by the high-statistics experimental data from BESIII. There have been a number of the $J/\psi$ exclusive decay channels measured with high precision, where contributions from $J^{PC}=0^{-+}$ can be clearly identified~\cite{BESIII:2012aa,Ablikim:2011pu,Ablikim:2010au}. It shows that in the vicinity of 1.4 GeV there is only one Breit-Wigner peak structure in the invariant mass spectrum for $\eta(1405/1475)$. Similar feature can be found in the radiative decays of $\psi'$. In the hadronic production channels, such as $e^+e^-\to \omega\eta\pi\pi$, $\phi\eta\pi\pi$, $\omega K\bar{K}\pi$, etc., there is also only one Breit-Wigner peak present in the invariant spectrum. The interesting observation is that the peak positions somehow are slightly different in exclusive channels. Further insights into this puzzling problem was gained from the measurement of isospin breaking effects into $J/\psi\to\gamma \eta(1405/1475)$ with $\eta(1405/1475)\to 3\pi$ at BESIII~\cite{BESIII:2012aa}, where the isospin breaking effects are found to be unexpectedly large, i.e. $\Gamma(\eta(1405)\to f_0\pi\to\pi^+\pi^-\pi^0)/\Gamma(\eta(1405)\to a_0\pi\to\eta\pi^0\pi^0)\sim 17.9\%$. This value is nearly one order of magnitude larger than that produced by the $a_0(980)-f_0(980)$ mixing. It was then discovered by the authors of Ref.~\cite{Wu:2011yx} that the significantly enhanced isospin breaking effects are caused by the so-called ``triangle singularity" (TS) mechanism. It was demonstrated in Ref.~\cite{Wu:2011yx} and later a detailed analysis~\cite{Wu:2012pg} that at the mass of $\eta(1405/1475)$ the non-vanishing coupling of the initial $\eta(1405/1475)$ to the intermediate $K^*\bar{K}+c.c.$ and then their rescatterings into $\pi f_0(980)$ by the exchange of a Kaon or anti-Kaon allow a perfect satisfaction of the TS condition. While the detailed analytical properties of the triangle diagrams will be discussed later, a simple way to picture the TS mechanism is that it corresponds to such a kinematic condition that all the internal states of the triangle loop can approach their on-shell condition simultaneously. As a consequence of such a leading singularity within the loop function, it will provide significant interferences in exclusive decays of $\eta(1405/1475)$ and produce the shift of peak positions of a single state in different channels and unexpectedly large isospin breaking effects in its decays into $3\pi$~\cite{Wu:2011yx,Wu:2012pg}. The dominance of the TS mechanism in $\eta(1405/1475)\to 3\pi$ was later confirmed by Ref.~\cite{Aceti:2012dj} in a chiral unitary approach.

Although the TS mechanism seems to be promising for understanding so-far all the existing puzzles about the $\eta(1405)$ and $\eta(1475)$ signals,  later according to Ref.~\cite{Achasov:2015uua}, the non-zero width of $K^*$ in the triangle loop integral can lead to significant suppressions of the decay rate. Therefore, the dominance of triangle diagrams in the isospin violated channel may become questionable. Implications of such a possibility suggests that the width effects due to the internal states should not be neglected and may lead to a significant impact on the role played by the TS mechanism. In order to clarify this, we carry out a coherent and quantitative investigation of the decays of $\eta(1405/1475)$ into $K\bar{K}\pi$, $\eta\pi\pi$ and $3\pi$ including the width effects in the TS mechanism.

As follows, in Sect. II we first introduce the TS conditions and discuss in detail the analytical properties of the triangle loop amplitude when non-zero $K^*$ width is considered. Then we will explore the decay patterns for $\eta(1405/1475)\to K\bar{K}\pi$, $\eta\pi\pi$ and $3\pi$, and clarify the role played by the TS mechanism in Sect. III. In particular, we will show that an additional transition process which enhances the $a_0(980)-f_0(980)$ mixing via the TS mechanism should contribute to $\eta(1405/1475)\to 3\pi$. The calculation results and discussions will be given in Sect. IV, and a conclusion will be given in Sect. V.

\section{Analytic properties of the triangle loop integral with non-zero widths}
\begin{figure}[H]
  \centering
  \includegraphics[width=2in]{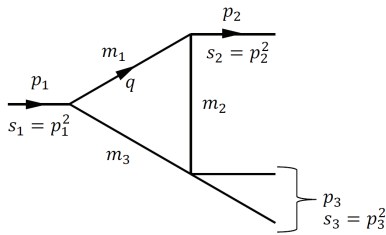}
  \caption{Typical triangle diagram with kinematic variables.}\label{fig-01}
\end{figure}

To understand the impact of unstable internal $K^*$ meson on the triangle loop integral of $\eta(1405/1475)$ (denoted by $\eta''$ in the following), we consider a typical triangle amplitude $I$ shown in Fig.~\ref{fig-01},
\begin{eqnarray}
I=-i\int\frac{d^4q}{(2\pi)^4}\frac{(2p_1-q)_{\mu}(-g^{\mu\nu}+\frac{q^{\mu}q^{\nu}}{q^2})(q-2p_2)_{\nu}}{(q^2-m_1^2+i m_1\Gamma)[(p_1-q)^2-m_2^2][(q-p_2)^2-m_3^2]},\label{eq1}
\end{eqnarray}
where the $m_1$, $m_2$ and $m_3$ are the masses for $K^*$, $K$ and $\bar{K}$, respectively. Since $K^*$ meson is unstable, a finite width $\Gamma$ has been introduced in its propagator. Due to the $P$-wave vertex and the polarization of $K^*$ meson, the amplitude $I$ is actually a tensor integral. However, this tensor integral can be reduced to a sum of scalar 3-point integral and some 2-point integrals~\cite{Achasov:2015uua}. By studying the analytical property of the 3-point and 2-point integrals, we can learn the property of the physical amplitude $I$. The TS condition applies to the physical amplitude $I$ where all the internal particles approach their on-shell condition simultaneously. In such a sense the reduction of $I$ into 3-point and 2-point loop integrals means that the manifestation of the physical TS contributions is given by the sum of the reduced loop integrals although some contributions are from the 2-point loops. Such a clarification is essential for the reason that we actually deal with the physical process instead of single loop integrals which are only part of the dynamics of the physical process. The kinematic condition that all the internal particles are on-shell determines the kinematics for all the reduced loops which have to be considered simultaneously. To be more specific, within the TS kinematics the two-body on shell condition in the 2-point integrals has been contained in the 3-body on-shell condition. With the above clarification, the TS contribution in this work is referred to the overall contributions from the physical integral $I$ instead of a reduced 3-point integral, and the influence of the finite width effects are also referred to its impact on the overall loop function.

Some more features about the TS loops should be pointed out before we proceed to the detailed analysis:
\begin{itemize}
\item The presence of the TS kinematics implies that the main contributions of the triangle loops come from the kinematic region near the on-shell condition for the internal particles. For physical processes where the internal motions of the internal particles can be treated non-relativistically the scalar triangle loop can be directly integrated out, and the leading logarithmic singularity can be explicitly extracted. In particular, for non-relativistic heavy meson loop transitions where the TS mechanism is present, the loop amplitudes can be analyzed in the non-relativistic effective field theory (NREFT) framework and a power-counting scheme can be established.

\item For loop transitions involving only light hadrons sometimes the non-relativistic approximation can hardly be justified. In such a case analysis of the triangle loop in the Mandelstam representation should be more appropriate. For most cases of the physical loop integrals, an empirical form factor has to be included to cut off ultraviolet divergence when the internal particles go off shell, which will inevitably introduce some model-dependence, although in general such an uncertainty can be under control. Even for convergent physical loops it is often checked that unphysical contributions from relatively large momentum transfers are reliably estimated and then removed~\cite{Xue:2017xpu}. For the physical triangle loop of Eq.~(\ref{eq1}), it converges with the choice of the vector propagator $(-g^{\mu\nu}+\frac{q^{\mu}q^{\nu}}{q^2})/(q^2-m_1^2+i m_1\Gamma)$ for the $K^* \ (\bar{K}^*)$. But in order to examine the sensitivities of the loop integrals to unphysical contributions from the ultraviolet region, we include a form factor ${\cal F}(q^2)$ and compare the numerical results for different cut-off parameters. The detailed discussions will be given later in Section III and IV.

\end{itemize}

\subsection{Analytical expression}
We first consider a typical scalar loop integral
\begin{eqnarray}
M&=&-i\int\frac{d^4q}{(2\pi)^4}\frac{1}{(q^2-m_1^2)((q-p_2)^2-m_2^2)((q-p_1)^2-m_3^2)} \ ,
\end{eqnarray}
where the notation is the same as that in Fig.~\ref{fig-01}.
Taking the Feynman parameterization, it can be expressed as
\begin{eqnarray}
M=-i\frac{d^4q}{(2\pi)^4}\int\int\int dx_1dx_2dx_3\frac{\delta(\Sigma_{i=1}^{3}x_i-1)}{D},
\end{eqnarray}
where $D$ is a homogeneous polynomial of $x_i$, i.e.
\begin{eqnarray}
D=\Sigma_{i=1}^{3}Y_{ij}x_ix_j, Y_{ij}=Y_{ji}.
\end{eqnarray}
In the case where the internal masses are real, the leading singularity of $M$ is determined by the Landau equation
\begin{eqnarray}
\frac{\partial D}{\partial x_i}=0\ \mbox{and}\ D=0.
\end{eqnarray}
This condition means that the extremum of $D$ touches 0. Physically, it means that the internal particles are stable and simultaneously become on-shell. The Landau equation gives $Y_{ij}x_j=0$. If this holds for $0<x_i<1$,  then there exist solutions for the Landau equation within the physical region which satisfies $\det{Y}=0$. This leads to the kinematic bounds for the triangle singularity.

In general, for the fixed external four-momentum squares $s_2$ and $s_3$ there are two solutions for $s_1$ which satisfy the Landau equation, i.e.
\begin{eqnarray}
s_1^-=(m_1+m_3)^2+\frac{1}{m_2^2}[(m_1^2+m_2^2-s_2)(s_3-m_2^2-m_3^2)-4m_2^2m_1m_3-\sqrt{\lambda[s_3,m_2^2,m_3^2]\lambda[s_2,m_1^2,m_2^2]}]\\
s_1^+=(m_1+m_3)^2+\frac{1}{m_2^2}[(m_1^2+m_2^2-s_2)(s_3-m_2^2-m_3^2)-4m_2^2m_1m_3+\sqrt{\lambda[s_3,m_2^2,m_3^2]\lambda[s_2,m_1^2,m_2^2]}],
\end{eqnarray}
where $\lambda[x,y,z]\equiv (x-y-z)^2-4yz$. Likewise, when $s_1$ and $s_2$ are fixed, we obtain two solutions for $s_3$, i.e.
\begin{eqnarray}
s_3^-=(m_2+m_3)^2+\frac{1}{m_1^2}[(m_1^2+m_2^2-s_2)(s_1-m_1^2-m_3^2)-4m_1^2m_2m_3-\sqrt{\lambda[s_1,m_1^2,m_3^2]\lambda[s_3,m_1^2,m_2^2]}]\\
s_3^+=(m_2+m_3)^2+\frac{1}{m_1^2}[(m_1^2+m_2^2-s_2)(s_1-m_1^2-m_3^2)-4m_1^2m_2m_3+\sqrt{\lambda[s_1,m_1^2,m_3^2]\lambda[s_3,m_1^2,m_2^2]}].
\end{eqnarray}

However, the conditions that $s_1=s_1^+$ with the fixed $s_2$ and $s_3$ or $s_3=s_3^+$ with the fixed $s_2$ and $s_1$ do not cause divergence because the corresponding solution $x_i$ is out of the region $(0,1)$. Only when $s_1$ ($s_3$) meets $s_1^-$ ($s_3^-$), the triangle singularity occurs within the physical region and can possibly produce detectable effects in experimental observables.

With the $s_2$ fixed (since it is an external particle), for any given $(m_2+m_3)^2<s_3<s_{3c}$, there is a $s_1^-$ lying within $((m_1+m_3)^2,s_{1c})$, where $s_{1c}$ and $s_{3c}$ are labeled as critical values for $s_1$ and $s_3$, beyond which the triangle singularity no longer exists. Taking the same notation as Ref.~\cite{Liu:2015taa}, these critical values are given by
\begin{eqnarray}
s_{1c}=(m_1+m_3)^2+\frac{m_3}{m_1}[(m_1-m_2)^2-s_2] \\
s_{3c}=(m_2+m_3)^2+\frac{m_3}{m_1}[(m_1-m_2)^2-s_2].
\end{eqnarray}

A caveat arising from the above discussion is that the internal particles are stable ones, i.e. they do not have a width in the propagator. In reality the nonvanishing coupling for $K^* (m_1)$ to pion ($s_2^{1/2}$) and kaon ($m_2$) demonds that the propagator for $K^*$ must contain an imaginary part. Therefore, a detailed investigation of the width effects in the TS mechanism is necessary and useful for a better understanding of the underlying dynamics.

To accommodate the width effects in the triangle loops, we consider complex masses for the internal particles. With the help of Spence function:
\begin{eqnarray}
Sp(z)&\equiv &-\int_0^1\frac{\ln{(1-zt)}}{t}dt,
\end{eqnarray}
which is also called dilogarithm function $Li_2$, as a special case for polylogarithm function $Li_n$ when $n=2$, an analytic expression of the transition matrix $M$ has been worked out by G.'t Hooft and M. Veltman~\cite{tHooft:1978jhc}:
\begin{eqnarray}
M&=&\frac{1}{16\pi^2}\int_0^1dy\frac{1}{N_1(y)}\{\ln{u_1(y)}-\ln{u_1(y^{(1)}_0)}\}\nonumber\\
&-&\frac{1}{16\pi^2}\int_0^1dy\frac{1}{N_2(y)}\{\ln{u_2(y)}-\ln{u_2(y^{(2)}_0)}\}\nonumber\\
&+&\frac{1}{16\pi^2}\int_0^1dy\frac{1}{N_3(y)}\{\ln{u_3(y)}-\ln{u_3(y^{(3)}_0)}\}\label{ampanalytic0}\\
&=&\frac{1}{16\pi^2}\frac{1}{c+2b\alpha}[ S(-\frac{d+e\alpha+2a+c\alpha}{c+2b\alpha},b,c+e,a+d+f)\nonumber\\
&&-S(-\frac{d+e\alpha}{(1-\alpha)(c+2b\alpha)},a+b+c,e+d,f)+S(\frac{d+e\alpha}{\alpha(c+2b\alpha)},a,d,f)],\label{ampanalytic}
\end{eqnarray}
where functions $u_i(y)$ have the following expressions:
\begin{eqnarray}
u_1(y)&\equiv & by^2+(c+e)y+a+d+f=s_3y^2+(m_2^2-m_3^2-s_3)y+m_3^2\\
u_2(y)&\equiv &(a+b+c)y^2+(e+d)y+f=s_2y^2+(m_2^2-m_1^2-s_2)y+m_1^2\\
u_3(y)&\equiv &ay^2+dy+f=s_1y^2+(m_3^2-m_1^2-s_1)y+m_1^2 \ ,
\end{eqnarray}
and $a, \ b, \ c, \ d, \ e, \ f$, and $\alpha$ are kinematic variables:
\begin{eqnarray}
&&a\equiv s_1, b\equiv s_3, c\equiv s_2-s_1-s_3, d\equiv m_3^2-m_1^2-s_1, e\equiv s_1-s_2+m_2^2-m_3^2, f\equiv m_1^2\\
&&\alpha\equiv \frac{-c\pm\sqrt{c^2-4ab}}{2b}=\frac{s_3+s_1-s_2}{2s_3}\pm\frac{\sqrt{\lambda[s_1,s_2,s_3]}}{2s_3}\\
&&c+2b\alpha=\pm\sqrt{\lambda[s_1,s_2,s_3]}
\end{eqnarray}
This analytic expression is valid for both real and complex internal masses. In Eq.~(\ref{ampanalytic0}) $N_i(y)$ are functions of the integration variable $y$:
\begin{eqnarray}
N_1(y)&\equiv &(c+2b\alpha)y+d+e\alpha+2a+c\alpha\\
N_2(y)&\equiv &(1-\alpha)(c+2b\alpha) y +d+e\alpha\\
N_3(y)&\equiv &-(c+2b\alpha)\alpha y+d+e\alpha,
\end{eqnarray}
and $y^{(i)}_0$ denotes the value of $y$ when $N_i=0$.

The function $S$ in Eq.~(\ref{ampanalytic}) can be written in terms of the Spence function with characteristic structures:
\begin{eqnarray}
S(y_0,a,b,c)&\equiv &\int_0^1dy\frac{1}{y-y_0}[\ln{(ay^2+by+c)}-\ln{(ay_0^2+by_0+c)}]\nonumber\\
&=&R(y_0,y_1)+R(y_0,y_2)\nonumber\\
&+&\left[\eta(1-y_1,1-y_2)-\eta(y_0-y_1,y_0-y_2)+\eta\left(a+Im[\frac{c}{a}],\frac{1}{a+Im[ay_0^2+by_0+c]}\right)\right]\ln{\frac{y_0-1}{y_0}}\label{defSi}
\end{eqnarray}
where
\begin{equation}
R(y_0,y_1)\equiv \int_0^1dy\frac{1}{y-y_0}[\ln{(y-y_1)}-\ln{(y_0-y_1)}] ,
\end{equation}
with
\begin{eqnarray}
y_1\equiv \frac{-b-\sqrt{b^2-4ac}}{2a}, \ y_2\equiv \frac{-b+\sqrt{b^2-4ac}}{2a}.
\end{eqnarray}
The $\eta$ function arises from
\begin{eqnarray}\label{lndecompose}
\ln{(z_1z_2)}=\ln{z_1}+\ln{z_2}+\eta(z_1,z_2),
\end{eqnarray}
with the argument in $\ln$ limited in $(-\pi,\pi)$.

The following features with the TS kinematics will help understand better the analytical properties of the triangle loop amplitude:
\begin{itemize}

\item In the vicinity of the TS kinematics the main contributions of the transition amplitude are given by the absorptive part. In particular, when the internal masses are real, the logarithmic divergence of the TS manifests itself in the absorptive part of scalar integral. So we will focus on the absorptive part in the analysis.

\item For the case that the internal masses are all real, the absorptive part can be derived analytically according to the Cutkosky rule. This allows us to examine the width effects on the triangle loop amplitude by comparing them with the Cutkosky rule result. In particular, the width dependence can be highlighted in the absorptive part of the amplitude.

\item For the physical case, namely the isospin-violating transition $\eta(1405/1475)\to f_0(980)\pi$, it should be noted that the dispersive part becomes negligible due to the cancellation between the charged and neutral triangle loop amplitudes. This actually leads to a rather model-independent behavior of the TS contributions to the isospin violations in $\eta(1405/1475)\to f_0(980)\pi$~\footnote{Note that even though form factors are often introduced to cut off divergence in the loop integral, in the TS kinematics the dependence of the form factors is relatively small due to the small virtuality for the coupling vertices. For the isospin-violating decay of $\eta(1405/1475)\to f_0(980)\pi$ the cancellation between the charged and neutral loop amplitudes will further reduce the model dependence. Detailed discussion and demonstration of such a consequence has been provided in Refs.~\cite{Wu:2011yx,Wu:2012pg}. }. As stressed at the beginning, the only thing left behind is the width effect that should be quantified by explicit and self-consistent calculations.

\end{itemize}
In the following sections we focus on the derivation of absorptive part of the scalar integral under the influence of the finite width of the intermediate states. We will provide detail analysis of $\Im{m_1}<0$ and $m_2, m_3\in R$, for the reason that the physical widths of the $K \ (\bar{K})$ by weak decays are much smaller than that of the $K^*$ meson. However, it is checked in the end that the same analytic expression is still valid when any of the internal state has a complex mass.

\subsection{The motion of singularities}

For convenience, we express the amplitude of Eq.~(\ref{ampanalytic}) in a concise form
\begin{eqnarray}
&&M=\frac{1}{16\pi^2}\frac{1}{c+2b\alpha}(S^{(1)}-S^{(2)}+S^{(3)}),\label{trans-simplify}\\
&&S^{(i)}=\Sigma_{j=1}^2R^{(i)}_j+\sigma^{(i)},\label{func-S-simplify}\\
&&R^{(i)}_j=Sp(z^{(i)}_{j1})-Sp(z^{(i)}_{j2})+T^{(i)}_j\equiv W^{(i)}_j+T^{(i)}_j,\label{func-R-simplify}\\
&&z^{(i)}_{1k}=\frac{y^{(i)}_k-1}{y^{(i)}_k-y^{(i)}_0}, \ \ \ z^{(i)}_{2k}=\frac{y^{(i)}_k}{y^{(i)}_k-y^{(i)}_0},\label{z-definition}
\end{eqnarray}
with
\begin{eqnarray}
&&S^{(1)}\equiv S\left(-\frac{d+e\alpha+2a+c\alpha}{c+2b\alpha},b,c+e,a+d+f\right),\\
&&S^{(2)}\equiv S\left(-\frac{d+e\alpha}{(1-\alpha)(c+2b\alpha)},a+b+c,e+d,f\right),\\
&&S^{(3)}\equiv S\left(\frac{d+e\alpha}{\alpha(c+2b\alpha)},a,d,f\right),\\
&&R^{(i)}_j\equiv R(y^{(i)}_0,y^{(i)}_j).
\end{eqnarray}
Our task in this subsection is to extract the imaginary part of the loop amplitude and investigate the movement of the singular kinematics manifested by the location of $y^{(i)}_k$ in the complex plane. The general condition for the TS requires that the following kinematic constraints are satisfied, i.e. $s_1>(m_1+m_3)^2$, $s_3>(m_2+m_3)^2$, and $\sqrt{s_2}<m_1-m_2$.
Moreover, the maximum allowed value for $s_1$ or $s_3$ is generally very close to the normal threshold. It allow us to make a substitution of $s_3$ by $(m_2+m_3)^2$ in some steps as a reasonable approximation. Therefore, in the following discussion of the finite width effects on the imaginary part of the amplitude, we can apply this approximation to simplify the deduction without loss of accuracy. The numerical result of this approximation compared with the exact one will be discussed in the end of this section.

To proceed, we will start with $W^{(i)}_j$ of which the imaginary part depends on the positions of $z^{(i)}_{jk}$ on the complex plane. The latter will then rely on the locations of $y^{(i)}_k$ and $y^{(i)}_0$ as given by Eq.~(\ref{z-definition}). Therefore, the motion of the TS can be illustrated by tracing the locations of $y^{(i)}_k$, where $y^{(i)}_k \ (k=1, \ 2)$ are defined as the roots of $u_i$, and we define the $y^{(i)}_1$ is the larger one of the two roots. In the case of finite $\Gamma$, $y^{(i)}_k$ may deviate slightly from real axis which will be our focus in this work.

According to the definition of $S^{(i)}$ (Eq.~\ref{defSi}), the divergence of $S^{(i)}$ occurs when $y^{(i)}_j=y^{(i)}_0$, i.e. the denominator of $z^{(i)}_{jk}=0$. Taking into account that $y^{(i)}_0$ and $y^{(i)}_j$ are complex functions of $s_1$ and $s_3$, and for sufficiently small width $\Gamma$ all the $\Gamma$ dependent terms only contribute to the imaginary parts of $y^{(i)}_j$ and $y^{(i)}_0$, the TS condition actually corresponds to $\Re{y^{(i)}_j}=\Re{y^{(i)}_0}$ and the imaginary part of the complex mass will push $y^{(i)}_j$ and $y^{(i)}_0$ away from the real axis.

In order to discuss how $y^{(i)}_k$ moves when the $K^*$ meson has a small but finite width, we need to first write down the explicit expression of $y^{(i)}_k$ as the roots of function $u_i(y)$ and then expand the $y^{(i)}_k$ with respect to $m_1^2$ (the mass square of the $K^*$) to the first order of $\Gamma$.

Since $u_1(y)$ has nothing to do with the complex mass $m_1$, $y^{(1)}_k$ is not affected by $\Gamma$. Thus, their locations on the real axis are
\begin{eqnarray}
y^{(1)}_{1}&=&\frac{1}{2s_3}[s_3+m_3^2-m_2^2+\sqrt{(m_2^2-m_3^2-s_3)^2-4s_3m_3^2}]=\frac{E^{(3)}_3+p^{(3)}_3}{\sqrt{s_3}}\label{y11real}\\
y^{(1)}_{2}&=&\frac{1}{2s_3}[s_3+m_3^2-m_2^2-\sqrt{(m_2^2-m_3^2-s_3)^2-4s_3m_3^2}]=\frac{E^{(3)}_3-p^{(3)}_3}{\sqrt{s_3}} \ . \label{y12real}
\end{eqnarray}
The only difference between $y^{(i)}_1$ and $y^{(i)}_2$ is the sign in front of their square roots. For simplicity we just need to show the expressions for $y^{(i)}_1$. The locations of $y^{(2)}_1$ and $y^{(3)}_1$ as the larger root of $u_2(y)$ and $u_3(y)$, respectively, are
\begin{eqnarray}
y^{(2)}_1&=&\frac{1}{2s_2}[(s_2+m_1^2-m_2^2)+\sqrt{(s_2+m_1^2-m_2^2)^2-4s_2m_1^2}]=\frac{E^{(2)}_1+p^{(2)}_1}{\sqrt{s_2}}\\
y^{(3)}_1&=&\frac{1}{2s_1}[(s_1+m_1^2-m_3^2)+\sqrt{(s_1+m_1^2-m_3^2)^2-4s_1m_1^2}]=\frac{E^{(1)}_1+p^{(1)}_1}{\sqrt{s_1}}.
\end{eqnarray}
Hence the variations of the locations of $y^{(2)}_1$ and $y^{(3)}_1$ in association with the presence of the finite width $\Gamma$ for the $K^*$  can be examined by
\begin{eqnarray}
-im_1\Gamma\frac{dy^{(2)}_1}{dm_1^2}&=&\frac{-im_1\Gamma}{2s_2}\left(1-\frac{E^{(2)}_2}{p^{(2)}_2}\right)\\
-im_1\Gamma\frac{dy^{(3)}_1}{dm_1^2}&=&\frac{-im_1\Gamma}{2s_1}\left(1-\frac{E^{(1)}_3}{p^{(1)}_3}\right).
\end{eqnarray}
Similarly, we have
\begin{eqnarray}
-im_1\Gamma\frac{dy^{(2)}_2}{dm_1^2}&=&\frac{-im_1\Gamma}{2s_2}\left(1+\frac{E^{(2)}_2}{p^{(2)}_2}\right)\label{dy22}\\
-im_1\Gamma\frac{dy^{(3)}_2}{dm_1^2}&=&\frac{-im_1\Gamma}{2s_1}\left(1+\frac{E^{(1)}_3}{p^{(1)}_3}\right) \ ,
\end{eqnarray}
where the kinematic variables are defined as follows:
\begin{eqnarray}
E^{(1)}_1&=&\frac{s_1+m_1^2-m_3^2}{2\sqrt{s_1}},\ E^{(1)}_3=\frac{s_1+m_3^2-m_1^2}{2\sqrt{s_1}},\nonumber\\
E^{(2)}_1&=&\frac{s_2+m_1^2-m_2^2}{2\sqrt{s_2}},\
E^{(2)}_2=\frac{s_2+m_2^2-m_1^2}{2\sqrt{s_2}},\
E^{(3)}_3=\frac{s_3+m_3^2-m_2^2}{2\sqrt{s_3}}, \nonumber\\
p^{(1)}_1&=&\frac{\lambda[s_1,m_1^2,m_3^2]^{\frac{1}{2}}}{2\sqrt{s_1}}, \ p^{(1)}_3=p^{(1)}_1,\nonumber\\ p^{(2)}_1&=&\frac{\lambda[m_1^2,m_2^2,s_2]^{\frac{1}{2}}}{2\sqrt{s_2}},\
p^{(2)}_2=p^{(2)}_1,\nonumber\\
p^{(3)}_3&=&\frac{\lambda[s_3,m_2^2,m_3^2]^{\frac{1}{2}}}{2\sqrt{s_3}} \ .
\end{eqnarray}
It should be noted that within the region $s_2<(m_1-m_2)^2$, the value of $E^{(2)}_2/p^{(2)}_2+1$ is always negative, which means that $y^{(2)}_2$ actually moves upward away from the real axis and $y^{(2)}_1$ moves downward away from the real axis with the increasing $\Gamma$. These expansions give the sign for the imaginary parts, according to which we can conclude that both $y^{(3)}_1$ and $y^{(2)}_2$ move upward and that both $y^{(2)}_1$ and $y^{(3)}_2$ moves downward under the influence of the increasing $\Gamma$.

Speaking of the real part of $y^{(i)}_k$, for $m_3>0$, $m_2>0$ and $s_3>(m_2+m_3)^2$, it can be proved that $0<y^{(1)}_2<y^{(1)}_1<1$. For $m_1>m_2>0$ and $s_2>0$, it can be easily verified that $y^{(2)}_1>y^{(2)}_2>1$. For $m_1>m_3$ and $s_1>(m_1+m_3)^2$, it can also be proved that $0<y^{(3)}_2<y^{(3)}_1<1$. Hence the locations of $y^{(i)}_j$ are quite clear, and we are to take a look at $y^{(i)}_0$. The real part of $y^{(i)}_0$ is slightly complicated. They are
\begin{eqnarray}
y^{(1)}_0&=&-\frac{1}{\lambda[s_1,s_2,s_3]^{1/2}}\left[-m_1^2+m_3^2+s_1+\frac{(m_2^2-m_3^2-s_3)(s_1-s_2+s_3+\lambda[s_1,s_2,s_3]^{1/2})}{2s_3}\right]\label{realy10}\nonumber\\
&=&-\frac{1}{\sqrt{s_1}p^1_{s_3}}[\sqrt{s_1}E^{(1)}_3-E^{(3)}_3(E^{(3)}_{s_1}+p^{(3)}_{s_1})]\label{y10real}\\
y^{(2)}_0&=&\frac{2s_3}{\lambda[s_1,s_2,s_3]^{1/2}(s_1-s_2-s_3+\lambda[s_1,s_2,s_3]^{1/2})}\nonumber\\
&&\times\left[-s_1-m_1^2+m_3^2+\frac{(m_2^2-m_3^2+s_1-s_2)(s_1-s_2+s_3+\lambda[s_1,s_2,s_3]^{1/2})}{2s_3}\right]\nonumber\\
&=&\frac{1}{2}\frac{1}{p^3_{s_2}(-E^3_{s_2}+p^3_{s_2})}[-2\sqrt{s_1}E^{(1)}_1+\frac{s_3+s_1-s_2-s_3+m_2^2-m_3^2}{\sqrt{s_3}}(E^{(3)}_{s_1}+p^{(3)}_{s_1})]\nonumber\\
&=&\frac{1}{p^3_{s_2}(-E^3_{s_2}+p^3_{s_2})}[-\sqrt{s_1}E^{(1)}_1+(E^{(3)}_{s_1}-E^{(3)}_3)(E^{(3)}_{s_1}+p^{(3)}_{s_1})]\\
y^{(3)}_0&=&\frac{2s_3}{\lambda[s_1,s_2,s_3]^{1/2}(s_1-s_2+s_3+\lambda[s_1,s_2,s_3]^{1/2})}\nonumber\\
&&\times\left[-s_1-m_1^2+m_3^2+\frac{(m_2^2-m_3^2+s_1-s_2)(s_1-s_2+s_3+\lambda[s_1,s_2,s_3]^{1/2})}{2s_3}\right]\nonumber\\
&=&\frac{1}{p^{(3)}_{s_1}(-E^{(3)}_{s_1}+p^{(3)}_{s_1})}[-\sqrt{s_1}E^{(1)}_1+(E^{(3)}_{s_1}-E^{(3)}_3)(E^{(3)}_{s_1}+p^{(3)}_{s_1})]\label{realy30},
\end{eqnarray}
but the derivatives, which can indicate their motion on the complex plane, are much simpler:
\begin{eqnarray}
-im_1\Gamma\frac{dy^1_0}{dm_1^2}&=&\frac{-im_1\Gamma}{[(s_2-s_1-s_3)^2-4s_1s_3]^{1/2}}=\frac{-im_1\Gamma}{2\sqrt{s_1}p^{(1)}_{s_3}}\\
-im_1\Gamma\frac{dy^2_0}{dm_1^2}&=&\frac{2 i s_3 m_1\Gamma}{[(s_2-s_1-s_3)^2-4s_1s_3]^{1/2}[s_1-s_2-s_3+\sqrt{(s_2-s_1-s_3)^2-4s_1s_3}]}\nonumber\\
&=&\frac{im_1\Gamma}{2\sqrt{s_1}p^{(1)}_{s_3}(-1+\frac{E^{(3)}_{s_1}+p^{(3)}_{s_1}}{\sqrt{s_3}})}\label{dy20}\\
-im_1\Gamma\frac{dy^3_0}{dm_1^2}&=&\frac{2is_3m_1\Gamma}{[(s_2-s_1-s_3)^2-4s_1s_3]^{1/2}[s_1-s_2+s_3+\sqrt{(s_2-s_1-s_3)^2-4s_1s_3}]}\nonumber\\
&=&\frac{im_1\Gamma}{2p^{(3)}_{s_1}(E^{(3)}_{s_1}+p^{(3)}_{s_1})},
\end{eqnarray}
where
\begin{eqnarray}
E^{(3)}_{s_1}=\frac{s_3+s_1-s_2}{2\sqrt{s_3}},\ p^{(3)}_{s_1}=\frac{\lambda[s_1,s_2,s_3]^{\frac{1}{2}}}{2\sqrt{s_3}},\ p^{(1)}_{s_3}=\frac{\lambda[s_1,s_2,s_3]^{\frac{1}{2}}}{2\sqrt{s_1}} \ .
\end{eqnarray}

With the explicit expressions for $y^{(i)}_0$ and $y^{(i)}_j$ we are now ready to explore how the TS moves under the influence of the finite width from the intermediate state $m_1$. As mentioned earlier, the TS occurs when the real parts of $y^{(i)}_0$ and $y^{(i)}_j$ equal to each other and function $S^{(i)}$ (Eq.~(\ref{defSi})) will become divergent.

\begin{figure}
  \centering
  \includegraphics[width=3.2in]{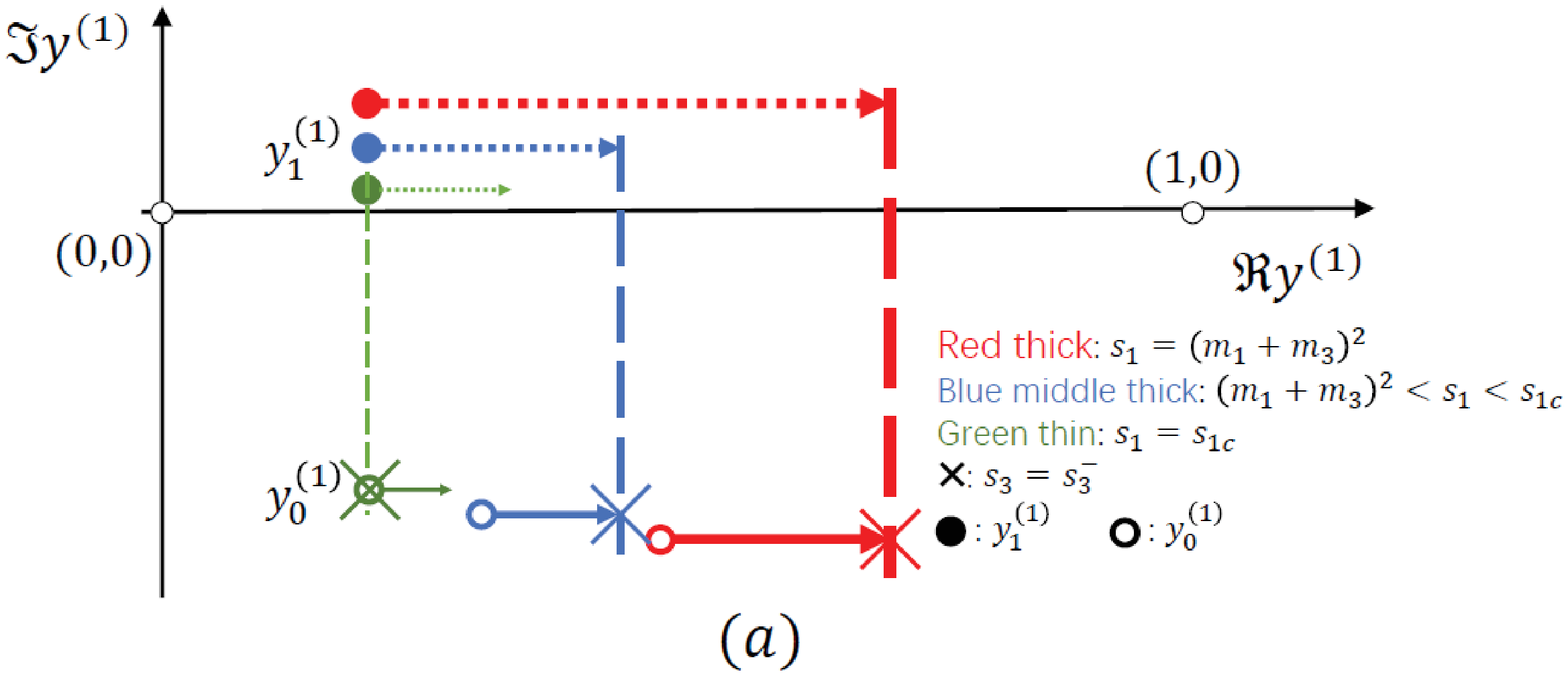}\\
  \includegraphics[width=3.3in]{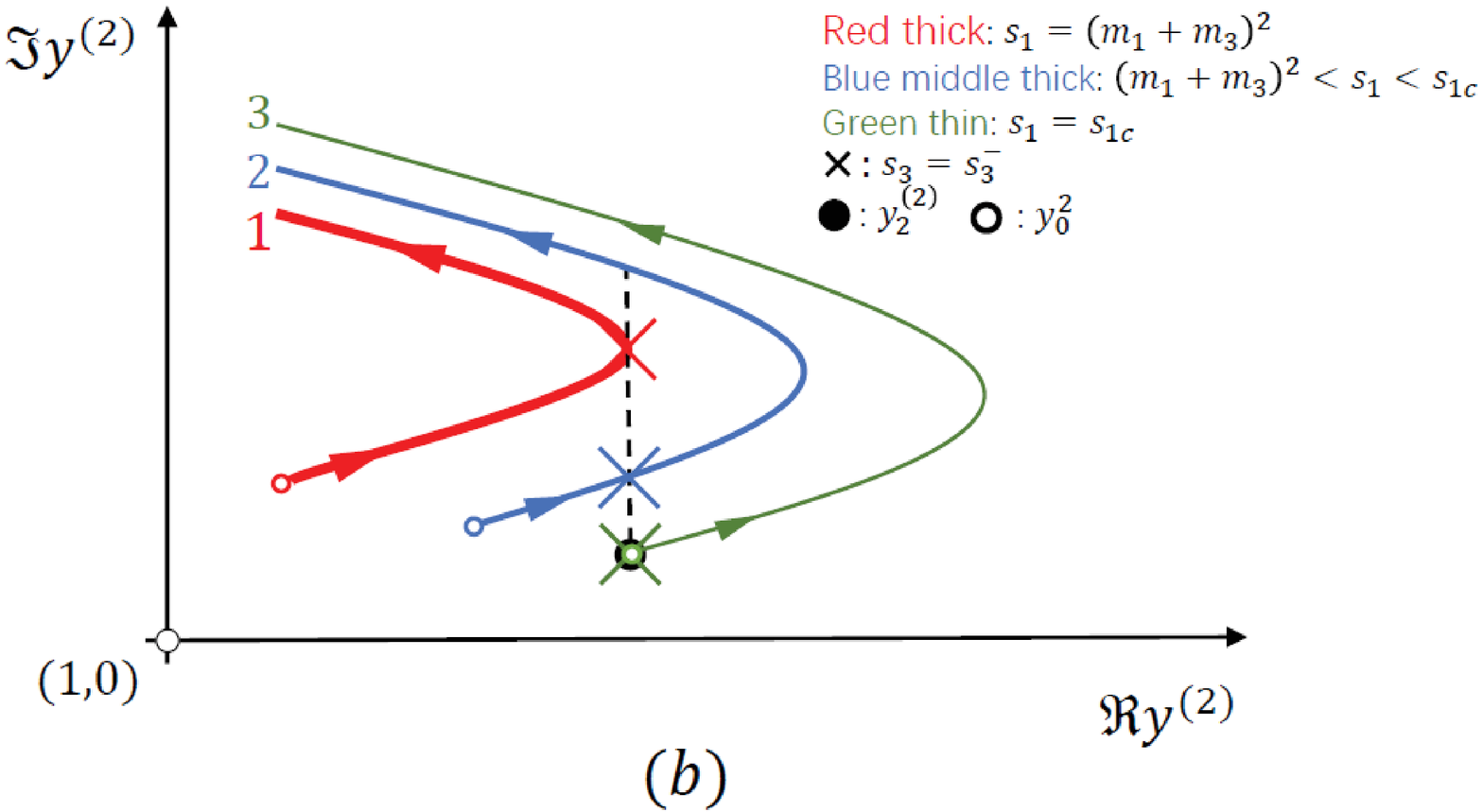}\\
  \includegraphics[width=3.2in]{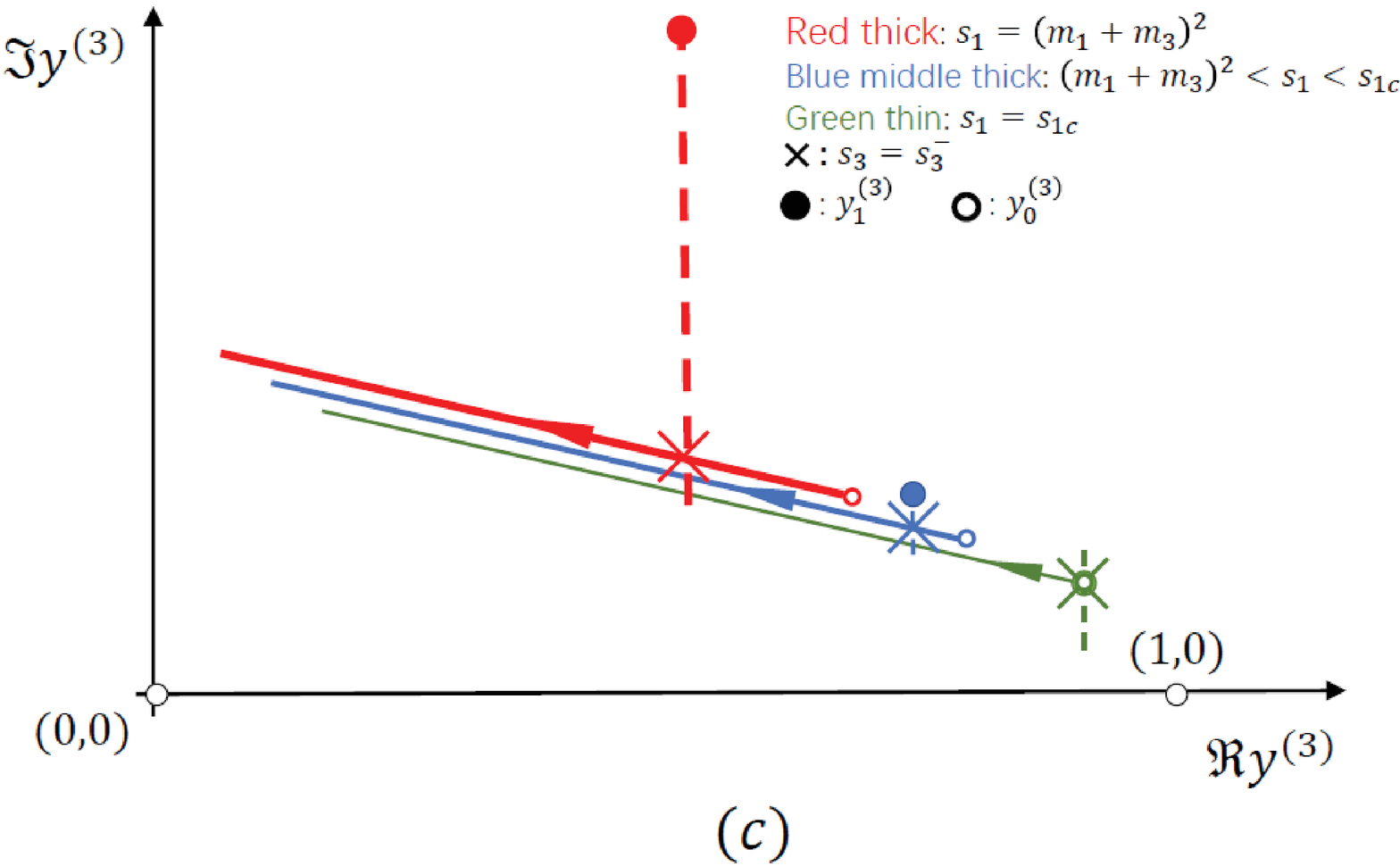}
  \caption{The motions of functions $y_0^{(i)}$ and $y_j^{(i)}$ ($j=1, \ 2$, and $i=1, \ 2, \ 3$) on the complex plane indicating the TS conditions via the Spence function variables $z_{1j}^{(i)}$ and $z_{2j}^{(i)}$. The detailed descriptions of these three pairing functions are given in the context. For a given value of $s_1$, the motions of the $y^{(i)}_0$ and $y^{(i)}_j$ with the variation of $s_3$ from the normal threshold $s_3=(m_2+m_3)^2$ to $s_3^-$ (Note $s_3^-=s_{3c}$ if $s_1=(m_1+m_3)^2$) are plotted. The dashed vertical lines indicate the TS conditions that the pairing functions overlap horizontally with $s_3=s_3^-$ for any $s_1\in [(m_1+m_3)^2,s_{1c}]$.
%with the regime between the solid dots (in (a) the solid dots will move towards the dashed line and match the value of $\Re{y^{(1)}_0}$) and crosses (indicating the positions where $\Re{y^{(1)}_1}=y^{(1)}_1=\Re{y^{(1)}_0}$) within the physical boundary.
%
%
%  In these three plots the solid dots denote $y_k^{(i)}$ and the solid circles denote $y_0^{(i)}$. These three plots correspond to the three pairings of $y_0^{(i)}$ and $y_k^{(i)}$.   The red colour represents that $s_1=(m_1+m_3)^2$. The blue colour represents that $(m_1+m_3)^2<s_1<s_{1c}$ and the green colour denotes that $s_1=s_{1c}$. The hollow circles denote the initial locations of $y^{(i)}_0$'s, at $s_3=(m_2+m_3)^2$. The solid lines with arrows denote the trajectories of $y^{(i)}_0$ as $s_3$ increases from $(m_2+m_3)^2$. The locations of $y^{(1)}_1$, $y^{(2)}_2$ and $y^{(3)}_1$ are marked by the solid circles. In (a), since the location of $y^{(1)}_1$ is moving as $s_3$ increases, we use dotted lines with arrows to show the trajectory. The dashed lines marks the occurrence of horizontal overlapping, which corresponds to the condition that $s_3=s_3^-$ for any $s_1\in [(m_1+m_3)^2,s_{1c}]$.
}\label{fig-02}
\end{figure}

In Fig.~\ref{fig-02} for a given value of $s_1$ within the physical range, the motions of the $y^{(i)}_0$ and $y^{(i)}_j$ with the variation of $s_3$ from the normal threshold $s_3=(m_2+m_3)^2$ to $s_3^-$ (Note $s_3^-=s_{3c}$ if $s_1=(m_1+m_3)^2$) are plotted for $y^{(i)}_0$ and $y^{(i)}_j$. Each of such trajectories is illustrated by a set of solid dots (with dotted arrows in (a)) and empty circles (with solid arrows). The dashed vertical lines indicate the fulfilled TS conditions with the regime between the solid dots (in (a) the solid dots will move towards the dashed line and match the value of $\Re{y^{(1)}_0}$) and crosses (indicating the positions where $\Re{y^{(1)}_1}=y^{(1)}_1=\Re{y^{(1)}_0}$) is within the physical boundary. Figure~\ref{fig-02} describes the following situations:
\begin{itemize}
\item In each plot the thick (red) arrow lines (solid and dotted) indicate the starting point of the trajectories of $y^{(i)}_j$ and $y^{(i)}_0$ when $s_3$ varies continuously from $(m_2+m_3)^2$ to $s_3^-$, while the corresponding $s_1$ of this red line is fixed at $s_1=(m_1+m_3)^2$.  When $s_3$ reaches $s_{3c}$ (note again, $s_3^-=s_{3c}$ if $s_1=(m_1+m_3)^2$), these two points overlap horizontally which is marked by the thick (red) dashed vertical line.

\item The middle-size (blue) arrow lines (solid and dotted) indicate the trajectories when $(m_1+m_3)^2<s_1<s_{1c}$.

\item The thin (green) arrow lines (solid and dotted) indicate the situation with $s_1=s_{1c}$. Then, $y^{(i)}_j$ and $y^{(i)}_0$ will have the same real parts and sit at the bound of the TS condition. With the increase of $s_1> s_{1c}$, these pairing functions in each plot are no longer overlapping and the kinematics move outside of the TS regime.

\item If $s_1<(m_1+m_3)^2$, as $s_3$ increases from $(m_2+m_3)^2$, although the starting point of $y^{(1)}_0$ is to the right of $y^{(1)}_1$, the $y^{(1)}_1$ would not be able to catch up with the $y^{(1)}_0$ before they get the same speed. Therefore, the valid region for $s_1$ where the horizontal overlapping can happen is $(m_1+m_3)^2<s_1<s_{1c}$. The crosses mark the positions of all possible overlaps between the pairing $y^{(i)}_j$ and $y^{(i)}_0$ for a given $s_1$. Once these two pairing functions meet these crosses simultaneously, namely, fulfill the TS condition, the Spence functions will be enhanced.
\end{itemize}

To be more specific, Fig.~\ref{fig-02}(a) illustrates the motions of $y^{(1)}_0$ and $y^{(1)}_1$. Note that $y^{(1)}_1$ is independent of $m_1$, thus, its location in the real axis only varies with $s_3$. The TS condition is fulfilled with the matching of $\Re{y^{(1)}_1}=y^{(1)}_1=\Re{y^{(1)}_0}$ which are indicated by vertical dashed lines with crosses.

In Fig.~\ref{fig-02} (b) the motion of $y^{(2)}_0$ and $y^{(2)}_2$ is illustrated. In this case the location of $y^{(2)}_2$ does not rely on $s_1$. It is marked by a solid (black) dot. When $s_1$ is fixed at $(m_1+m_3)^2$, as $s_3$ varies from $(m_2+m_3)^2$, the trajectory of $y^{(2)}_0$ is represented by the thick (red) solid line with the empty (red) circle as the starting point. At this kinematic point, the horizontal overlapping of $y^{(2)}_0$ and $y^{(2)}_2$ only occurs when $s_3=s_{3c}$, which is marked by the (red) cross. If $s_1$ is set to be a larger value, the trajectory of $y^{(2)}_0$ (noted by the middle-thick (blue) solid line with arrow) may overlap horizontally with $y^{(2)}_0$ twice, but only the first overlap is responsible for the TS. The first overlap occurs exactly when $s_3=s_3^-$, and it is marked by the (blue) cross. If $s_1$ is fixed at $s_{1c}$, only when $s_3=(m_2+m_3)^2$ can the $y^{(2)}_0$ and the $y^{(2)}_2$ overlap as indicated by the thin (green) solid line with arrow. The upper overlap does not correspond to the TS eventually. Therefore, the valid range of the TS corresponds to the line segment from the (red) cross on the thick (red) solid line to the (green) one on the thin (green) solid line.

The motions of $y^{(3)}_0$ and $y^{(3)}_1$ are plotted in Fig.~\ref{fig-02} (c). The condition that $s_1=(m_1+m_3)^2$ is also marked by a set of dots (solid and empty) with a thick (red) arrow line. The vertical dashed lines with crosses mark the TS condition when $s_3$ varies. In this case, as $s_3$ increases from the normal threshold, the empty (red) circle ($y^{(3)}_0$) moves along the thick (red) arrow and horizontally overlaps with the solid (red) dot ($y^{(3)}_1$). If $s_1$ is some value between $(m_1+m_3)^2$ and $s_{1c}$, the location of $y^{(3)}_1$ is represented by the solid dot in the middle (blue), and the trajectory of $y^{(3)}_0$ as $s_3$ increases from $(m_2+m_3)^2$ is shown by the middle-think (blue) solid line with arrow. The empty (blue) circle is the corresponding starting point. When $s_3$ reaches $s_3^-$, the $y^{(3)}_0$ and $y^{(3)}_1$ overlap horizontally, which is marked by the middle (blue) cross. However, if $s_1=s_{1c}$, only when $s_3=(m_2+m_3)^2$ can the $y^{(3)}_0$ and $y^{(3)}_1$ overlap. In such a case, the position of $y^{(3)}_1$ is coincident with the starting point of $y^{(3)}_0$, which is marked by the right (green) cross. If $s_1$ is smaller than the normal threshold, the $y^{(3)}_1$ will rapidly run away from the real axis, so that there is no longer enhancement. Therefore, the valid region for the occurrence of the TS ($s_1=s_1^-, s_3=s_3^-$) corresponds to the line segment from the right (green) cross to the left (red) one.

A crucial fact is that when $s_1=s_1^-$, the overlaps of $y^{(1)}_0$ and $y^{(1)}_1$, of $y^{(2)}_0$ and $y^{(2)}_2$, and of  $y^{(3)}_0$ and $y^{(3)}_1$ occur simultaneously, which means that $S^{(1)}$, $S^{(2)}$ and $S^{(3)}$ simultaneously become enhanced (or maximized). This exactly corresponds to the TS condition for the triangle loop.

There may be some other horizontal overlaps, which however, are not responsible for the TS. They happen either when $s_1=s_1^+$ or when $s_3=s_3^+$, and these effects eventually cancel out when sum over all the $S^{(i)}$ in the final result of $\Im{M}$. We do not show these in Fig.~\ref{fig-02} since they are not the focus of this analysis. However, the motions of $y^{(i)}_j$ and $y^{(i)}_0$ in these energy regions are useful for extracting the absorptive part. As a brief summary for our analysis based on the Spence function,  the occurrence of TS is recognized as the simultaneous (horizontally) overlaps of these singular points on the complex plane.

\subsection{The imaginary part of $W^{(i)}_j$}
To extract the imaginary part from the Spence functions, we adopt the following formula
\begin{eqnarray}
\Im{Sp(a\pm ib)}=\pm\pi\Theta(a-1)\ln{a}\mp\int_0^1\frac{dt}{t}\arctan{\frac{bt}{at-1}},\label{imsp}
\end{eqnarray}
where $a$ is a real number and $b$ is a positive real number. If $b$ is much smaller than $a$, the second term is to the order of $b$ and can be neglected. In this case, we obtain the following approximation
\begin{eqnarray}
\Im{Sp(a\pm ib)}=\pm\pi\Theta(a-1)\ln{a}.
\end{eqnarray}

We first discuss the sign of the real and imaginary part of $z^{(i)}_{jk}$. One should keep in mind that only the $z^{(i)}_{jk}$ with positive real part can contribute to the imaginary part of the Spence function, and then to the imaginary part of $W^{(i)}_j$. The sign of the imaginary part of $z^{(i)}_{jk}$ can be determined by the relative positions of $y^{(i)}_k$ and $y^{(i)}_0$. For instance, considering $W^{(1)}_1$
\begin{eqnarray}
\Im{W^{(1)}_1}=\Im{[Sp(z^{(1)}_{11})-Sp(z^{(1)}_{21})]},
\end{eqnarray}
where
\begin{eqnarray}
z^{(1)}_{11}=\frac{y^{(1)}_1-1}{y^{(1)}_1-y^{(1)}_0}, \ z^{(1)}_{21}=\frac{y^{(1)}_1}{y^{(1)}_1-y^{(1)}_0}.
\end{eqnarray}
In the energy region $s_1<s_1^-$, since $y^{(1)}_1\in R$ and $0<y^{(1)}_1<1$, the numerator of $z^{(1)}_{11}$ has a negative real part. Since $y^{(1)}_0$ moves downward under the influence of $\Gamma$, the denominator of $z^{(1)}_{11}$ has a small and positive imaginary part. Thus, in this region for $s_1$, the denominator of $z^{(1)}_{11}$ has an argument being slightly smaller than $\pi$. Consequently, $z^{(1)}_{11}$ has a small positive imaginary part, which results in $\Im{Sp(z^{(1)}_{11})}=\pi\Theta(\Re[z^{(1)}_{11}]-1)\ln{\Re{z^{(1)}_{11}}}$. Since $\Re{y^{(1)}_0}<1$ and $\Gamma$ is small, we obtain $\Re{z^{(1)}_{11}}\sim |z^{(1)}_{11}|>1$. It allows us to omit the $\Theta$ function, and then have $\Im{Sp(z^{(1)}_{11})}=\pi\ln{\Re{z^{(1)}_{11}}}$. For $z^{(1)}_{21}$, since $\Re{z^{(1)}_{21}}<0$, this term has nothing to do with $\Im[W^{(1)}_1]$. Hence, we obtain the explicit imaginary part for $W^{(1)}_1$ when $s_1<s_1^-$,
\begin{eqnarray}
\Im{W^{(1)}_1}=\pi\ln{|z^{(1)}_{11}|}.
\end{eqnarray}

Similarly, we can work out the explicit imaginary parts for all other $W^{(i)}_j$ in each energy region of $s_1$ with $s_3>(m_2+m_3)^2$, which are shown in Tab.~\ref{imwij}. To obtain the summed contribution of $\Im{W^{(i)}_j}$ to $\Im{M}$, we define $W_{tot}\equiv W^{(1)}_1+W^{(1)}_2-W^{(2)}_1-W^{(2)}_2+W^{(3)}_1+W^{(3)}_2$, then we have the results for $\Im{W_{tot}}$ in each energy region, which are listed in Tab.~\ref{imwtot}.
\begin{table}
\centering
  \caption{The imaginary part of $W^{(i)}_j$ in different energy regions.}\label{imwij}
\begin{tabular}{|c|c|c|c|c|}
  \hline\hline
  Energy & $s_1<s_1^-$ & $s_1^-<s_1<s_{1c}$ & $s_{1c}<s_1<s_1^+$ & $s_1^+<s_1$ \\
  \hline
  $\Im{W^{(1)}_1}$ & $\pi\ln{|z^{(1)}_{11}|}$  & $\pi\ln{|z^{(1)}_{21}|}$ & $\pi\ln{|z^{(1)}_{21}|}$ & $\pi\ln{|z^{(1)}_{21}|}$ \\
  \hline
  $\Im{W^{(1)}_2}$ & $\pi\ln{|z^{(1)}_{12}|}$  & $\pi\ln{|z^{(1)}_{12}|}$ & $\pi\ln{|z^{(1)}_{12}|}$ & $\pi\ln{|z^{(1)}_{22}|}$ \\
  \hline
  $\Im{W^{(2)}_1}$ & $\pi\ln{|z^{(2)}_{11}|}-\pi\ln{|z^{(2)}_{21}|}$  & $\pi\ln{|z^{(2)}_{11}|}-\pi\ln{|z^{(2)}_{21}|}$ & $\pi\ln{|z^{(2)}_{11}|}-\pi\ln{|z^{(2)}_{21}|}$ & $\pi\ln{|z^{(2)}_{11}|}-\pi\ln{|z^{(2)}_{21}|}$ \\
  \hline
  $\Im{W^{(2)}_2}$ & $\pi\ln{|z^{(2)}_{12}|}-\pi\ln{|z^{(2)}_{22}|}$  & 0 & 0 & $-\pi\ln{|z^{(2)}_{12}|}+\pi\ln{|z^{(2)}_{22}|}$ \\
  \hline
  $\Im{W^{(3)}_1}$ & $\pi\ln{|z^{(3)}_{11}|}$  & $\pi\ln{|z^{(3)}_{21}|}$ & $-\pi\ln{|z^{(3)}_{21}|}$ & $-\pi\ln{|z^{(3)}_{11}|}$ \\
  \hline
  $\Im{W^{(3)}_2}$ & $-\pi\ln{|z^{(1)}_{12}|}$  & $-\pi\ln{|z^{(1)}_{12}|}$ & $-\pi\ln{|z^{(3)}_{12}|}$ & $-\pi\ln{|z^{(3)}_{12}|}$ \\
  \hline\hline
\end{tabular}
\end{table}

\begin{table}
\centering
  \caption{Summed contributions of $W^{(i)}_j$ to the imaginary part of $M$ in different kinematic regions.}\label{imwtot}
\begin{tabular}{|c|c|c|c|c|}
  \hline\hline
  Energy & $s_1<s_1^-$ & $s_1^-<s_1<s_{1c}$ & $s_{1c}<s_1<s_1^+$ & $s_1^+<s_1$ \\
  \hline
  $\Im{W_{tot}}$ & $\pi\ln{|\frac{z^{(1)}_{11}z^{(1)}_{12}z^{(2)}_{21}z^{(2)}_{22}z^{(3)}_{11}}{z^{(2)}_{11}z^{(2)}_{12}z^{(3)}_{12}}|}$ & $\pi\ln{|\frac{z^{(1)}_{21}z^{(1)}_{12}z^{(2)}_{21}z^{(3)}_{21}}{z^{(2)}_{11}z^{(3)}_{12}}|}$ & $\pi\ln{|\frac{z^{(1)}_{21}z^{(1)}_{12}z^{(2)}_{21}}{z^{(2)}_{11}z^{(3)}_{21}z^{(3)}_{12}}|}$ & $\pi\ln{|\frac{z^{(1)}_{21}z^{(1)}_{22}z^{(2)}_{21}z^{(2)}_{12}}{z^{(2)}_{11}z^{(2)}_{22}z^{(3)}_{11}z^{(3)}_{12}}|}$ \\
  \hline\hline
\end{tabular}
\end{table}

\subsection{The imaginary part of $T^{(i)}_j$}
The term $T^{(i)}_j$ in Eq.~(\ref{func-R-simplify}) is given by
\begin{eqnarray}
T^{(i)}_j=\ln{\frac{1-y^{(i)}_0}{y^{(i)}_j-y^{(i)}_0}}\left[\ln{1-y^{(i)}_j}-\ln{y^{(i)}_0-y^{(i)}_j}\right]-\ln{\frac{-y^{(i)}_0}{y^{(i)}_j-y^{(i)}_0}}\left[\ln{-y^{(i)}_j}-\ln{y^{(i)}_0-y^{(i)}_j}\right].
\end{eqnarray}
With the help of Eq.~(\ref{lndecompose}), $T^{(i)}_j$ is reduced to
\begin{eqnarray}
T^{(i)}_j&=&\ln{\frac{y^{(i)}_0-1}{y^{(i)}_0-y^{(i)}_j}}\left[\ln{\frac{y^{(i)}_j-1}{y^{(i)}_j-y^{(i)}_0}} -\eta\left(1-y^{(i)}_j,\frac{1}{y^{(i)}_0-y^{(i)}_j}\right)\right] -\ln{\frac{y^{(i)}_0}{y^{(i)}_0-y^{(i)}_j}}\left[\ln{\frac{y^{(i)}_j}{y^{(i)}_j-y^{(i)}_0}}-\eta\left(-y^{(i)}_j,\frac{1}{y^{(i)}_0-y^{(i)}_j}\right)\right]\nonumber\\
&=&\ln{r^{(i)}_{1j}}\left[\ln{r^{(i)}_{2j}}-\eta\left(1-y^{(i)}_j,\frac{1}{y^{(i)}_0-y^{(i)}_j}\right)\right] -\ln{r^{(i)}_{3j}}\left[\ln{r^{(i)}_{4j}}-\eta\left(-y^{(i)}_j,\frac{1}{y^{(i)}_0-y^{(i)}_j}\right)\right].
\end{eqnarray}
If $\eta$ is non-zero, $\eta$ will be purely imaginary. The imaginary part of $T^{(i)}_j$ is given by
\begin{eqnarray}
\Im[T^{(i)}_j]&=&\Im{\ln{r^{(i)}_{1j}}}\ln{|r^{(i)}_{2j}|}-\Im{\ln{r^{(i)}_{3j}}}\ln{|r^{(i)}_{4j}|}+\Im{\ln{r^{(i)}_{2j}}}\ln{|r^{(i)}_{1j}|}-\Im{\ln{r^{(i)}_{4j}}}\ln{|r^{(i)}_{3j}|}\nonumber\\
&-&\ln{|r^{(i)}_{1j}|}\Im{\eta\left(1-y^{(i)}_j,\frac{1}{y^{(i)}_0-y^{(i)}_j}\right)} +\ln{|r^{(i)}_{3j}|}\Im{\eta\left(-y^{(i)}_j,\frac{1}{y^{(i)}_0-y^{(i)}_j}\right)}.\label{imTij}
\end{eqnarray}
Note that only when $\Re{r^{(i)}_{jk}}$ is negative, can $r^{(i)}_{jk}$ contribute to $\Im{\ln{r^{(i)}_{jk}}}$. In each energy region, among all the 24 elements of $r^{(i)}_{jk}$, only twelve of them contain a negative real part. For simplicity, we define
\begin{eqnarray}
\eta^{(i)}_{1j}\equiv\eta\left(1-y^{(i)}_j,\frac{1}{y^{(i)}_0-y^{(i)}_j}\right)\\
\eta^{(i)}_{2j}\equiv\eta\left(-y^{(i)}_j,\frac{1}{y^{(i)}_0-y^{(i)}_j}\right) \ .
\end{eqnarray}
Then the contributive $\eta^{(i)}_{jk}$ are
\begin{eqnarray}
\eta^{(1)}_{22}=-2\pi i, \ \eta^{(2)}_{12}=\eta^{(2)}_{22}=2\pi i\Theta(s_{1c}-s_1), \ \eta^{(3)}_{21}=2\pi i\Theta(s_1-s_{1c}).\label{etaijk}
\end{eqnarray}
Referring to Tab.~\ref{imlnrijk} and Eq.~(\ref{etaijk}), we can derive the imaginary parts of every $T^{(i)}_j$ based on Eq.~(\ref{imTij}), which is shown in Tab.~\ref{imTijresult}. To sum over all imaginary parts of $T^{(i)}_j$, we define $T_{tot}\equiv T^{(1)}_1+T^{(1)}_2-T^{(2)}_1-T^{(2)}_2+T^{(3)}_1+T^{(3)}_2$, with $\Im{T_{tot}}$ in each energy region shown in Tab.~\ref{imTtot}. Combining $\Im{T_{tot}}$ and $\Im{W_{tot}}$ one notices that the sum of $\Im{T_{tot}}$ and $\Im{W_{tot}}$ is not continuous as $s_1$ varies. This is natural since the imaginary part of $\sigma^{(i)}$ has to be included in order to compensate this discontinuity.
\begin{table}
\centering
\caption{Non-zero elements of $\Im{\ln{r^{(i)}_{jk}}}$ with the explicit index $ijk$ in different energy regions.}\label{imlnrijk}
\begin{tabular}{|c|c|c|c|}
  \hline\hline
   $s_1<s_1^-$ & $s_1^-<s_1<s_{1c}$ & $s_{1c}<s_1<s_1^+$ & $s_1^+<s_1$ \\
   \hline
  111($-\pi$) & 112($-\pi$) & 112($-\pi$) & 121($\pi$)  \\
  112($-\pi$) & 121($\pi$) & 121($\pi$) & 122($\pi$) \\
  141($-\pi$) & 131($\pi$) & 131($\pi$) & 131($\pi$) \\
  142($-\pi$) & 142($-\pi$) & 142($-\pi$) & 132($\pi$) \\
  211($-\pi$) & 211($-\pi$) & 211($-\pi$) & 211($-\pi$) \\
  212($-\pi$) & 222($\pi$) & 222($-\pi$) & 212($\pi$) \\
  231($-\pi$) & 231($-\pi$) & 231($-\pi$) & 231($-\pi$) \\
  232($-\pi$) & 242($\pi$) & 242($-\pi$) & 232($\pi$) \\
  311($-\pi$) & 312($\pi$) & 312($\pi$) & 311($\pi$) \\
  312($\pi$) & 321($\pi$) & 321($-\pi$) & 312($\pi$) \\
  341($-\pi$) & 331($\pi$) & 331($-\pi$) & 341($\pi$) \\
  342($\pi$) & 342($\pi$) & 342($\pi$) & 342($\pi$) \\
  \hline\hline
\end{tabular}
\end{table}

\begin{table}
\centering
\caption{Imaginary part of $T^{(i)}_j$ in different regions.}\label{imTijresult}
\begin{tabular}{|c|c|c|c|c|}
  \hline\hline
  Energy & $s_1<s_1^-$ & $s_1^-<s_1<s_{1c}$ & $s_{1c}<s_1<s_1^+$ & $s_1^+<s_1$ \\
  \hline
  $\Im{T^{(1)}_1}$ & $\pi\ln{|\frac{r^{(1)}_{31}}{r^{(1)}_{21}}|}$ & $\pi\ln{|\frac{r^{(1)}_{11}}{r^{(1)}_{41}}|}$ & $\pi\ln{|\frac{r^{(1)}_{11}}{r^{(1)}_{41}}|}$ & $\pi\ln{|\frac{r^{(1)}_{11}}{r^{(1)}_{41}}|}$ \\
  $\Im{T^{(1)}_2}$ & $\pi\ln{|\frac{1}{r^{(1)}_{22}r^{(1)}_{32}}|}$ & $\pi\ln{|\frac{1}{r^{(1)}_{22}r^{(1)}_{32}}|}$ & $\pi\ln{|\frac{1}{r^{(1)}_{22}r^{(1)}_{32}}|}$ & $\pi\ln{|\frac{r^{(1)}_{12}}{r^{(1)}_{42}(r^{(1)}_{32})^2}|}$ \\
  $\Im{T^{(2)}_1}$ & $\pi\ln{|\frac{r^{(2)}_{41}}{r^{(2)}_{21}}|}$ & $\pi\ln{|\frac{r^{(2)}_{41}}{r^{(2)}_{21}}|}$ & $\pi\ln{|\frac{r^{(2)}_{41}}{r^{(2)}_{21}}|}$ & $\pi\ln{|\frac{r^{(2)}_{41}}{r^{(2)}_{21}}|}$ \\
  $\Im{T^{(2)}_2}$ & $\pi\ln{|\frac{r^{(2)}_{42}(r^{(2)}_{32})^2}{r^{(2)}_{22}(r^{(2)}_{12})^2}|}$ & $\pi\ln{|\frac{r^{(2)}_{32}}{r^{(2)}_{12}}|}$ & $\pi\ln{|\frac{r^{(2)}_{32}}{r^{(2)}_{12}}|}$ & $\pi\ln{|\frac{r^{(2)}_{22}}{r^{(2)}_{42}}|}$ \\
  $\Im{T^{(3)}_1}$ & $\pi\ln{|\frac{r^{(3)}_{31}}{r^{(3)}_{21}}|}$ & $\pi\ln{|\frac{r^{(3)}_{11}}{r^{(3)}_{41}}|}$ & $\pi\ln{|\frac{r^{(3)}_{41}(r^{(3)}_{31})^2}{r^{(3)}_{11}}|}$ & $\pi\ln{|r^{(3)}_{21}r^{(3)}_{31}|}$ \\
  $\Im{T^{(3)}_2}$ & $\pi\ln{|\frac{r^{(3)}_{22}}{r^{(3)}_{32}}|}$ & $\pi\ln{|\frac{r^{(3)}_{22}}{r^{(3)}_{32}}|}$ & $\pi\ln{|\frac{r^{(3)}_{22}}{r^{(3)}_{32}}|}$ & $\pi\ln{|\frac{r^{(3)}_{22}}{r^{(3)}_{32}}|}$ \\
  \hline\hline
\end{tabular}
\end{table}

\begin{table}
\centering
\caption{$\Im{T_{tot}}$ in different energy regions.}\label{imTtot}
\begin{tabular}{|c|c|c|c|c|}
  \hline\hline
  Energy & $s_1<s_1^-$ & $s_1^-<s_1<s_{1c}$ & $s_{1c}<s_1<s_1^+$ & $s_1^+<s_1$ \\
  \hline
  $\Im{T_{tot}}$ & $\pi\ln{\left|\frac{r^{(1)}_{31}r^{(2)}_{41}r^{(2)}_{42}(r^{(2)}_{32})^2r^{(3)}_{31}r^{(3)}_{22}} {r^{(1)}_{21}r^{(1)}_{22}r^{(1)}_{32}r^{(2)}_{21}r^{(2)}_{22}(r^{(2)}_{12})^2r^{(3)}_{21}r^{(3)}_{32}}\right|}$ & $\pi\ln{\left|\frac{r^{(1)}_{11}r^{(2)}_{32}r^{(2)}_{41}r^{(3)}_{11}r^{(3)}_{22}} {r^{(1)}_{22}r^{(1)}_{32}r^{(1)}_{41}r^{(2)}_{12}r^{(2)}_{21}r^{(3)}_{32}r^{(3)}_{41}}\right|}$ & $\pi\ln{\left|\frac{r^{(1)}_{11}r^{(2)}_{32}r^{(2)}_{41}r^{(3)}_{22}(r^{(3)}_{31})^2r^{(3)}_{41}} {r^{(1)}_{22}r^{(1)}_{32}r^{(1)}_{41}r^{(2)}_{12}r^{(2)}_{21}r^{(3)}_{11}r^{(3)}_{32}}\right|}$ & $\pi\ln{\left|\frac{r^{(1)}_{11}r^{(1)}_{12}r^{(2)}_{22}r^{(2)}_{41}r^{(3)}_{21}r^{(3)}_{22}r^{(3)}_{31}} {(r^{(1)}_{32})^2r^{(1)}_{41}r^{(1)}_{42}r^{(2)}_{21}r^{(2)}_{42}r^{(3)}_{32}}\right|}$ \\
  \hline\hline
\end{tabular}
\end{table}

\subsection{The imaginary part of $\sigma^{(i)}$}

Since the $\eta(a,\frac{1}{a})$ term is zero, the expression for $\sigma^{(i)}$ is
\begin{eqnarray}
\sigma^{(i)}=[\eta(1-y^{(i)}_1,1-y^{(i)}_2)-\eta(y^{(i)}_0-y^{(i)}_1,y^{(i)}_0-y^{(i)}_2)]\ln{\frac{y^{(i)}_0-1}{y^{(i)}_0}} \ ,\nonumber\\
\Im{\sigma^{(i)}}=\frac{1}{i}[\eta(1-y^{(i)}_1,1-y^{(i)}_2)-\eta(y^{(i)}_0-y^{(i)}_1,y^{(i)}_0-y^{(i)}_2)] \ln{|\frac{y^{(i)}_0-1}{y^{(i)}_0}|}\equiv\frac{1}{i}\delta\eta^{(i)}\ln{|\frac{y^{(i)}_0-1}{y^{(i)}_0}|} \ ,
\end{eqnarray}
with
\begin{eqnarray}
\delta\eta^{(1)}&=&-2\pi i\Theta(s_1-s_{1c}) \, \\
\delta\eta^{(2)}&=&2\pi i\Theta(s_{1c}-s_1) \ ,\\
\delta\eta^{(3)}&=&0 \ .
\end{eqnarray}
To understand how to derive this, we can take $\delta\eta^{(1)}$ as an example. Since $1-y^{(1)}_1$ and $1-y^{(1)}_2$ are both real and positive, the first $\eta$ in $\delta\eta^{(1)}$ is zero. Considering the second $\eta$, since $y^{(1)}_0$ has negative imaginary part, both $y^{(1)}_0-y^{(1)}_1$ and $y^{(1)}_0-y^{(1)}_2$ have negative imaginary parts. If $s_1$ increases to such a value, at which $y^{(1)}_0$ is closer to $y^{(1)}_2$ than $y^{(1)}_1$, then the product $(y^{(1)}_0-y^{(1)}_1)(y^{(1)}_0-y^{(1)}_2)$ crosses negative real axis giving a positive imaginary part. This required value for $s_1$ can be obtained by setting $2y^{(1)}_0=y^{(1)}_1+y^{(1)}_2$, which gives
\begin{eqnarray}
s_1=\frac{s_3^2+s_3(2m_1^2-m_2^2-m_3^2-s_2)+s_2(m_2^2-m_3^2)}{s_3+m_2^2-m_3^2},
\end{eqnarray}
which equals to $s_{1c}$ when $s_3=(m_2+m_3)^2$. Hence, in this case the second $\eta$ term picks up $+2\pi i\Theta(s_1-s_{1c})$, which gives $\delta\eta^{(1)}=-2\pi i\Theta(s_1-s_{1c})$.

Summing up all the imaginary parts from $\sigma^{(i)}$, we have
\begin{eqnarray}
\Im{\sigma_{tot}}\equiv\Im{\sigma^{(1)}}-\Im{\sigma^{(2)}}+\Im{\sigma^{(3)}}=-2\pi\Theta(s_{1c}-s_1)\ln{\left|\frac{y^{(1)}_0-1}{y^{(1)}_0}\right|} -2\pi\Theta(s_1-s_1^+)\ln{\left|\frac{y^{(2)}_0-1}{y^{(2)}_0}\right|}.
\end{eqnarray}

\subsection{Reduction of $\Im{M}$}
The imaginary part of $M$ thus becomes
\begin{eqnarray}
\Im{M}=\frac{1}{16\pi^2\lambda[s_1,s_2,s_3]^{1/2}}(\Im{W_{tot}}+\Im{T_{tot}}+\Im{\sigma_{tot}})\label{imMfinal}.
\end{eqnarray}
In principle, Eq.~(\ref{imMfinal}) can be used to directly calculate the imaginary part of scalar loop integral in the case where the internal mass has a small negative imaginary part. In the case where $m_2$ or $m_3$ has imaginary part, one can reconsider the motions of poles in complex plane, redetermine the proper sign of each parameter to obtain new result, which will be identical to Eq.~(\ref{imMfinal}). That is to say, Eq.~(\ref{imMfinal}) can be used to do calculations when any of the internal mass has a small imaginary part. In the case where $\Gamma=0$, it will be shown later that this expression is numerically identical to the $\Im{M}$ calculated directly from the Cutkosky rule.

\begin{figure}
  \centering
  \includegraphics[width=3in]{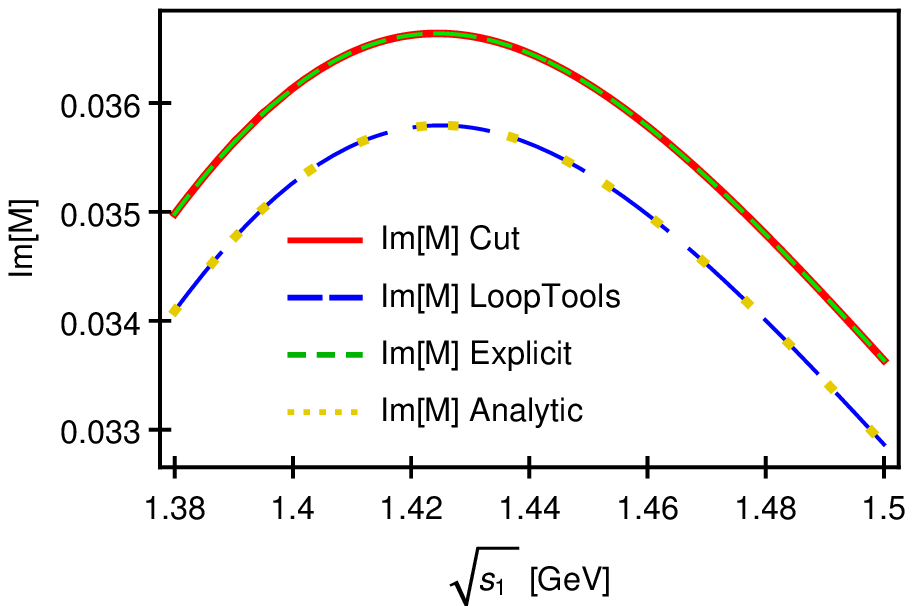}
  \includegraphics[width=3in]{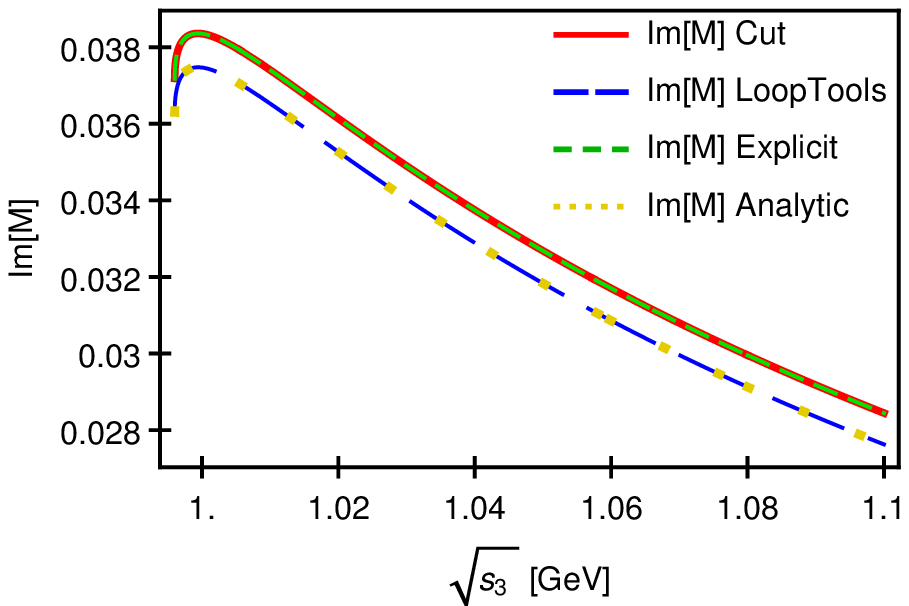}\\
  \includegraphics[width=3in]{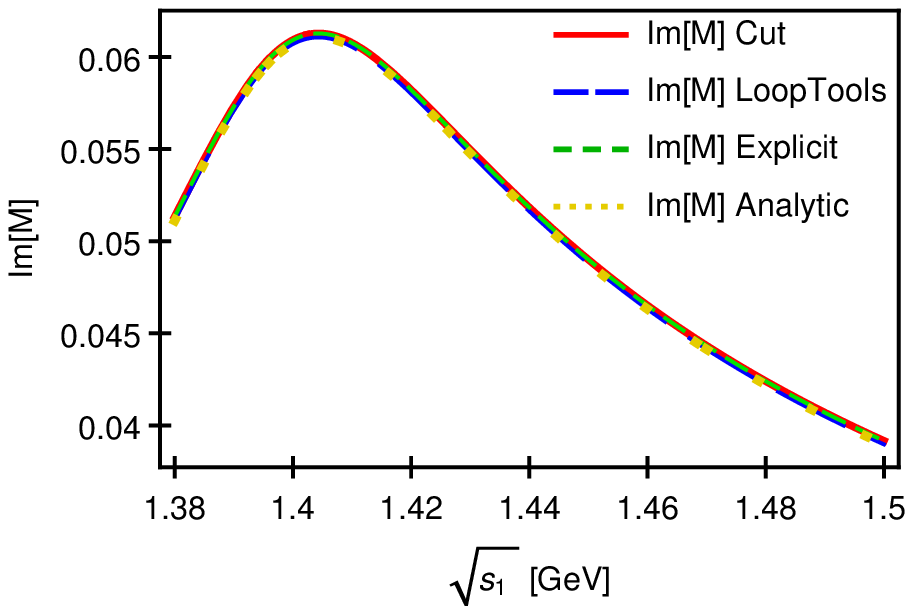}
  \includegraphics[width=3in]{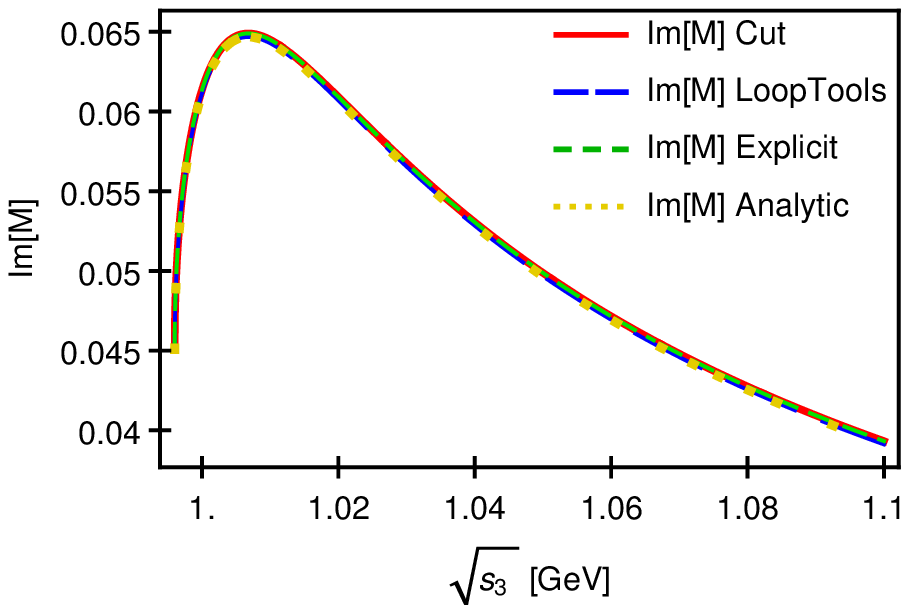}\\
  \includegraphics[width=3in]{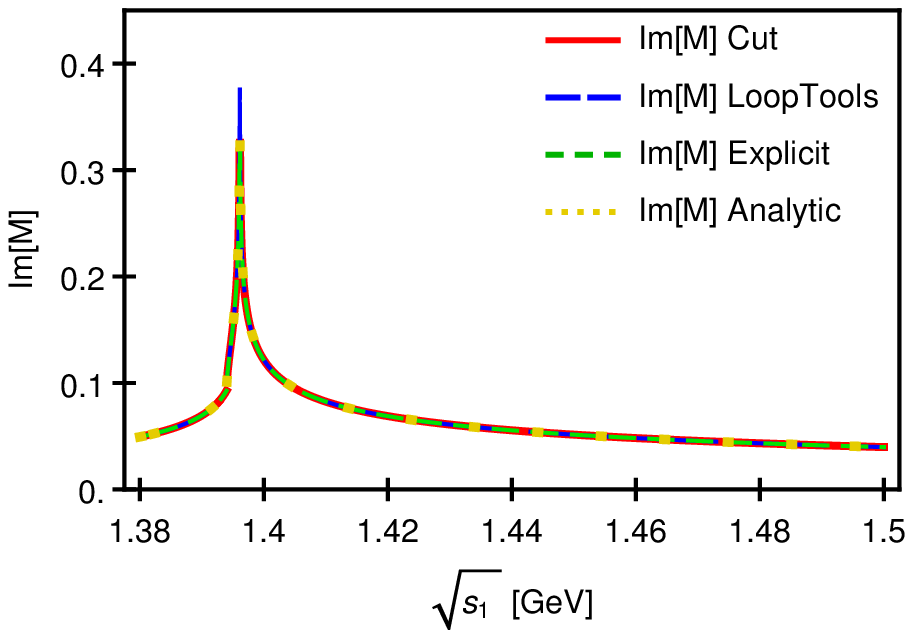}
  \includegraphics[width=3in]{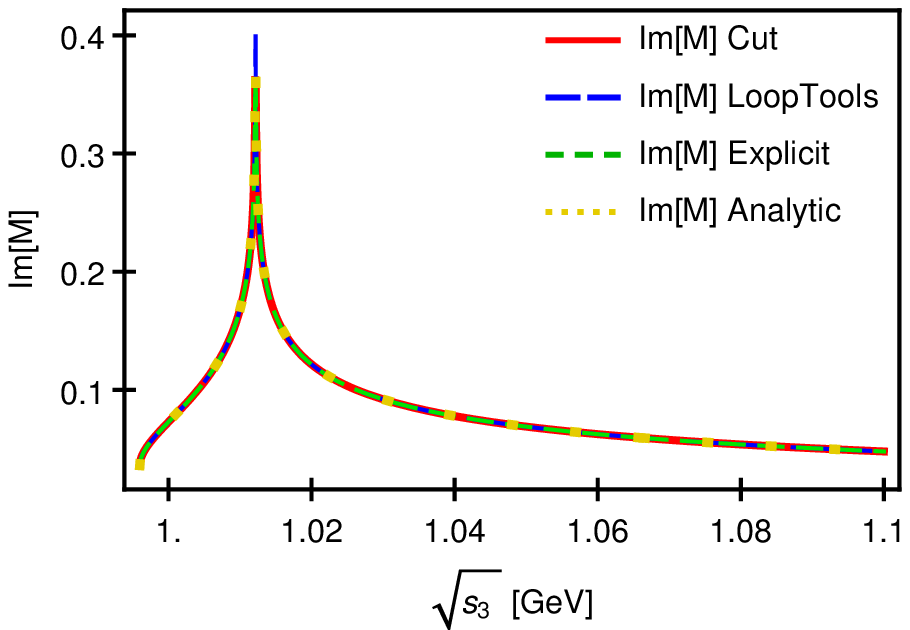}
  \caption{Calculations of the imaginary part of scalar loop corresponding to $\eta''\to K^*\bar{K}+c.c.\to K\bar{K}\pi\to f_0(980)\pi$. The plots on the left panel from upper to lower correspond to the $\Im{M}$ dependence of $\sqrt{s_1}$ at $\sqrt{s_3}=1$ GeV with $\Gamma=200$, 50 and 0 MeV, respectively. The plots on the right panel from upper to lower correspond to the $\Im{M}$ dependence of $\sqrt{s_3}$ at $\sqrt{s_1}=1.4$ GeV with $\Gamma=200$, 50 and 0 MeV, respectively.
  The blue solid, thick red solid, green dashed and orange dashed lines are calculated using Eq.~(\ref{imMcut}), LoopTools, Eq.~(\ref{imMfinal}) and Eq.~(\ref{ampanalytic}), respectively.}
  \label{fig-03}
\end{figure}

Although the logarithmic dependence is evident from Eq.~(\ref{imMfinal}), its form is still very complicated. For convenience, we denote the imaginary part derived by the Cutkosky rule by $f_c$. Before doing any reduction, it is very interesting to notice that by replacing the internal mass $m_1^2$ (here $m_1$ should be a real mass) in $f_c$ with its complex form $m_1^2-im_1\Gamma$, we obtain the same result as what is calculated by Eq.~(\ref{imMfinal}). In Fig.~\ref{fig-03}, we show the imaginary part of the scalar loop corresponding to $\eta''\to K^*\bar{K}+c.c.\to K\bar{K}\pi\to f_0(980)\pi$ with and without the $K^*$ width. Different approaches give the same value for $\Gamma=0$. Visible difference between these methods appear when $\Gamma=50$ MeV. But the thick (red) solid line which is given by the replacement of $m_1^2$ with $m_1^2-im_1\Gamma$, and the thin (green) solid line which is given by Eq.~(\ref{imMfinal}), always match even when $\Gamma$ is up to $200$ MeV.

The disagreement between Eq.~(\ref{ampanalytic}) (yellow dot-dashed line) and Eq.~(\ref{imMfinal}) (green dashed line) is due to several aspects. Firstly, when the width grows, the step function is no longer a good approximation, and a continuous function should be used to describe the parameters. Secondly, some substitutions using $s_3=(m_2+m_3)^2$ should be carefully dealt with. Thirdly, it is not a good approximation to omit the second term in Eq.~(\ref{imsp}). Instead, Eq~(\ref{imsp}) is suggested to be replaced by
\begin{eqnarray}
\Im{Sp(z)}=D_2(z)-\arg{(1-z)}\ln{z},
\end{eqnarray}
where $D_2$ is the Bloch-Wigner function~\cite{DZagier:1990MathAnn}, which a real analytic function in the cut plane $C\setminus\{0,1\}$.

Despite of this disagreement, the matching of thick (red) solid and (green) dashed lines encourages us to believe that the reduced result of Eq.~(\ref{imMfinal}) actually takes exactly the same form as $f_c$ with complex internal masses. This equivalence is non-trivial, for the fact that if $f_c$ is regarded as a function of internal masses, it is real so that it must be nonanalytic everywhere in the complex plane of internal masses, which makes it hard to be continued into the complex plane. A possible reason is that Eq.~(\ref{imMfinal}) and $f_c$ are the imaginary parts of two analytic functions of internal mass, respectively, then the uniqueness of analytic functions ensures their equivalence. However, this equivalence shows that the $f_c$ with complex internal masses can be regarded as the actual value of $\Im{M}$, given the width of internal particle is not too big to break the assumptions made in the previous discussion.

This conjecture has profound consequences. Since the analytic form of $f_c$ is much simpler than Eq.~(\ref{imMfinal}), if the equivalence holds, one can use $f_c$ with complex masses to isolate the $\Gamma$ dependence, which may provide the answer to how sensitive the absorptive part of $M$ is to $\Gamma$.

Now we can concentrate on the process $\eta''\to K^*\bar{K}+c.c.\to K\bar{K}\pi\to f_0(980)\pi$. If the equivalence is true, under the circumstance where internal masses all have a small imaginary part, the absorptive part of the corresponding scalar integral is
\begin{eqnarray}
\Im{M}=\frac{1}{32\pi\sqrt{s_1}p_{\pi}^{(s_1)}}\ln{\left|\frac{(a_1+1)}{(a_1-1)}\frac{(a_2+1)}{(a_2-1)}\right|},\label{imMcut}
\end{eqnarray}
where
\begin{eqnarray}
a_1&=&\frac{-s_2+\frac{(s_1+m^2-m_k^2)(s_1+s_3-s_2)}{2s_1}}{2\sqrt{[\frac{(s_1+m_k^2-m^2)^2}{4s_1}-m_k^2][-s_2+\frac{(s_1+s_3-s_2)^2}{4s_1}]}}, \nonumber\\
a_2&=&\frac{-m^2+m_k^2+s_2+\frac{1}{2}(s_3+s_2-s_1)}{2\sqrt{[-s_2+\frac{(s_3+s_2-s_1)^2}{4s_3}][-m_k^2+\frac{s_3}{4}]}},
\end{eqnarray}
with $m$ ($\Re{m}=m_1$) and $m_k$ being the masses of $K^*$ and $K$ mesons, respectively. In the case where internal masses are real, as $s_1$ approaches the abnormal threshold $s_1^-$, both $a_1$ and $a_2$ approaches 1, which leads to a logarithmic divergence. When internal masses are taken to be complex, as $s_1\to s_1^-$, $a\to 1+iO(\Gamma)$, so that the divergence disappear and the $\Im{M}$ is expected to be like $\ln{\frac{\beta^2}{\Gamma^2}}$.

\begin{figure}
  \centering
  \includegraphics[width=3in]{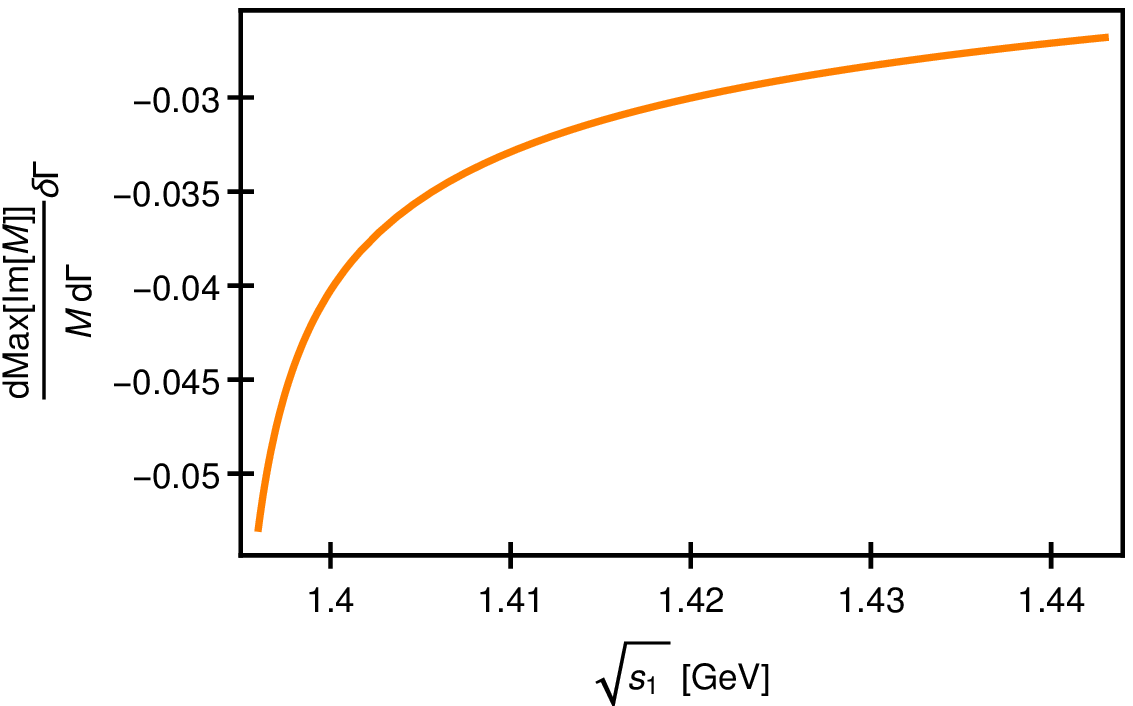}
  \includegraphics[width=3in]{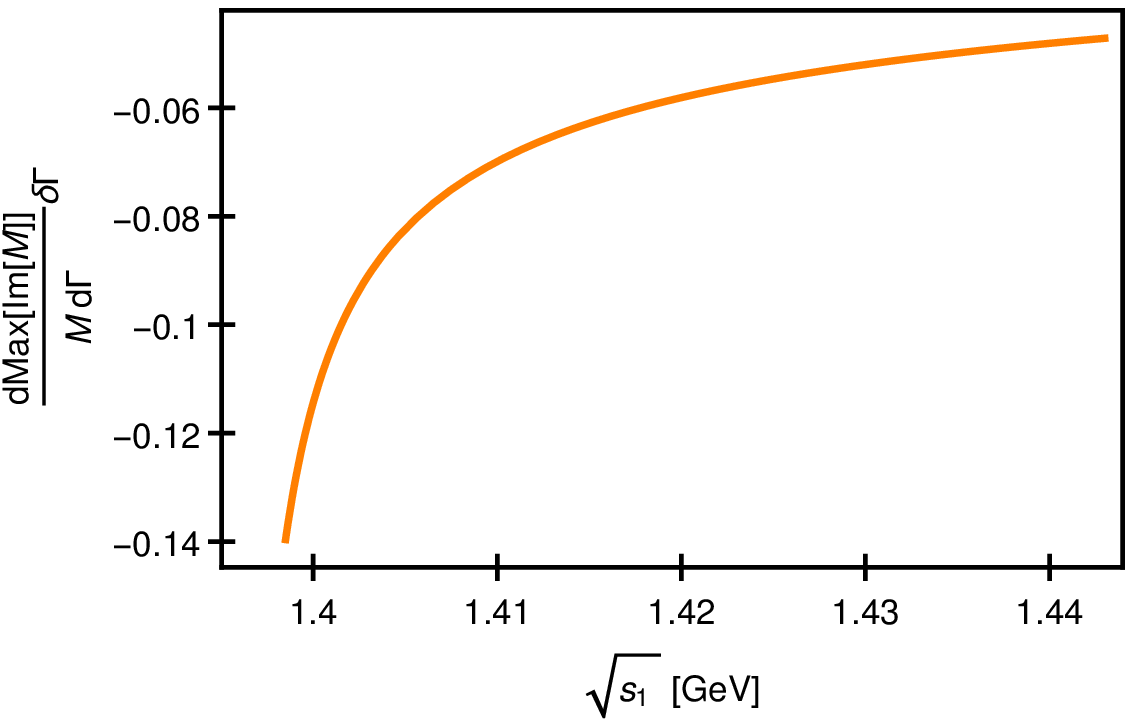}
  \caption{Variation of $\max{\Im{M}}$ when $\Gamma=10$ MeV (left) and $50$ MeV (right), with $\delta\Gamma=\Gamma/10$ MeV.}\label{fig-04}
\end{figure}

In the case where $m^2$ has an imaginary part $-im_1\Gamma$, we can expand $a_1$ and $a_2$ to the first order of $\Gamma$, and take the limit $s_1\to s_1^-$ to see how the maximum value of $\Im{M}$ becomes divergent as $\Gamma\to 0$. The expansion is not difficult so we just write down the result:
\begin{eqnarray}\label{maximM}
\max{\Im{M}}&=&\frac{1}{32\pi\sqrt{s_1}p_{\pi}^{(s_1)}}\ln{\frac{\beta^2}{\Gamma^2}} \ ,
\end{eqnarray}
with
\begin{eqnarray}\label{beta-func}
\beta^2&\equiv &\frac{1}{\sqrt{s_1}m^2}\frac{32s_1 (p_K^{(s_1)})^3p_{\pi}^{(s_1)}p_{\pi}^{(s_3)}p_K^{(s_3)}}{E_K^{(s_1)}(2E_f^{(s_1)}E_K^{(s_1)}-s_3)-2E_f^{(s_1)}(p_K^{(s_1)})^2},
\end{eqnarray}
where the kinematic variables are defined as follows:
\begin{eqnarray}
p_{\pi}^{(s_1)}&\equiv &\frac{\lambda[s_1,s_2,s_3]^{1/2}}{2\sqrt{s_1}}, \ p_{\pi}^{(s_3)}\equiv\frac{\lambda[s_3,s_2,s_1]^{1/2}}{2\sqrt{s_3}}, \
p_K^{(s_1)}\equiv \frac{\lambda[s_1,m^2,m_k^2]^{1/2}}{2\sqrt{s_1}},\nonumber\\
p_K^{(s_3)}&\equiv &\sqrt{\frac{s_3}{4}-m_k^2}, \
E_K^{(s_1)}\equiv \frac{s_1+m_k^2-m^2}{2\sqrt{s_1}}, \
E_f^{(s_1)}\equiv \frac{s_1+s_3-s_2}{2\sqrt{s_1}}, \ s_2=m_{\pi}^2.
\end{eqnarray}
Since some of the $\Gamma$ terms are contained in terms like $\lambda[s_1,m^2,m_k^2]$ with the complex masses, the closer to normal threshold the $s_1$ is, the smaller the convergence radius would be, which means this approximation becomes worse when $s_1$ is closer to $(m+m_k)^2$. To see whether the amplitude is sensitive to $\Gamma$ as it varies a small value in the vicinity around its physical value, we can calculate the logarithmic derivative of $\Im{M}$, i.e.
\begin{eqnarray}
\frac{1}{\max{\Im{M}}}\frac{d}{d\Gamma}\max{\Im{M}}=-\frac{1}{\Gamma\ln{\frac{\beta}{\Gamma}}},
\end{eqnarray}
and the results for $\delta\Gamma=\Gamma\times 10\%$ MeV with $\Gamma=10$ MeV and $50$ MeV, respectively, are illustrated by Fig.~\ref{fig-04}. From Fig.~\ref{fig-04}, we can conclude that, if $\Gamma$ varies $10\%$, the typical variation of $\max{\Im{M}}$ is around $4\%$ for $\Gamma=10$ MeV and $10\%$ for $\Gamma=50$ MeV, which are rather small effects.

\begin{figure}
  \centering
  \includegraphics[width=4in]{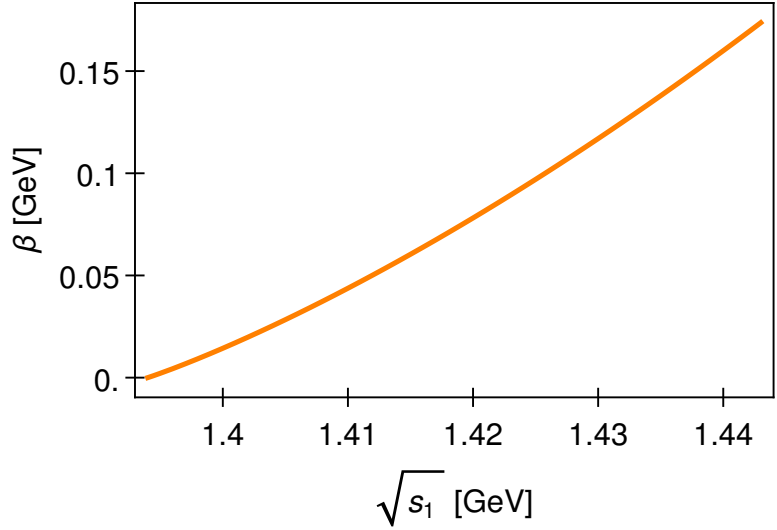}\\
  \caption{The value of $\beta$ of the scalar loop containing $K^*K\bar{K}$ as internal particles.}\label{fig-05}
\end{figure}

Naively, one may consider that the $\Gamma$ dependence is described by $\Gamma/\Re{m}$. On the contrary, the quantity $\beta$ in Eq.~(\ref{beta-func}) should be identified as the energy scale to describe the strength of $\Gamma$ dependence. Figure~\ref{fig-05} shows that within the region $(\Re{m}+m_k)^2<s_1<s_{1c}$, the value of $\beta$ is much smaller than $m$ ($\sim 0.9$ GeV).

\begin{figure}
  \centering
  \includegraphics[width=3in]{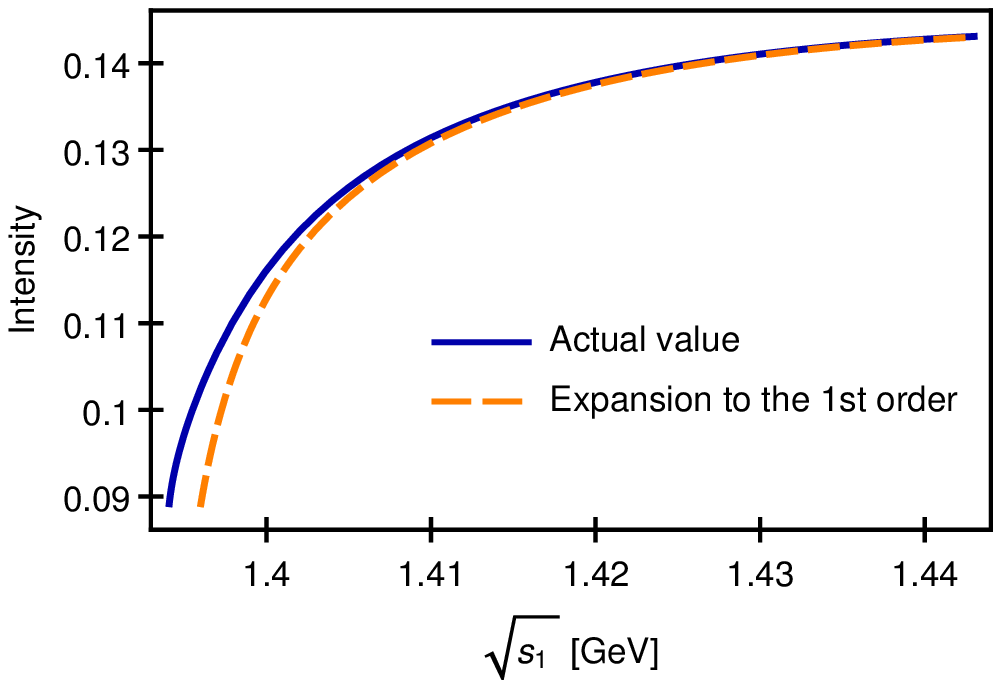}
  \includegraphics[width=3in]{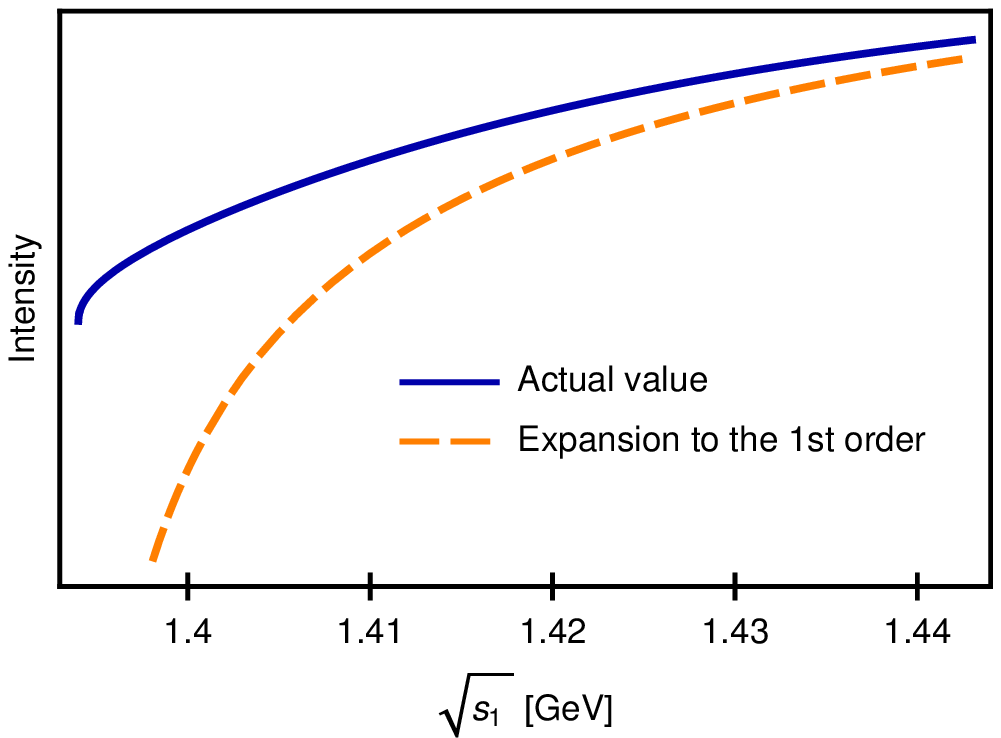}
  \caption{The maximum value of $\Im{M}$ as a function of $s_1$. The blue solid line is the value calculated using Eq.~(\ref{imMcut}) and the orange dashed line is the approximated value using Eq.~(\ref{maximM}). The left panel corresponds to $\Gamma=10$ MeV, while the right one corresponds to $\Gamma=50$ MeV.}\label{fig-06}
\end{figure}

Figure~\ref{fig-06} shows that, when $\Gamma=10$ MeV, the approximation holds well enough. When $\Gamma=50$ MeV, the approximation breaks down for small values of $s_1$. It shows that although $\Gamma=50$ MeV is almost $1/18$ of $\Re{m}=m_1$, compared to the scale $\beta$, it is actually a large value and an expansion to high order of $\Gamma$ is required for small value of $s_1$. This feature suggests that for relatively narrow widths the treatment of Ref.~\cite{Pavao:2017kcr} can also provide a reliable evaluation of the TS contributions. Figure~\ref{fig-06} also shows that when $s_1$ goes closer to $s_{1c}$, the amplitude becomes larger. Hence, it can be expected that the larger the allowed kinematic region for the TS, the stronger the enhancement would be. This is consistent with the analysis of Ref.~\cite{Liu:2015taa}.

\section{The triangle singularity mechanism in $\eta''$ decay}
\subsection{Experimental aspects}
In this section we will revisit the radiative decay $J/\psi \to \gamma\eta(1405/1475)$ with the $\eta(1405/1475)$ decays into $K\bar{K}\pi$, $\eta\pi\pi$ and $3\pi$ with the TS mechanism considered. The focus is to quantify the width effects based on the formulation developed in the previous section and present a coherent analysis of these exclusive channels on the same basis. For abbreviation, we denote $\eta(1405/1475)$ by $\eta''$ as follows.

We are going to adopt the $J/\psi$ radiative decay data from the BESIII Collaboration in this analysis. As pointed out in the Introduction, the present PDG~\cite{Patrignani:2016xqp} lists $\eta(1405)$ and $\eta(1475)$ as different states appearing in different channels. However, the high-statistic data analyses at BESIII so far have not observed two separated states in any exclusive channel. In order to keep self-consistent we only consider the $J/\psi$ radiative decays and assume that there is only one state $\eta''$ contributing. We then examine all observables with the implementation of the TS mechanism as a self-consistent check of our one-state scenario. This strategy is the same as that in Ref.~\cite{Wu:2012pg} except that we will quantify the width effects in the loop function here and point out an additional contribution to the isospin violation channel.

We adopt the PDG averaged combined branching ratios $BR(J/\psi\to\gamma\eta''\to\gamma K\bar{K}\pi)=(2.8\pm 0.6)\times 10^{-3}$ and $BR(J/\psi\to\gamma\eta''\to\gamma a_0\pi\to\gamma \eta\pi^0\pi^0)=(8.40\pm 1.75)\times 10^{-5}$~\cite{Patrignani:2016xqp} to constrain the parameters in this analysis. Note that the main weights for extracting these averaged branching ratios are from the BESIII measurements. One obvious feature is that in both processes there is only one state of $(I=0, \ J^{PC}=0^{-+})$ has been observed around 1.4$\sim$1.48 GeV. Assuming that these two signals are from the same state, the much larger branching ratio for the $\gamma K\bar{K}\pi$ channel than $\gamma\eta\pi\pi$ should be one of the important observables for our understanding of the structure of $\eta''$.

The branching ratio for the isospin violating channel $J/\psi\to\gamma\eta''\to \gamma f_0(980)\pi\to\gamma\pi^+\pi^-\pi^0$ is measured to be $(1.50\pm0.11\pm0.11)\times 10^{-5}$ by BESIII~\cite{BESIII:2012aa}. Its relative branching ratio fraction to $J/\psi\to\gamma\eta''\to \gamma a_0(980)\pi\to\gamma\eta\pi^0\pi^0$ is found to be,
\begin{eqnarray}
\frac{BR(\eta''\to f_0(980)\pi\to\pi^+\pi^-\pi^0)}{BR(\eta''\to a_0(980)\pi\to\eta\pi^0\pi^0)}=17.9\pm4.2\%,
\end{eqnarray}
which is about one order of magnitude larger than the isospin breaking effects given by the $a_0(980)-f_0(980)$ mixing. As studied in Refs.~\cite{Wu:2011yx,Wu:2012pg}, the $a_0(980)-f_0(980)$ mixing mechanism can only account for a few percent of isospin breaking effects. Thus, other mechanism is required to compensate such a significant deficit.

It is also interesting to note that the $a_0(980)-f_0(980)$ mixing measured in $J/\psi\to\phi\eta\pi$ and $\chi_{c1}\to 3\pi$ at BESIII is at the order of 1\%~\cite{Ablikim:2018pik}, which is consistent with the theoretical prediction of Refs.~\cite{Wu:2007jh,Wu:2008hx} and can be regarded as a direct evidence for additional mechanism that strongly enhances the isospin breakings in $\eta''\to f_0(980)\pi\to 3\pi$. Note that the TS condition is not fulfilled in $J/\psi\to\phi\eta\pi$ and $\chi_{c1}\to 3\pi$. Taking into account this fact, the role played by the TS in $\eta''$ decays is rather natural and easy to test~\cite{Wu:2011yx,Wu:2012pg}.

\subsection{The amplitudes for $\eta''\to f_0\pi\to\pi^+\pi^-\pi^0$}

\begin{figure}[H]
  \centering
  \includegraphics[width=2in]{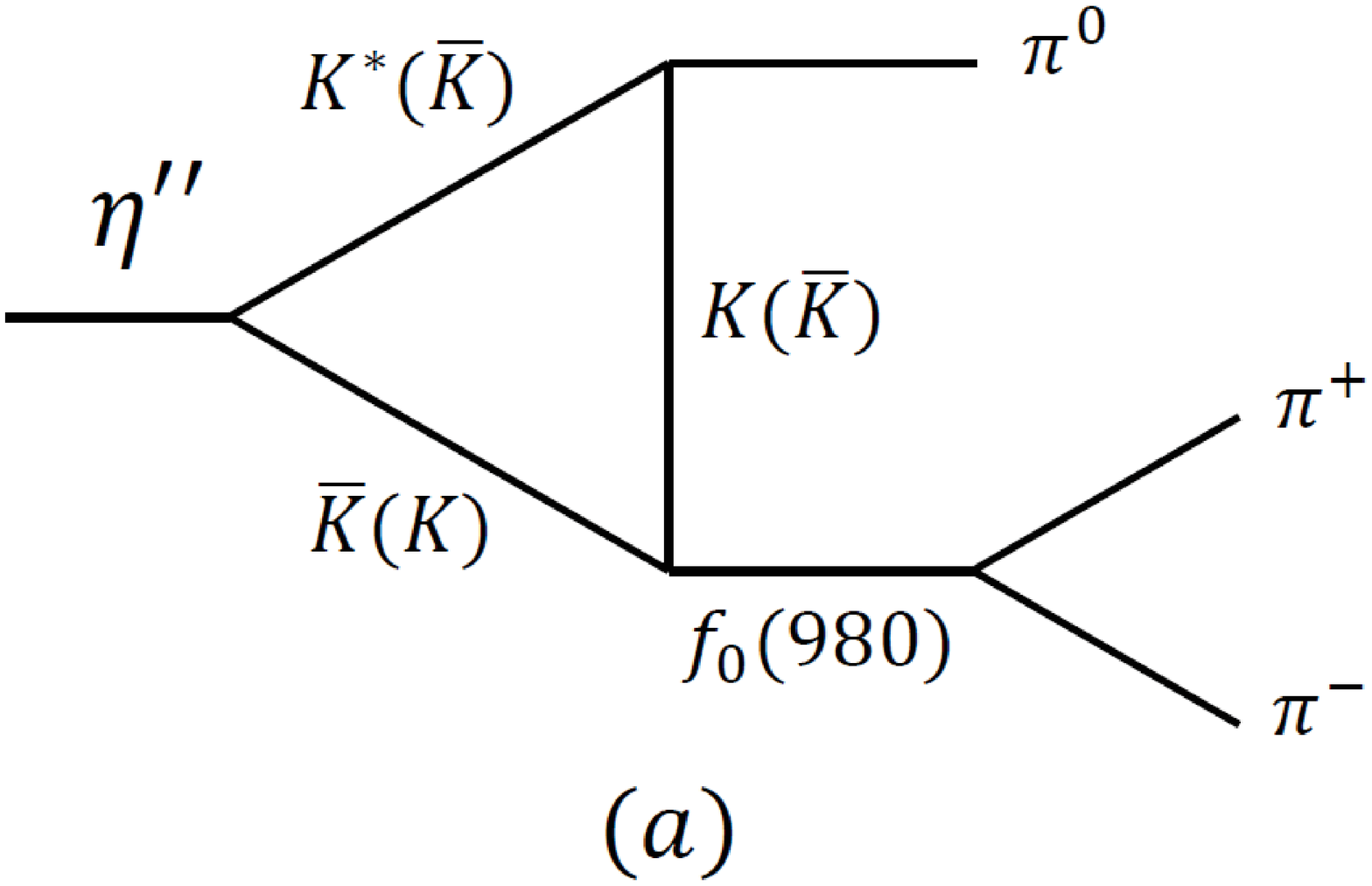}
  \includegraphics[width=2in]{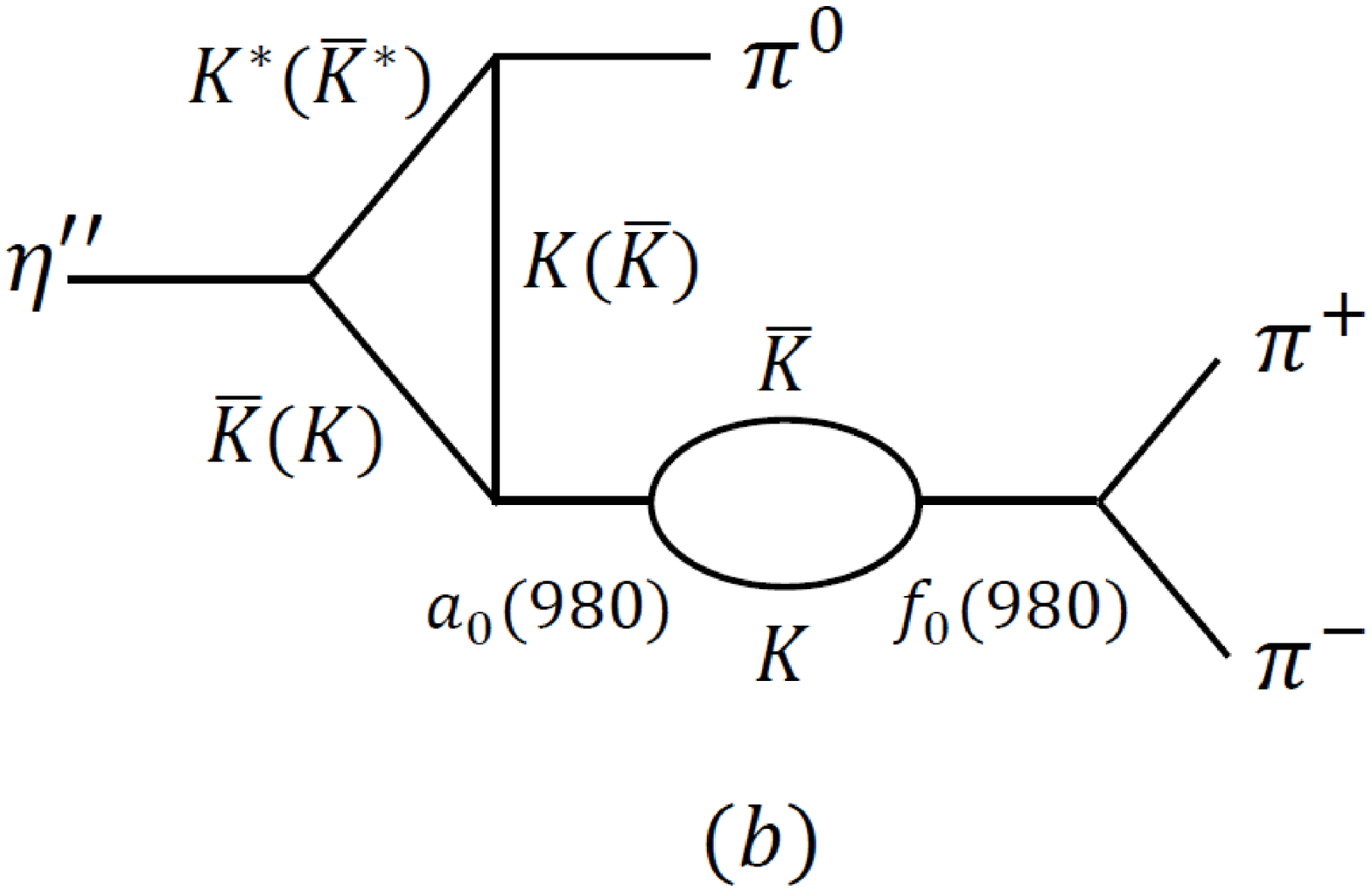}
  \includegraphics[width=2in]{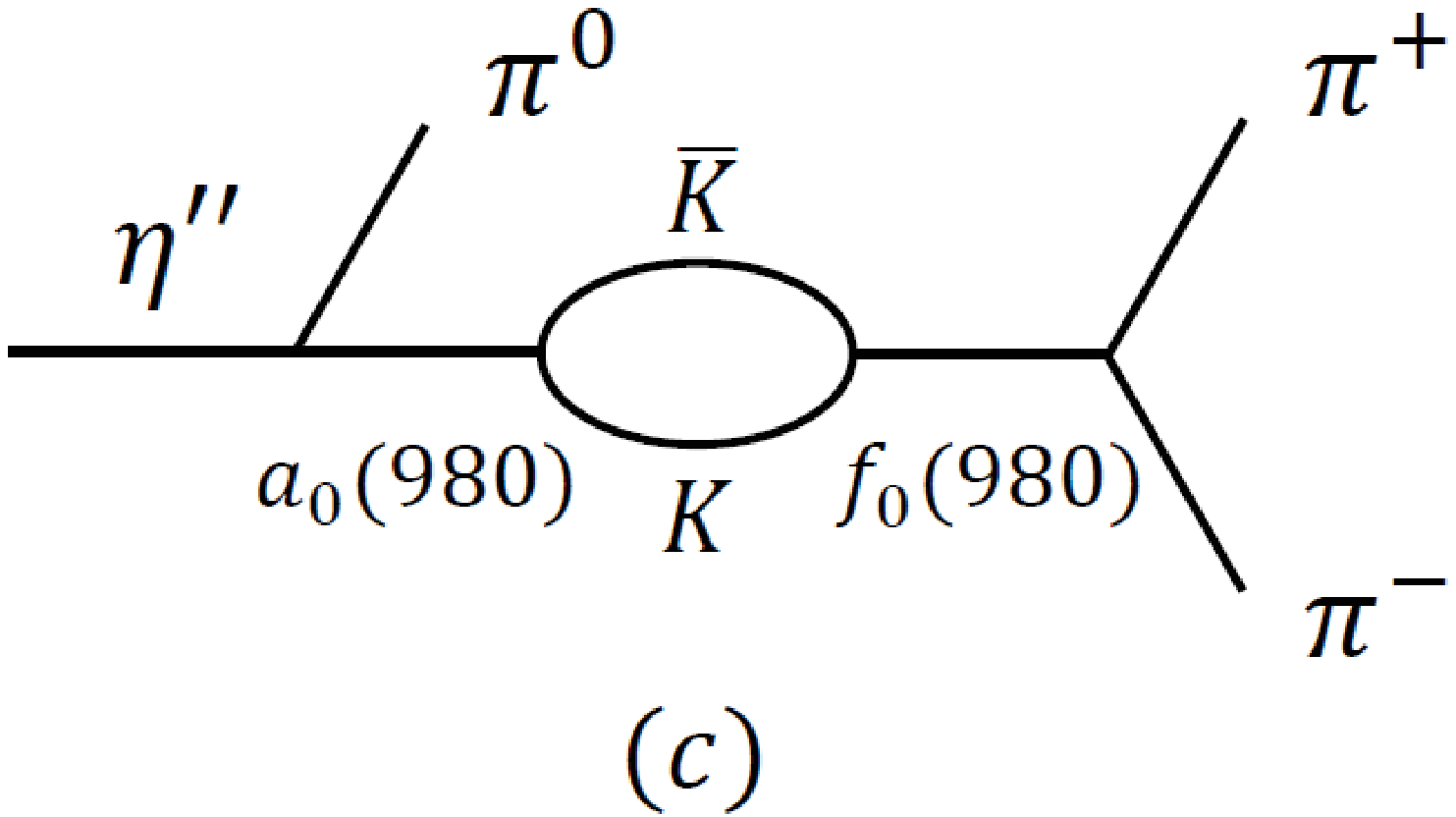}
  \caption{Isospin violating mechanisms for $\eta''\to \pi^+\pi^-\pi^0$. (a) describes the direct isospin breaking via the TS mechanism, (b) describes the TS contribution to the production of the $a_0(980)$ which then mixes with the $f_0(980)$ via the $K\bar{K}$ loops, and (c) represents the tree-level contributions to the intermediate $a_0(980)$ production, which will then mixing with the $f_0(980)$ via the $K\bar{K}$ loops. }\label{fig07}
\end{figure}

Taking into account the TS mechanism, the isospin violating decay of $\eta''\to 3\pi$ can be accounted for by Fig.~\ref{fig07}, where Fig.~\ref{fig07} (a) and (b) describe the TS mechanism which can contribute via two processes, and Fig.~\ref{fig07} (c) describes the $a_0(980)-f_0(980)$ mixing mechanism via the leading tree-level diagram. Namely, in Fig.~\ref{fig07} (a) the isospin violation arises from the mass difference between the neutral and charged triangle loops, and in Fig.~\ref{fig07} (b) the triangle loop provides an enhanced production mechanism for $a_0(980)$ which then mixes with $f_0(980)$ to contribute to the isospin breaking effects. It should be emphasized that although these two processes both are categorized as the TS mechanism, they represent different dynamics contributing to the significant isospin breakings.

It should be addressed that a coherent study including Fig.~\ref{fig07} (b) and (c) has not been carried out. In Refs.~\cite{Wu:2011yx,Wu:2012pg} process of Fig.~\ref{fig07} (a) is considered with the coupling vertices determined by experimental data. It can be regarded as a leading approximation for the vertex renormalization. In addition, since the dispersive parts from the charged and neutral loop amplitudes cancel each other, the dominance of the absorptive parts largely reduces the model-dependence in the estimate of the loop contributions. But the width effects were not estimated therein. In Ref.~\cite{Achasov:2015uua} the authors investigated the width effects arising from Fig.~\ref{fig07} (a), while the process of Fig.~\ref{fig07} (c) is estimated by the data for $\eta''\to a_0(980)\pi$ which is also a leading approximation for the coupling renormalization for the $\eta'' a_0(980)\pi$ vertex. However, the experimental data for $\eta''\to a_0(980)\pi$ still contain large uncertainties either in $\eta''\to K\bar{K}\pi$ or $\eta\pi\pi$. In Ref.~\cite{Aceti:2012dj}, similar processes as Fig.~\ref{fig07} (a) and (c) are considered and the authors confirm the irreplaceable contributions from the TS mechanism. However, the width effects are not investigated. In order to obtain reliable results for the role played by the TS mechanism, we are going to coherently include Fig.~\ref{fig07} (a) (b) and (c), and the width effects within the triangle loops. Meanwhile, a combined analysis of $\eta''\to K\bar{K}\pi$ and $\eta\pi\pi$ in association with the isospin violating decay of $\eta''\to 3\pi$ is necessary.

To proceed, the effective Lagrangian for vector-pseudoscalar-pseudoscalar ($VPP$) vertex is defined as
\begin{eqnarray}
L_{VPP}=ig_{VPP}Tr[V^{\mu}(\partial_{\mu}PP-P\partial_{\mu}P)] \ ,\label{lagrangianvpp}
\end{eqnarray}
where $V$ and $P$ stand for the fields for the flavor SU(3) multiplets, respectively, and they have the following expressions:
\begin{equation}
P=
\begin{pmatrix}%[1.3]
    \frac{\pi^0}{\sqrt{2}}+\frac{\eta}{\sqrt{6}}  &  \pi^+ &   K^+ \\
    \pi^- &  -\frac{\pi^0}{\sqrt{2}}+\frac{\eta}{\sqrt{6}}  &   K^0  \\
    K^-    &  \bar{K}^0 &  -\frac{2}{\sqrt6}\eta  \\
\end{pmatrix} \ ,
\end{equation}
and
\begin{equation}
V=
\begin{pmatrix}%[1.3]
    \frac{\rho^0}{\sqrt{2}}+\frac{\omega}{\sqrt{2}}  &  \rho^+ &   K^{*+} \\
    \rho^- &  -\frac{\rho^0}{\sqrt{2}}+\frac{\omega}{\sqrt{2}}  &   K^{*0}  \\
    K^{*-}    &  \bar{K}^{*0} &  \phi  \\
\end{pmatrix} \ .
\end{equation}

The overall coupling strength $g_{VPP}$ in Eq.~(\ref{lagrangianvpp}) for different channels can be determined by experimental data. For instance, with the partial decay width of $K^{*}\to K\pi$~\cite{Patrignani:2016xqp}, the coupling$g_{K^{*0}K^0\pi^0}\equiv g_{VPP}/\sqrt{2}=3.20$ can be determined, and then we extract $g_{VPP}=4.53$. The relative signs for $g_{\eta'' K^*\bar{K}}$ in different exclusive coupling channels are given by the charge conjugation symmetry, i.e.
\begin{eqnarray}
|\eta''\rangle &=& \frac{1}{\sqrt{2}}(|K^{*0}\bar{K}^0\rangle-|\bar{K}^{*0}K^0\rangle),\nonumber\\
|\eta''\rangle &=& \frac{1}{\sqrt{2}}(|K^{*+}K^-\rangle-|K^{*-}K^+\rangle).
\end{eqnarray}
The triangle amplitude for Fig.~\ref{fig07} (a) can then be expressed as
\begin{eqnarray}
M_{tri1}&=& 2g_{\eta'' K^{*0}\bar{K}^0}g_{K^{*0}K^0\pi^0}g_{f_0K^{0}\bar{K}^0}g_{f_0\pi^+\pi^-}[I_1(s_1,s_3)-I_2(s_1,s_3)]\frac{i}{s_3-m_{f_0}^2+\Pi_{f_0(s_3)}},\label{eqtri1}
\end{eqnarray}
where the function $I_{1,2}(s_1,s_3)$ has the same expression as Eq.~(\ref{eq1}) and the subscriptions 1 and 2 denote the neutral and charged triangle loop integrals, respectively.
Note that $I_{1,2}(s_1,s_3)$ are complex functions of $s_1$ and $s_3$. When the kinematic reflection is considered, $s_3$ should be replaced by the invariant mass squared of the corresponding 2-body system.

Equation~(\ref{eq1}) converges with the choice of the gauge for the vector meson. However, since we are dealing with a hadronic loop which involves non-fundamental particles it is possible that unphysical contributions can be present in the dispersive part of the amplitude and should be subtracted. To examine such a possibility we include a form factor ${\cal F}(q^2)$ in the numerical calculations. By changing the different values for the cut-off parameter, we are able to examine uncertainties arising from unphysical contributions from the ultraviolet kinematic region:
\begin{eqnarray}
{\cal F}(q^2)\equiv \prod_{R={K^*, K, \bar{K}}}\frac{\Lambda_R^2-m_R^2}{\Lambda_R^2-p_R^2(q^2)},\label{form-factor-monopole}
\end{eqnarray}
where $p_R$ with $R=K^*, \ K, \ \bar{K}$ are the four-momenta of the corresponding intermediate particles and $\Lambda_R\equiv m_R+\alpha\Lambda_{QCD}$ (with $\Lambda_{QCD}\simeq 200\sim 300$ MeV) is the cut-off energy. This choice of form factor is not perfect but convenient and sufficiently efficient. It implies that the vertex couplings will be suppressed as long as the propagators become off-shell towards the space-like region. One notices that the monopole form factor does not guarantee that the amplitude will be suppressed when the internal particles go off shell towards the space-like region. However, it can be easily checked that in the vicinity of the TS kinematics, in particular, when at least two particles are on shell, the third one will mostly stay in the space-like region. The kinematics when one state moves into the time-like region and the other two stay in the space-like region can still be efficiently cut-off when the kinematics deviate away from the TS condition. Such a situation can be recognized by the following decomposition:
\begin{eqnarray}
\frac{m_R^2-\Lambda_R^2}{p_R^2-\Lambda_R^2}\frac{1}{p_R^2-m_R^2}=\frac{1}{p_R^2-m_R^2}-\frac{1}{p_R^2-\Lambda_R^2} \ .
\end{eqnarray}
To be more specific, since $\Lambda_R$ is sufficiently larger than $m_R$ the artificial pole contribution from $p_R^2=\Lambda_R^2$ is far away from the vicinity of $p_R^2=m_R^2$. Therefore, the artificial pole is not going to contribute to the physical value of the loop amplitude. Meanwhile, when $p_R^2$ is sufficiently far away from the physical pole of $m_R^2$, the cancellation in Eq.~(\ref{form-factor-monopole}) is obvious and traceable.

For the processes of Fig.~\ref{fig07} (b) and (c) both processes contribute to the production of $a_0(980)$ and then $a_0(980)-f_0(980)$ mixing leads to the isospin violation. Note that in Fig.~\ref{fig07} (b) the neutral and charged triangle loop amplitudes will constructively add to each other. The contributions from the dispersive amplitude within the triangle loops will suffer from the uncertainties arising from the regularization of the loop divergence. This is very different from Fig.~\ref{fig07} (a) where the dispersive parts will cancel and the model-dependence is strongly suppressed. We will come back to this point in the numerical calculations in the next section.

For the $a_0(980)-f_0(980)$ mixing we adopt the unitarized coupled-channel approach for their mixing via the intermediate $K\bar{K}$ meson loops and extract their propagators with the full loop corrections following the prescription of Ref.~\cite{Achasov:1980gu,Achasov:2004uq}. This will allow us to directly compare with the result of Ref.~\cite{Achasov:2015uua}.
The general form of the propagators for $f_0(980)$ and $a_0(980)$ is written as the following:
\begin{equation}
\frac{i}{D}=\frac{i}{s-m_0^2+\Pi(s)}=\frac{i}{s-m^2-\Re{\Pi(m^2)}+\Pi(s)}\equiv\frac{i}{s-m_0^2+\Sigma_{a,b}\Pi_{ab(s)}},
\end{equation}
with
\begin{equation}
i\Pi_{ab}\equiv (ig_{ab})^2\int\frac{d^4q}{(2\pi)^4}\frac{i^2}{(q^2-m_a^2)((q-k)^2-m_b^2)},
\end{equation}
where $m$ is the physical mass defined at $\Re{\Pi(m^2)}=0$,
$k^2\equiv s$, and the subscription $ab$ denotes the final state particles that $a_0(980)$ and $f_0(980)$ can decay into, i.e. $ab=\{\eta\pi, K^0\bar{K}^0, K^+K^-\}$ for $a_0(980)$, and $ab=\{\pi^0\pi^0, \pi^+\pi^-, K^0\bar{K}^0, K^+K^-\}$ for $f_0(980)$. Function $\Pi(s)$ is the self-energy correction arising from the intermediate meson loops and $i\Pi_{ab}$ is the two-point loop function. The $g_{ab}$ is the coupling of $a_0(980)\to a+b$ or $f_0(980)\to a+b$, of which the adopted values are listed in Tab.~\ref{tab-coupling}.

The interference among the amplitudes in Fig.~\ref{fig07} has an interesting implication of the partial decay width for $\eta''\to \eta\pi\pi$. It is noticeable that the branching ratio of $\eta''\to\eta\pi\pi$ is much smaller than that of $\eta''\to K\bar{K}\pi$. So far, this question has not been seriously investigated because of the inconsistent treatment of the signals in these two channels, namely, they are treated as two irrelevant states $\eta(1405)$ and $\eta(1475)$. In our proposal their relative strength should contain dynamical information about the TS mechanism. One can see later that it can serve as a self-consistent examination of our solution.

Denoting the amplitudes of Fig.~\ref{fig07} (b) and (c)  by $M_{trimix}$ and $M_{treemix}$, respectively, we can write down their explicit expressions as follows:
\begin{eqnarray}\label{trimix-eq}
M_{trimix}&=&-2g_{\eta'' K^{*0}\bar{K}^0}g_{K^{*0}K^0\pi^0}g_{a_0K^0\bar{K}^0}g_{a_0K^0\bar{K}^0}g_{f_0K^0\bar{K}^0}g_{f_0\pi^+\pi^-}\nonumber\\
&&\times[I_1(s_1,s_3)+I_2(s_1,s_3)]\frac{i}{s_3-m_{a_0}^2+\Pi_{a_0(s_3)}}(B_1-B_2)\frac{i}{s_3-m_{f_0}^2+\Pi_{f_0(s_3)}} ,
\end{eqnarray}
and
\begin{eqnarray}\label{treemix-eq}
M_{treemix}&=&g_{\eta'' a_0\pi}g_{a_0K^0\bar{K}^0}g_{f_0K^0\bar{K}^0} g_{f_0\pi^+\pi^-}\frac{i}{s_3-m_{a_0}^2+\Pi_{a_0(s_3)}}(B_1-B_2)\frac{i}{s_3-m_{f_0}^2+\Pi_{f_0(s_3)}},
\end{eqnarray}
where $B_1$ and $B_2$ are the loop functions for the neutral and charged kaon loops in the $a_0(980)-f_0(980)$ mixing:
\begin{eqnarray}
B_1&\equiv & \int\frac{d^4q}{(2\pi)^4}\frac{i}{q^2-m_{K^0}^2}\frac{i}{(p_3-q)^2-m_{\bar{K}^0}^2}, \nonumber\\
B_2&\equiv & \int\frac{d^4q}{(2\pi)^4}\frac{i}{q^2-m_{K^+}^2}\frac{i}{(p_3-q)^2-m_{K^-}^2}.
\end{eqnarray}
In Eqs.~(\ref{trimix-eq}) and (\ref{treemix-eq}) the relations $g_{f_0K^0\bar{K}^0}=g_{f_0K^+K^-}$, $g_{a_0K^0\bar{K}^0}=-g_{a_0K^+K^-}$ and $g_{\eta'' K^{*0}\bar{K}^0}=g_{\eta'' K^{*+}K^-}$ are implied for the vertex coupling constants.

\subsection{The amplitudes for $\eta''\to K\bar{K}\pi$ and $\eta\pi\pi$}

\begin{figure}[H]
  \centering
  \includegraphics[width=2in]{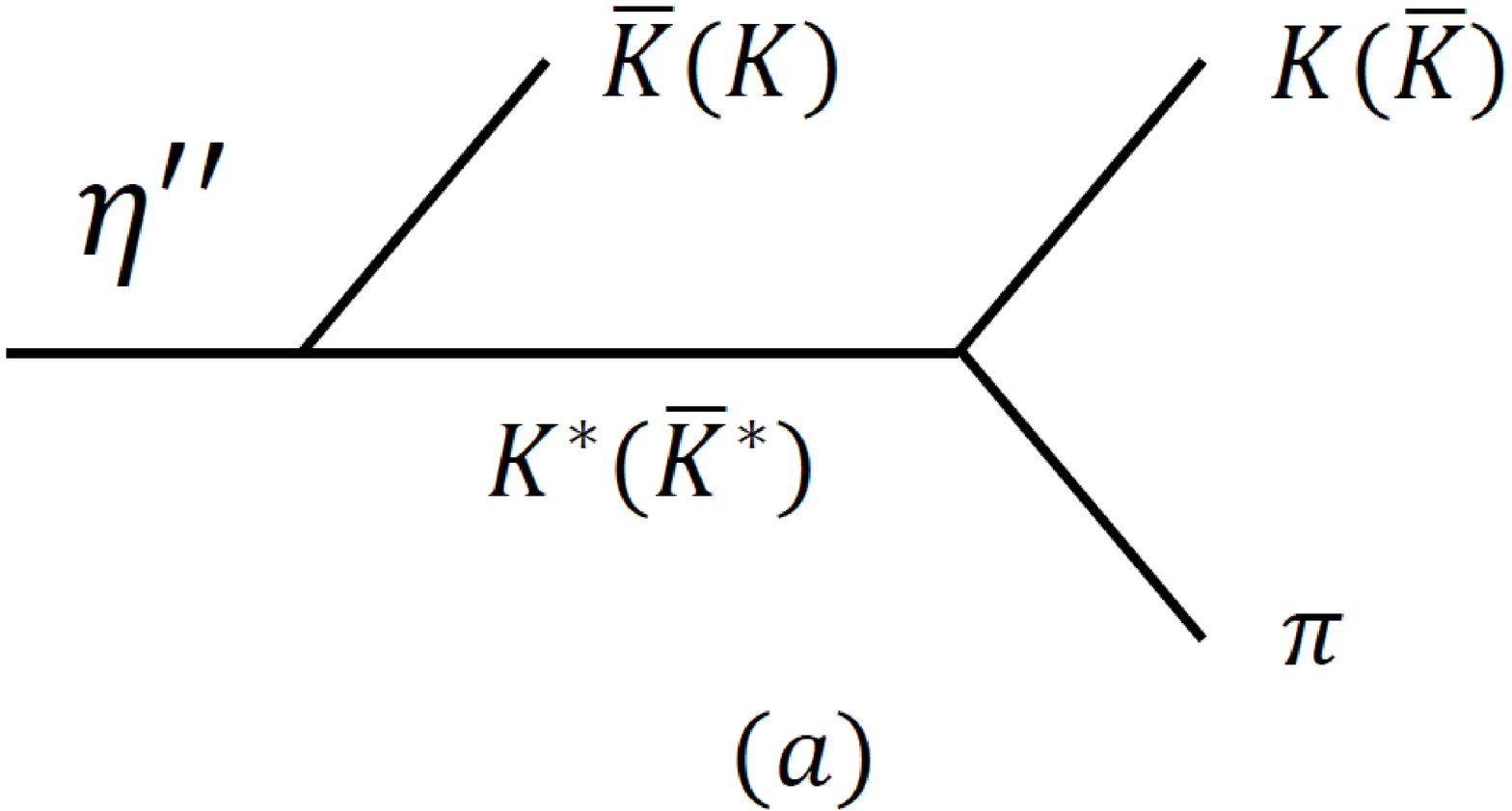}
  \includegraphics[width=2in]{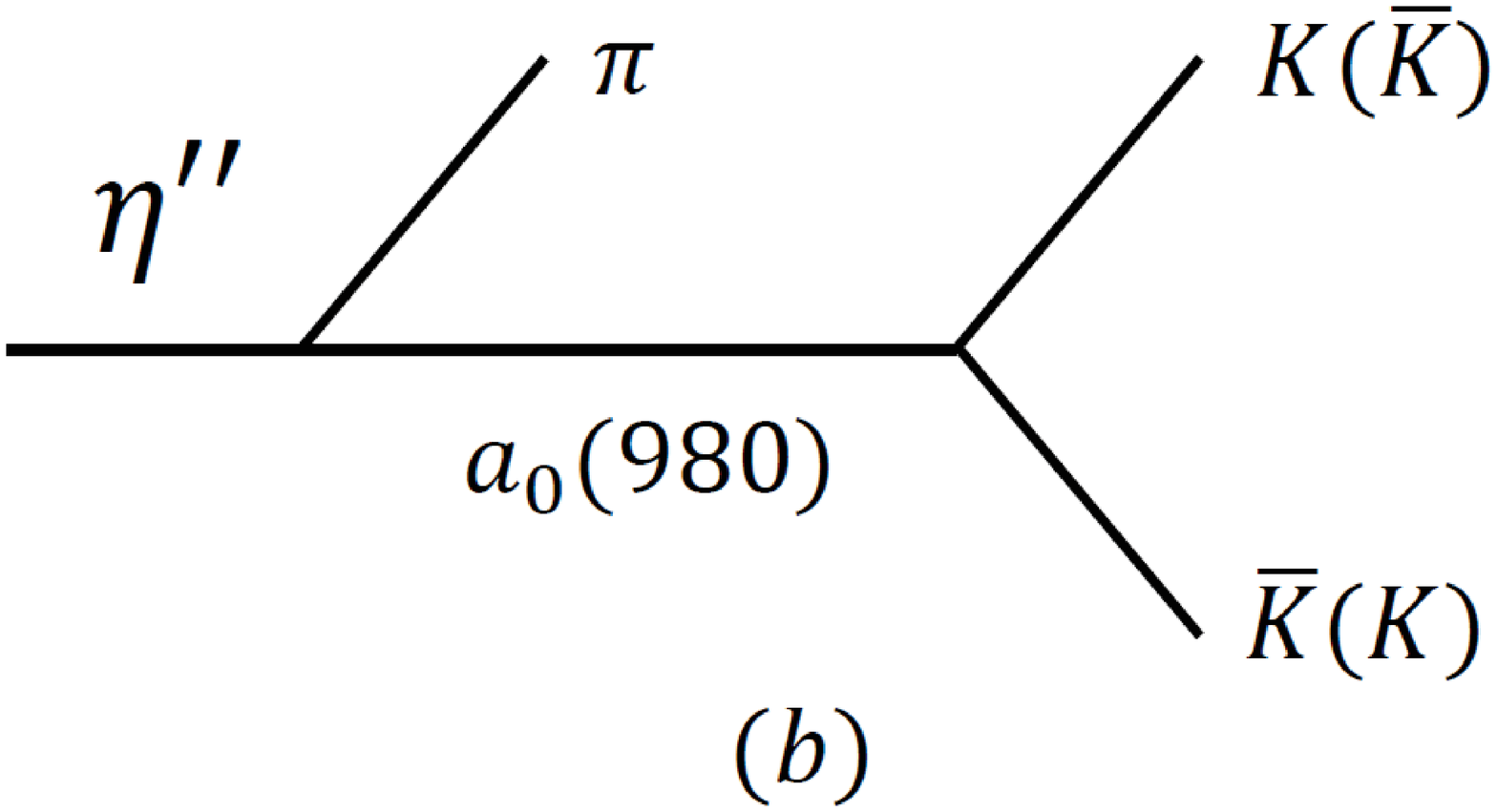}
  \includegraphics[width=2in]{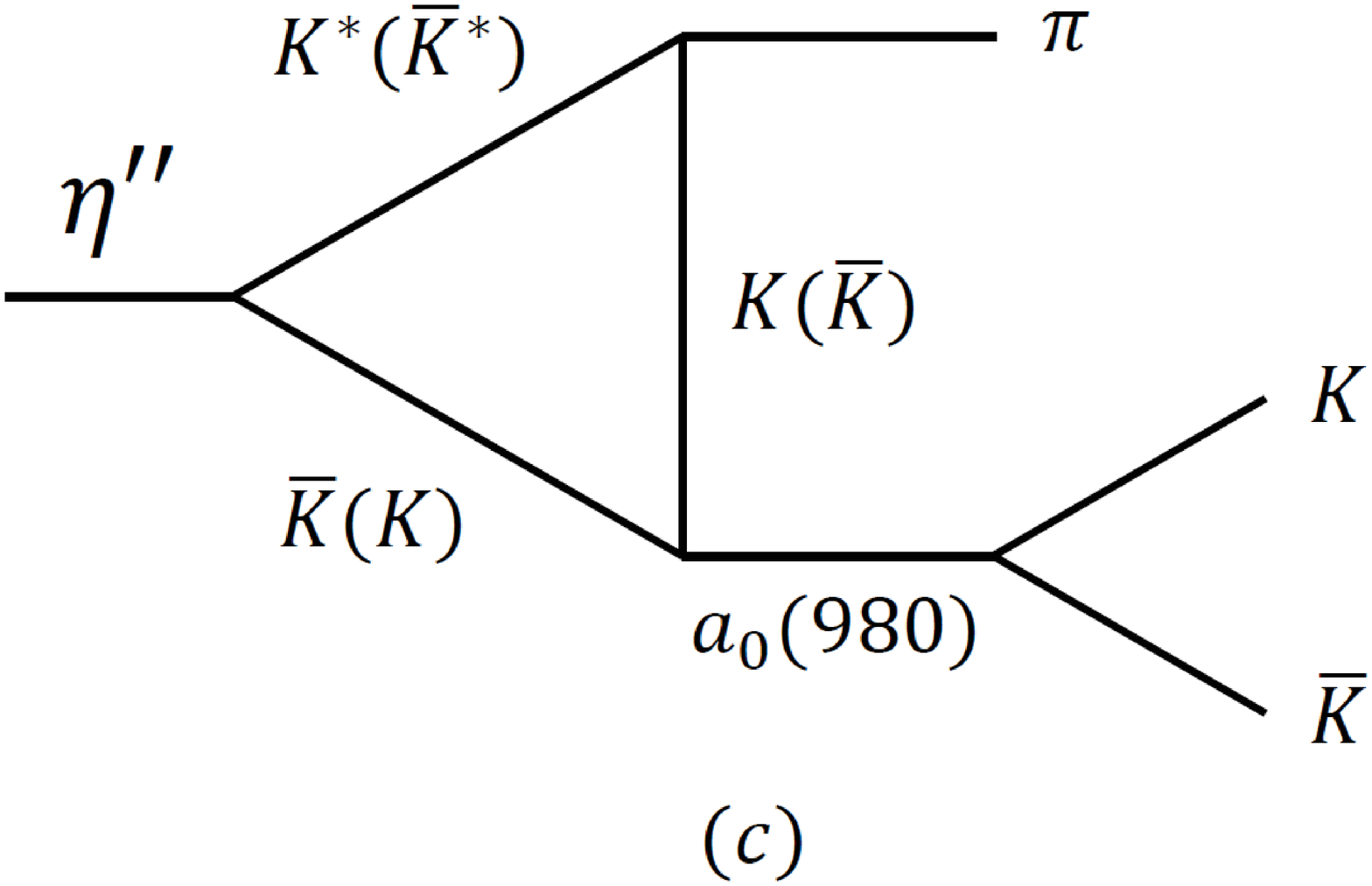}
  \caption{Transition mechanisms for $\eta''\to K\bar{K}\pi$. (a) describes the tree-level transition of $\eta''\to K^*\bar{K}+c.c.\to K\bar{K}\pi$, (b) describes $\eta''\to a_0(980)\pi\to K\bar{K}\pi$, and (c) describes the TS mechanism which enhances the production of the intermediate $a_0(980)\pi$.}\label{fig08}
\end{figure}

A coherent study of the decays of $\eta''\to K\bar{K}\pi$ and $\eta\pi\pi$ is required to quantify the contributions from the TS mechanism, and also to determine the parameters introduced in the formulation. For $\eta''\to K\bar{K}\pi$ the decay can occur via the processes illustrated by Fig.~\ref{fig08}, where (a) and (b) describe the tree-level transitions via the intermediates two-body decays $\eta''\to K^*\bar{K}+c.c.$ and $a_0(980)\pi$, respectively, and (c) describes the contribution from the TS mechanism. Note that the couplings of $g_{\eta''K^*\bar{K}}$ and $g_{\eta'' a_0K\bar{K}}$ are bare couplings to be determined by experimental data with the dominant transition mechanisms included.

The amplitude for Fig.~\ref{fig08} (a) has the following form:
The $M_{tree1}$ is the sum of the amplitude of $K^*\bar{K}$ channel and its charge conjugate:
\begin{eqnarray}
M_{tree1}&=&-g_{\eta'' K^{*0}\bar{K}^0}g_{K^{*0}K^0\pi^0}(2p_1-q)_{\mu}\frac{i(-g^{\mu\nu}+\frac{q^{\mu}q^{\nu}}{q^2})}{q^2-m_{K^*}^2+i m_{K^*}\Gamma_{K^*}(q^2)}(q-2p_2)_{\nu}\nonumber\\
&-&g_{\eta'' \bar{K}^{*0}K^0}g_{\bar{K}^{*0}\bar{K}^0\pi^0}(2p_1-q')_{\mu}\frac{i(-g^{\mu\nu}+\frac{q'^{\mu}q'^{\nu}}{q'^2})}{q'^2-m_{K^*}^2+i m_{K^*}\Gamma_{K^*}(q'^2)}(q'-2p_2)_{\nu},
\end{eqnarray}
where the $q$ and $q'$ are the invariant mass of $K^*$ and $\bar{K}^{*}$, respectively. And for Fig.~\ref{fig08} (b) it reads
\begin{eqnarray}
M_{tree2}&=&-g_{\eta'' a_0\pi^0}g_{a_0K^0\bar{K}^0}\frac{i}{s_{a_0}-m_{a_0}^2+\Pi_{a_0(s_(a_0))}}.
\end{eqnarray}
The loop amplitude of Fig.~\ref{fig08} (c) reads
\begin{eqnarray}
M_{tri3}&=&2g_{\eta'' K^{*0}\bar{K}^0}g_{K^{*0}K^0\pi^0}g_{a_0K^{0}\bar{K}^0}g_{a_0K^0\bar{K}^0}[I_1(s_1,s_3)+I_2(s_1,s_3)]\frac{i}{s_3-m_{a_0}^2+\Pi_{a_0(s_3)}} \ ,
\end{eqnarray}
where the triangle loop functions $I_{1,2}(s_1,s_3)$ have been defined in the previous subsection.

\begin{figure}[H]
  \centering
  \includegraphics[width=2in]{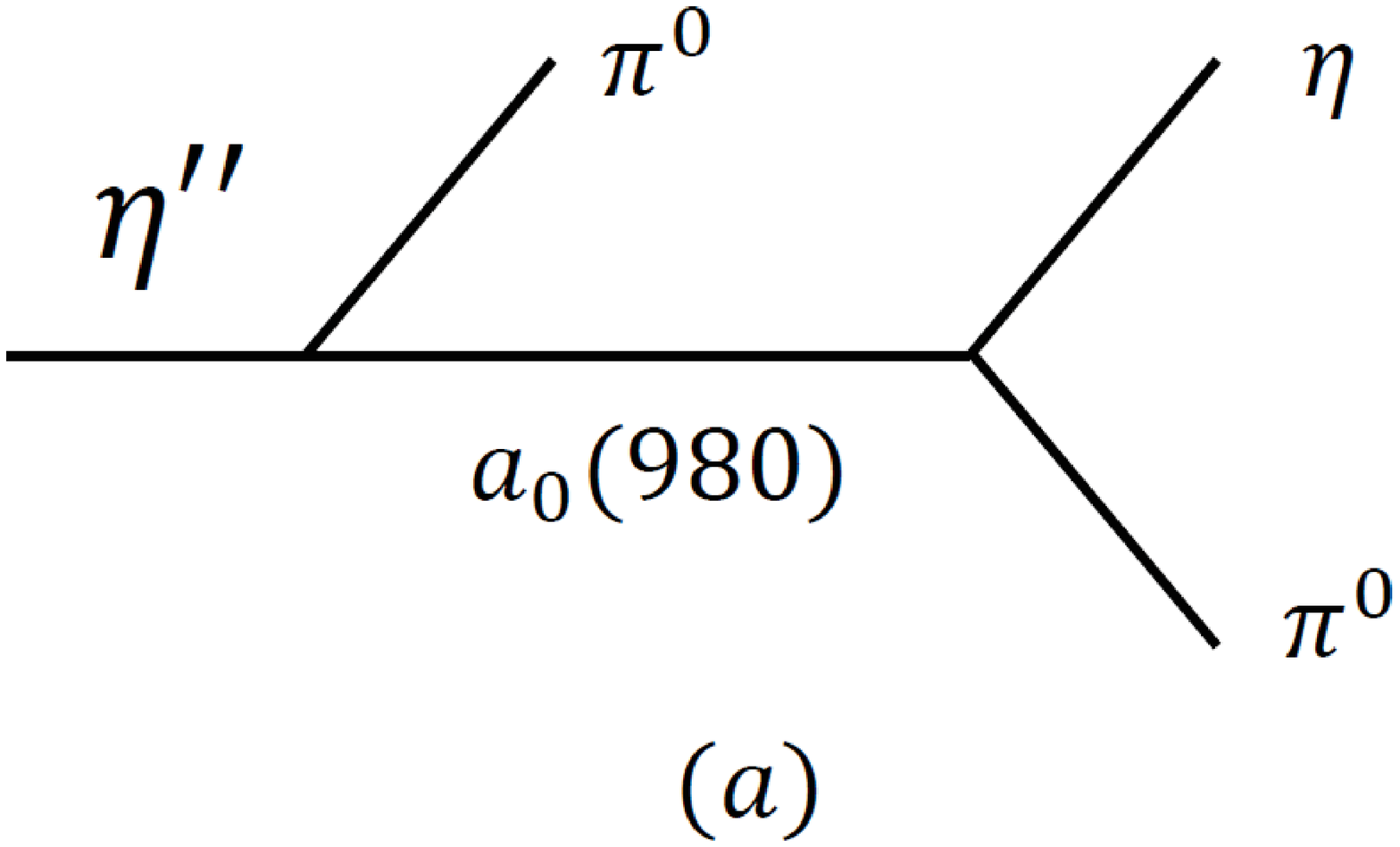}
  \includegraphics[width=2in]{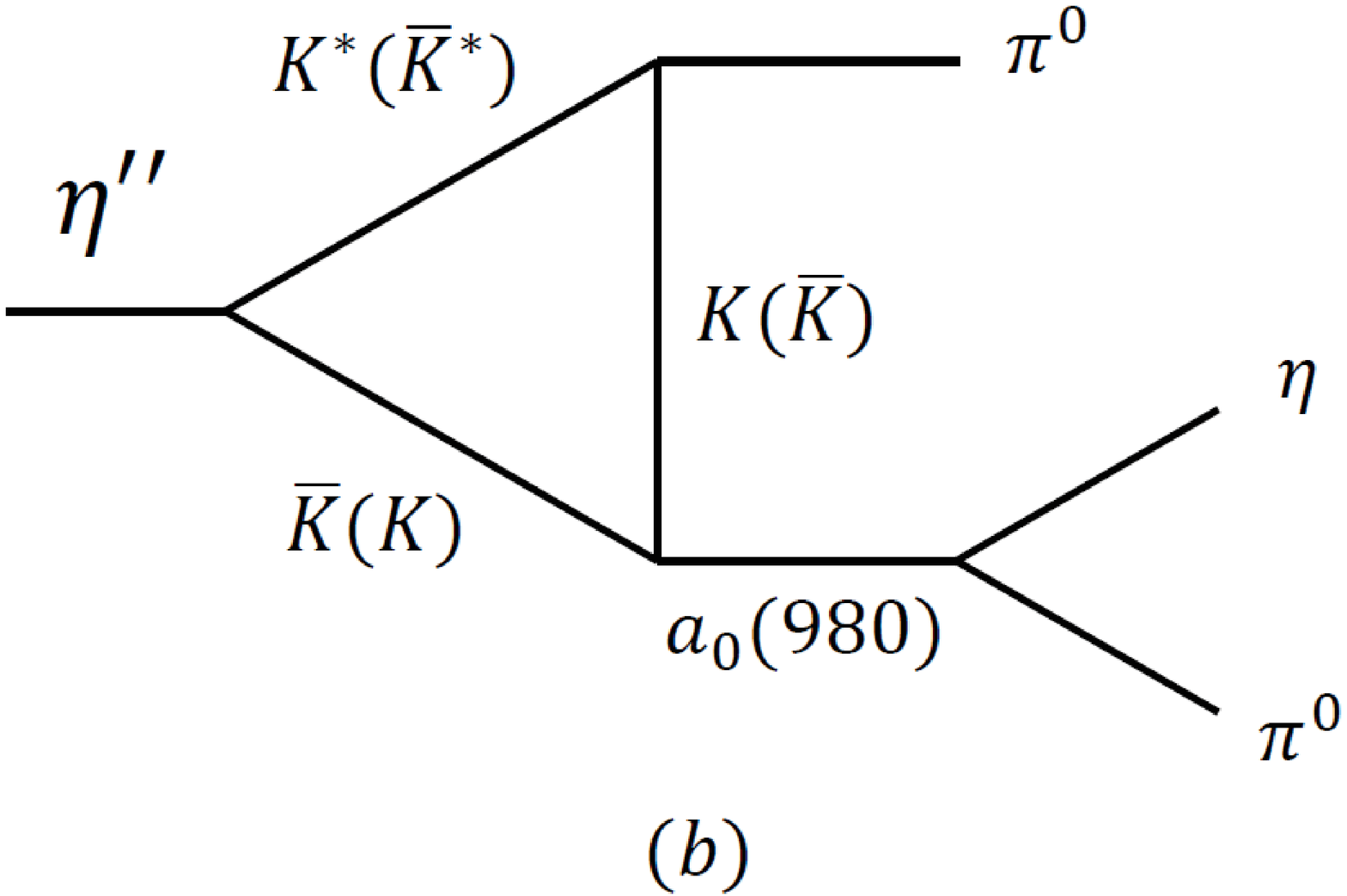}
  \caption{Transition mechanisms for $\eta''\to\eta\pi\pi$. (a) describes the tree-level transition of $\eta'' \to a_0(980)\pi\to\eta\pi\pi$, and (b) describes the TS mechanism which enhances the production of the intermediate $a_0(980)\pi$.}\label{fig09}
\end{figure}

Similarly, the decay of $\eta''\to \eta\pi\pi$ is described by Fig.~\ref{fig09}, where (a) is the tree-level transition via the intermediate $a_0(980)\pi$ and (b) describes the contribution from the TS mechanism. Again, we can see that the bare vertex couplings $g_{\eta''K^*\bar{K}}$ and $g_{\eta'' a_0K\bar{K}}$ are the same as those appearing in Figs.~\ref{fig07} and \ref{fig08}. The corresponding amplitudes for Fig.~\ref{fig09} (a) and (b) are as follows:
\begin{eqnarray}
M_{tree3}=-\frac{1}{\sqrt{2}}g_{\eta'' a_0\pi^0}g_{a_0\eta\pi^0}\left(\frac{i}{s_{12}-m_{a_0}^2+\Pi_{a_0(s_{12})}}+\frac{i}{s_{13}-m_{a_0}^2+\Pi_{a_0(s_{13})}}\right),
\end{eqnarray}
and
\begin{eqnarray}
M_{tri2}&=&\frac{2}{\sqrt{2}}g_{\eta'' K^{*0}\bar{K}^0}g_{K^{*0}K^0\pi^0}g_{a_0K^{0}\bar{K}^0}g_{a_0\eta\pi^0}[I_1(s_1,s_3)+I_2(s_1,s_3)]\nonumber\\
&&\times\left[\frac{i}{s_{12}-m_{a_0}^2+\Pi_{a_0(s_{12})}}+\frac{i}{s_{13}-m_{a_0}^2+\Pi_{a_0(s_{13})}}\right],
\end{eqnarray}

\subsection{Coupling constants and relative phases}

In Tab.~\ref{tab-coupling} all the vertex couplings are listed. Couplings $g_{a_0\eta\pi}=3.02$ GeV and $g_{a_0K^+K^-}=-g_{a_0K^0\bar{K}^0}=2.24$ GeV are determined by the KLOE Collaboration in $\phi\to\gamma a_0(980)\pi\to\gamma\pi\eta$~\cite{Aloisio:2002bsa}, and $g_{f_0K^+K^-}$ and $g_{f_0\pi^+\pi^-}$ are based on the data of Ref.~\cite{Aloisio:2002bt}. We note that the couplings from the KLOE Collaboration are chosen for consistency. Since the unitarized propagators for $a_0$ and $f_0$ are adopted, the couplings to $K\bar{K}$ should be extracted based on the same parametrization scheme. In fact, with the same couplings given in Tab.~\ref{tab-coupling}, significant difference between the Flatte propagator and the unitarized propagator can be identified. In our calculation the $g_{\eta''K^*\bar{K}}$ and the $g_{\eta''a_0\pi}$ are the bare couplings determined by the combined analysis of the decays of $\eta''\to K\bar{K}\pi$, $\eta\pi\pi$ and $3\pi$. Note that the corresponding vertex appears in all these processes. The physical coupling $\tilde{g}_{\eta''a_0\pi}$ can be extracted by the combined contributions from the tree-level amplitude and TS mechanism in $\eta''\to a_0\pi$. We also note that since experimentally the branching ratio of $\eta''\to a_0\pi$ is much smaller than that for $\eta''\to K\bar{K}^*+c.c.$, we do not expect that the rescattering of $\eta''\to a_0\pi\to K\bar{K}^*+c.c.$ has significant renormalization contributions to the coupling of $g_{\eta''K^*\bar{K}}$. Namely, we approximate  $\tilde{g}_{\eta''K^*\bar{K}}\simeq g_{\eta''K^*\bar{K}}$. It should be cautioned that the determination of the couplings $\tilde{g}_{\eta''K^*\bar{K}}$ and $\tilde{g}_{\eta''a_0\pi}$ will strongly depend on the experimental accuracies of the corresponding branching ratios. For the coherent analysis done in this work it is sufficient for us to demonstrate the patterns arising from the TS mechanism. Therefore, we only list the central values for these two couplings in Tab.~\ref{tab-coupling}.

\begin{table}
\centering
\caption{Coupling constants and phase angles determined in the combined analysis of $\eta''\to K\bar{K}\pi$, $\eta\pi\pi$ and $3\pi$. The coupling $\tilde{g}_{\eta''a_0\pi}$ is defined with the TS contributions taken into account.}\label{tab-coupling}
\begin{tabular}{|c|c|}
  \hline
  Parameters & Values \\
  \hline
  $g_{f_0K^+K^-}$ & $5.92\pm 0.13$ GeV~\cite{Aloisio:2002bt} \\
  \hline
  $g_{f_0\pi^+\pi^-}$ & $2.96\pm 0.12$ GeV~\cite{Aloisio:2002bt} \\
  \hline
  $g_{a_0K^+K^-}$ & $2.24\pm 0.11$ GeV ~\cite{Aloisio:2002bsa}\\
  \hline
  $g_{a_0\eta\pi}$ & $3.02\pm 0.35$ GeV ~\cite{Aloisio:2002bsa}\\
  \hline
  $g_{\eta''K^*\bar{K}}$ & 3.97\\
  \hline
  $g_{\eta''a_0\pi}$ & 0.72 GeV\\
  \hline
  $\tilde{g}_{\eta''a_0\pi}$ & 0.50 GeV\\
  \hline
  $\{\phi_a,\phi_b\}$ & $\{60^{\circ}, -40^{\circ}\}$\\
  \hline
\end{tabular}
\end{table}

Hence, we collect the amplitudes and express the total amplitudes for each channel as follows:
\begin{eqnarray}
M_{tot(\pi^+\pi^-\pi^0)}&= &M_{tri1}+M_{trimix}+M_{treemix}\label{eqmtot1}\\
M_{tot(K^0\bar{K}^0\pi^0)}&=&M_{tree1}+M_{tree2}+M_{tri3}\\
M_{tot(\eta\pi^0\pi^0)}&= &M_{tree3}+M_{tri2}.\label{eqmtot3}
\end{eqnarray}
The partial widths for $\eta''\to f_0\pi\to\pi^+\pi^-\pi^0$, $\eta''\to a_0\pi\to\eta\pi\pi$ and $\eta''\to K\bar{K}\pi$ channels are
\begin{eqnarray}
&&\Gamma_{\eta''\to\pi^+\pi^-\pi^0}=\frac{1}{2\sqrt{s}}\int d\Phi_{\pi^+\pi^-\pi^0} |M_{tot(\pi^+\pi^-\pi^0)}|^2\nonumber\\
&&\Gamma_{\eta''\to K\bar{K}\pi}=6\Gamma_{K^0\bar{K}^0\pi^0}=\frac{6}{2\sqrt{s}}\int d\Phi_{K^0\bar{K}^0\pi} |M_{tot(K^0\bar{K}^0\pi^0)}|^2\\
&&\Gamma_{\eta''\to\eta\pi\pi}=3\Gamma_{\eta\pi^0\pi^0}=\frac{3}{2\sqrt{s}}\int d\Phi_{\eta\pi^0\pi^0} |M_{tot(\eta\pi^0\pi^0)}|^2\nonumber,
\end{eqnarray}
where $\Phi_{abc}$ is the phase space of $\eta''\to abc$. For $J/\psi$ radiative decay, the following expressions are adopted
\begin{eqnarray}
&&\Gamma_{J/\psi\to\gamma \eta''\to\gamma ABC}=\int d\sqrt{s}\frac{2s}{\pi}\frac{\Gamma_{J/\psi\to\gamma \eta''(\sqrt{s})}\Gamma_{\eta''\to ABC(\sqrt{s})}}{(s-m_{\eta''}^2)^2+m_{\eta''}^2\Gamma_{\eta''}^2}\label{eqjpsi}\\
&&\Gamma_{J/\psi\to\gamma\eta''}\sim \frac{(m_{J/\psi}^2-s)^3}{48\pi^2 m_{J/\psi}^2},
\end{eqnarray}
where $\Gamma_{\eta''}\simeq\Gamma_{K\bar{K}\pi}+\Gamma_{\eta\pi\pi}$, $m_{\eta''}=1.42$ GeV and $m_{J/\psi}=3.1$ GeV.

There are four types of couplings, corresponding to $\eta''\to K^*\bar{K}$, $\eta''\to a_0\pi$, $V\to PP$, and $S\to PP$ ($S$ denotes $a_0$ or $f_0$). By taking the $g_{VPP}$ as a real coupling in the SU(3)-flavor symmetry, we introduce three energy independent phase factors $e^{i\phi_1}$, $e^{i\phi_2}$ and $e^{i\phi_3}$ to the couplings $g_{\eta'' K^{*0}\bar{K}^0}$, $g_{\eta'' a_0\pi}$ and $g_{a_0K^0\bar{K^0}}$, respectively. Namely, these phase factors are shared by the SU(3)-flavor multiplets because of flavor symmetry. For instance, in respect to the real coupling $g_{VPP}$ we express $g_{a_0K^+K^-}=-g_{a_0K^0\bar{K}^0}=2.24e^{i\phi_3}$ GeV and $g_{f_0K^+K^-}=g_{f_0K^0\bar{K}^0}=5.92e^{i\phi_3}$ GeV. By examining the transition processes of Figs.~\ref{fig07}, \ref{fig08} and \ref{fig09}, one notices that there are only two independent phase angles needed in the combined analysis. For convenience, we set $\phi_2=0$, and redefine the two remained independent phase angles as the linear combination of $\phi_1$ and $\phi_3$, i.e. $\phi_a\equiv\phi_1-\phi_2-\phi_3=\phi_1-\phi_3$, and $\phi_b\equiv\phi_1-\phi_2+\phi_3=\phi_1+\phi_3$.

A feature arising from the experimental measurements of the $\eta''\to K\bar{K}\pi$ and $\eta\pi\pi$ is that the branching ratio for $K\bar{K}\pi$ is much larger than that for $\eta\pi\pi$. In the combined analysis, such a difference is determined by the relative coupling strength between $g_{\eta'' a_0\pi}$ and $g_{\eta'' K^{*0}\bar{K}^0}$ and the relative phase factors which are treated as parameters and to be constrained by experimental data. Our strategy of determining these couplings and relative phase angles is as follows: We first require that the relative branching ratio fractions between $K\bar{K}\pi$ and $\eta\pi\pi$ satisfy the experimental data, and then the sum of the partial widths from these two channels can contribute to up to 50 MeV of the total width. Although both the total width and partial decay widths of $\eta''$ have large uncertainties, it is sufficient for understanding nearly all the existing puzzling questions about the $\eta(1405/1475)$. With the cut-off parameter $\alpha \Lambda_{QCD}=500 \ \text{MeV}$, i.e. $\Lambda_R=m_R+0.5 \ \text{GeV}$, the couplings and phase angles are determined and listed in Tab.~\ref{tab-coupling}.

\section{Results and discussions}

With the determined couplings in Tab.~\ref{tab-coupling}, the relative partial widths are listed in Tab.~\ref{tab1}, where the dominance of the $K^*\bar{K}$ ($M_{tree1}$) channel in $\eta''$ decay is evident. For the $K\bar{K}\pi$ final state the relatively small contributions from the intermediate $a_0\pi$ provide a small destructive cancellation which leads to ${\Gamma_{K^*\bar{K}}}/{\Gamma_{K\bar{K}\pi}}=1.06$. This is also an indication that a partial wave analysis is need for the combined analysis of all these three decay channels into $K\bar{K}\pi$, $\eta\pi\pi$ and $3\pi$. We will see later that the interference is more significant in the $\eta\pi\pi$ channel.

\begin{figure}[H]
  \centering
  \includegraphics[width=3in]{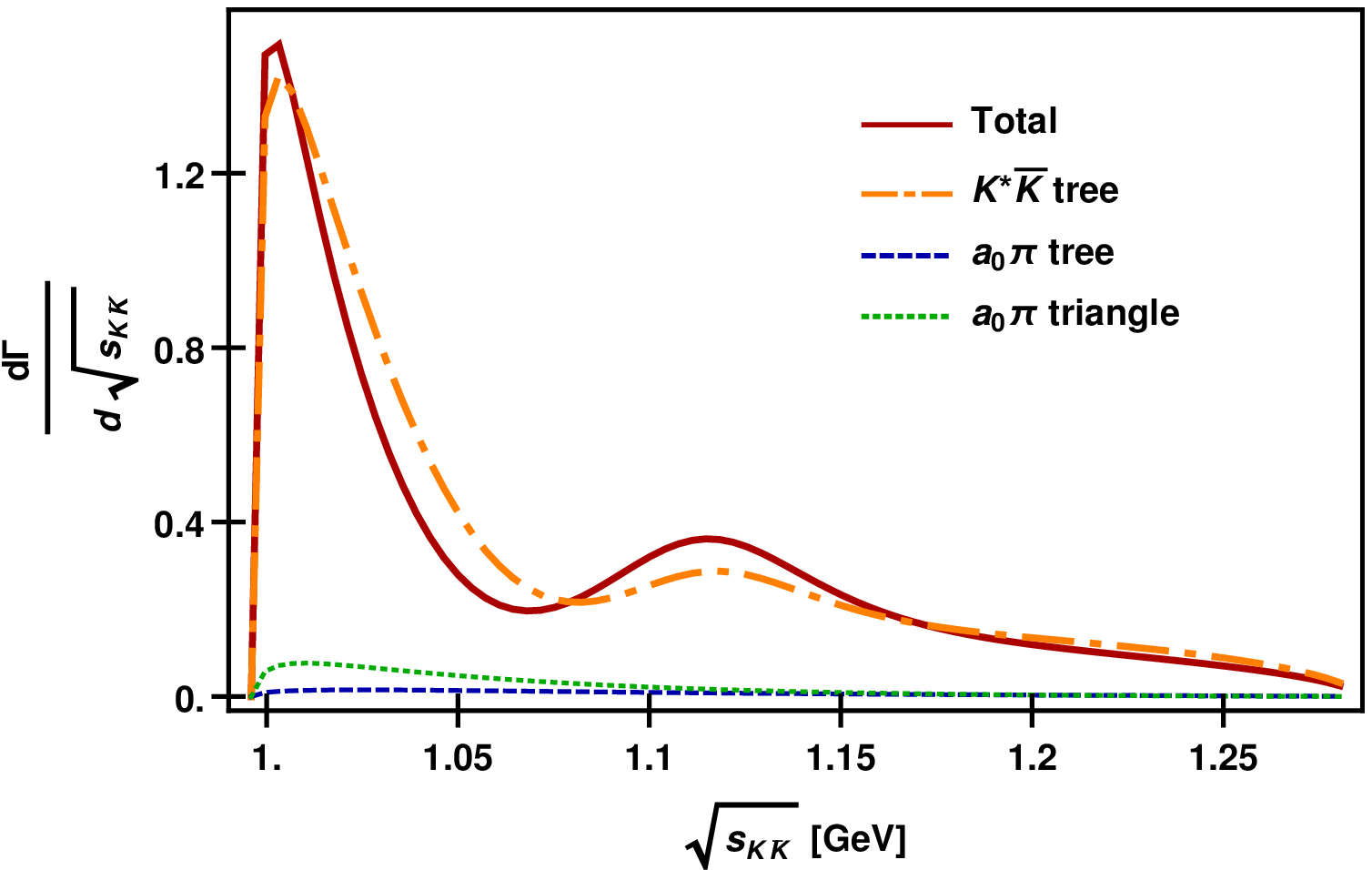}
  \includegraphics[width=3in]{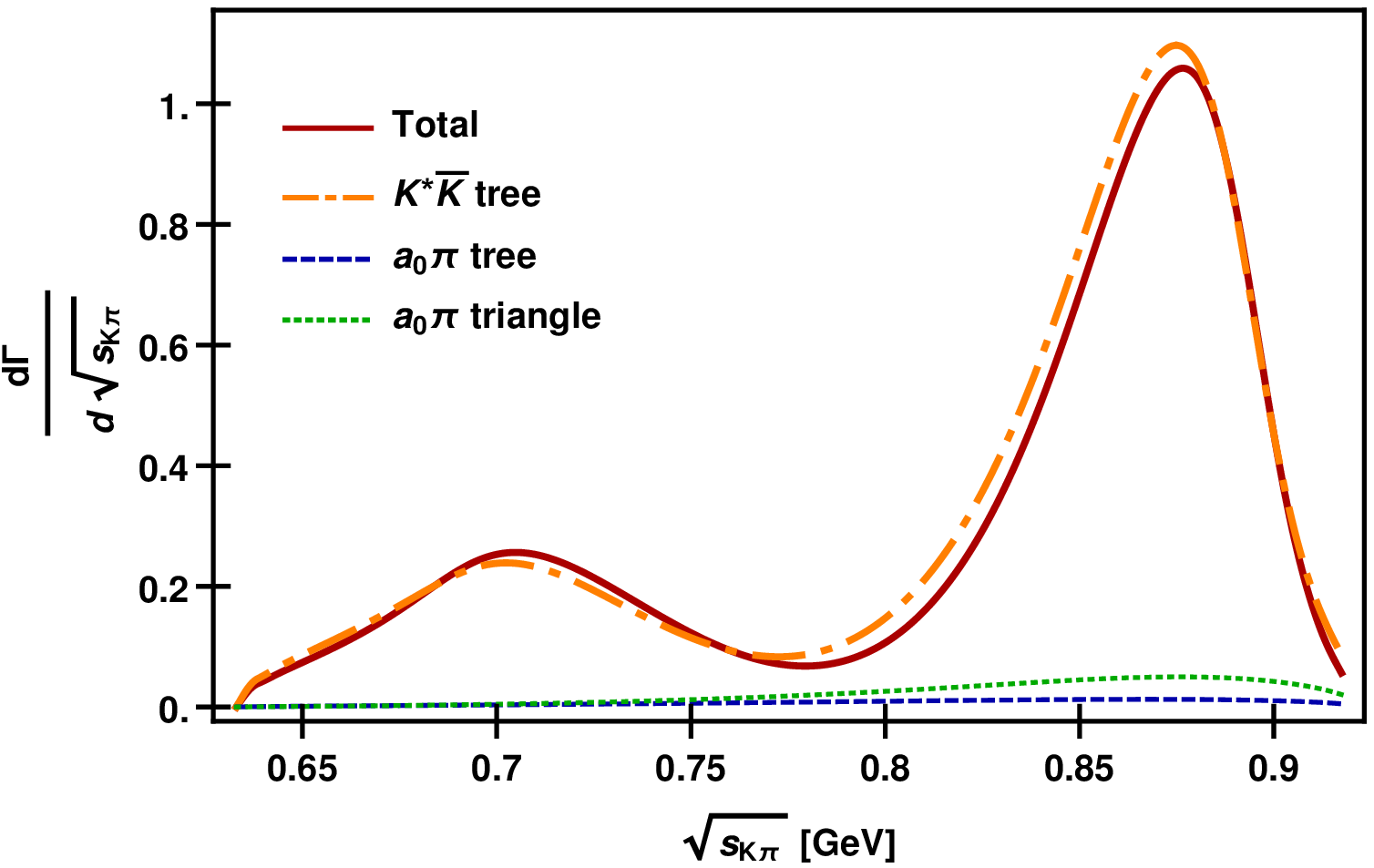}
  \caption{Invariant mass spectra of $K\bar{K}$ (left panel) and $K\pi$ (right panel) at $\sqrt{s}=1.42$ GeV in $\eta''\to K\bar{K}\pi$. The full calculations are denoted by the red solid lines. The blue dashed lines are the contributions from the tree-level $a_0(980)\pi$ amplitude. The green dotted lines denote the contribution of triangle diagrams, and the orange dot-dashed lines are from tree diagrams of $K^*\bar{K}$.}\label{fig10}
\end{figure}

The $K\bar{K}$ and $K\pi$ spectra at $\sqrt{s}=1.42$ GeV are shown in Fig.~\ref{fig10}. In the left panel of Fig.~\ref{fig10}, since $K^*\bar{K}$ is dominant in the $K\bar{K}\pi$ channel, the peak of the spectrum (red line) near 1 GeV is not generated by the pole of $a_0(980)$, but a projection of the peaks of $K^*$ and $\bar{K}^*$ in the Dalitz plot into the $K\bar{K}$ spectrum. The $K^*$ peak is evident in the right panel of Fig.~\ref{fig10}. It is worth mentioning that the intensity of triangle diagrams in $K\bar{K}\pi$ channel is very small, although it becomes compatible with the tree diagram for $\eta''\to a_0\pi$ in $\eta''\to \eta\pi\pi$.

\begin{table}
\centering
\caption{The partial widths of $\eta''$ decays calculated for different amplitudes noted by the brackets. The second row of the experiment branching ratio is calculated by $BR(J/\psi\to\gamma\eta''\to\gamma a_0\pi\to\gamma\eta\pi^0\pi^0)=(8.5\pm1.75)\times 10^{-5}$~\cite{BESIII:2012aa} and $BR(J/\psi\to\gamma\eta''\to\gamma K\bar{K}\pi)=(2.8\pm0.6)\times 10^{-3}$~\cite{Patrignani:2016xqp}. The first row of the experiment branching ratio is taken from Ref.~\cite{Baillon:1967zz}, but is not used in this work. }\label{tab1}

\begin{tabular}{|c|c|c|c|c|c|}
\hline
Channels & Tree diagrams [MeV] & Triangle loop [MeV] & Total [MeV] & BR from experiments & BR from our work \\
\hline
$\eta''\to K\bar{K}\pi$ & $\begin{tabular}{c} 48.7 ($M_{tree1}$) \\ 1.10 ($M_{tree2}$) \end{tabular}$ & 3.27 ($M_{tri3}$) & 45.7 & $\frac{\Gamma_{K^*\bar{K}}}{\Gamma_{K\bar{K}\pi}}=0.5\pm0.1$ & $\frac{\Gamma_{K^*\bar{K}}}{\Gamma_{K\bar{K}\pi}}=1.06$\\
\hline
$\eta''\to a_0\pi\to\eta\pi\pi$ & 8.7 ($M_{tree3}$) & 15.5 ($M_{tri2}$) & 4.25 & $\frac{\Gamma_{a_0\pi\to\eta\pi^0\pi^0}}{\Gamma_{K\bar{K}\pi}}=3.04^{+0.16}_{-0.11}\%$ & $\frac{\Gamma_{a_0\pi\to\eta\pi^0\pi^0}}{\Gamma_{K\bar{K}\pi}}=3.03\%$\\
\hline
$\eta''\to\pi^+\pi^-\pi^0$ & 0.0041 ($M_{treemix}$) & $\begin{tabular}{c} 0.186 ($M_{tri1}$) \\ 0.012 ($M_{trimix}$) \end{tabular}$ & 0.242 & $\frac{\Gamma_{\pi^+\pi^-\pi^0}}{\Gamma_{\eta\pi^0\pi^0}}=(17.9\pm4.2)\%$ & $\frac{\Gamma_{\pi^+\pi^-\pi^0}}{\Gamma_{\eta\pi^0\pi^0}}=17.1\%$ \\
\hline
\end{tabular}

\end{table}

\begin{figure}[H]
  \centering
  \includegraphics[width=5in]{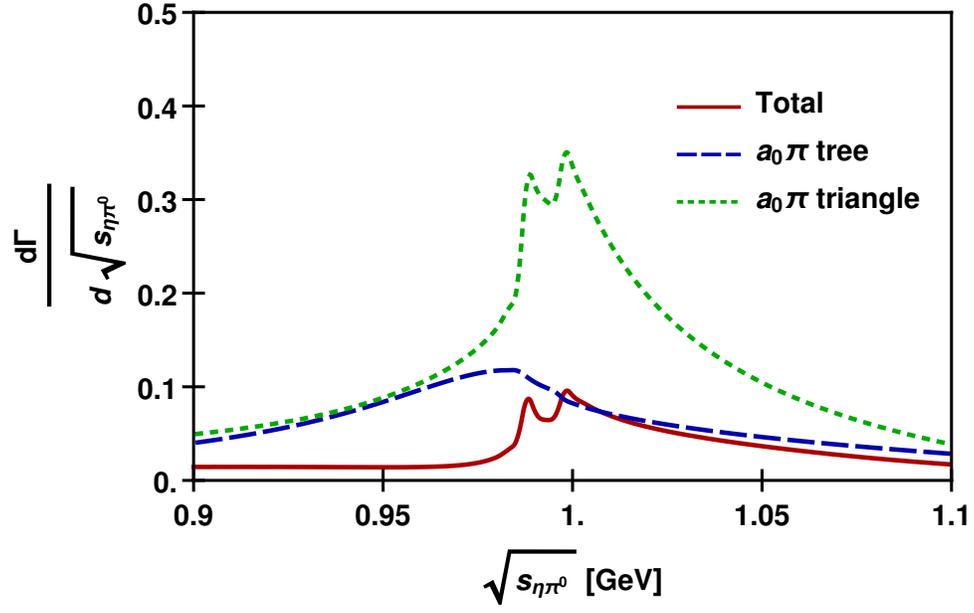}
  \caption{The invariant mass spectrum of $\eta\pi^0$ at $\sqrt{s}=1.42$ GeV in $\eta''\to \eta\pi\pi$. The full calculation is denoted by the red solid line.  The green dotted and blue dashed lines represent contributions from the triangle and tree-level diagrams, respectively.}\label{fig11}
\end{figure}

For the $\eta''\to\eta\pi\pi$ channel the exclusive contributions from $M_{tri2}$ and $M_{tree3}$ are both larger than the experimental values. It requires a destructive interference between these two amplitudes which is crucial for understanding the rather small branching ratio for $\eta''\to\eta\pi\pi$. To illustrate this destructive interference, the $\eta\pi^0$ mass spectrum at $\sqrt{s}=1.42$ GeV is plotted in Fig.~\ref{fig11} with the exclusive and coherent contributions from $M_{tri2}$ and $M_{tree3}$. It is noticeable that the destructive interference leads to a nontrivial structure for the $a_0(980)$ lineshape. In particular, there is a double-peak structure arising from the $K^+K^-$ and $K^0\bar{K}^0$ thresholds of the triangle amplitude $M_{tri2}$. This prediction can be examined by future experiments with higher energy resolution.

It is helpful to understand why the contribution of triangle diagrams in $\eta\pi\pi$ channel is one order of magnitude larger than that in $K\bar{K}\pi$ channel. On the one hand, this may be due to a larger phase space for $\eta''\to\eta\pi\pi$. On the other hand, the most contributive part of a triangle diagram comes form the vicinity of the pole of $a_0(980)$, which is in the middle of the phase space for $\eta''\to\eta\pi\pi$, but at the boundary for $\eta''\to K\bar{K}\pi$.

Due to the presence of the TS mechanism the effective coupling $\tilde{g}_{\eta a_0\pi}$ should include contributions from the tree diagram and the TS, which yields $\tilde{g}_{\eta'' a_0\pi}=0.5$ GeV. It can be compared with the bare coupling $g_{\eta'' a_0\pi}=0.72$ GeV extracted from the overall analysis. We also note that the overall analysis gives the bare coupling $g_{\eta'' K^*\bar{K}}=3.97$ which is slightly larger than that adopted in Refs.~\cite{Wu:2011yx,Wu:2012pg} since there is a weak destructive interference in the $K\bar{K}\pi$ channel in our result. However, the dominance of the $K^*\bar{K}$ in $\eta''\to K\bar{K}\pi$ is evident.

Since the two-body decay of $\eta''\to a_0\pi$ has not been explicitly measured, we estimate the relative branching ratio fraction $R_{\eta \pi^+\pi^-/K\bar{K}\pi}\equiv BR(J/\psi\to\gamma\eta''\to \gamma a_0\pi\to\gamma\eta\pi^+\pi^-)/BR(J/\psi\to\gamma\eta''\to\gamma K\bar{K}\pi)$ with the help of Eq.~(\ref{eqjpsi}). The integration ranges for the $\eta''\to K\bar{K}\pi$ and $\eta''\to\eta\pi^+\pi^-$ channels in Eq.~(\ref{eqjpsi}) are of $\sqrt{s}\in $[1.14 GeV, 1.65 GeV] and [0.82 GeV, 1.6 GeV], respectively. Then, the calculation gives $R_{\eta \pi^+\pi^-/K\bar{K}\pi}=6.6\%$ (at $m_{\eta''}=1.42$ GeV). This value is slightly smaller than the PDG averaged value of which the uncertainties are still large:

\begin{eqnarray}
\frac{Br(J/\psi\to\gamma\eta''\to\gamma\eta\pi^+\pi^-)}{Br(J/\psi\to\gamma\eta''\to\gamma K\bar{K}\pi)}=\frac{(3.0\pm0.5)\times10^{-4}}{(2.8\pm0.6)\times10^{-3}}\simeq (7.35\sim 15.9) \% \ .
\end{eqnarray}

Notice that partial wave analyses for these channels are still unavailable, and possible contributions from the $\sigma\eta$ channel to the $\eta\pi\pi$ final state are not excluded. Future improvement of the experimental data is strongly desired.

\begin{figure}[H]
  \centering
  \includegraphics[width=5in]{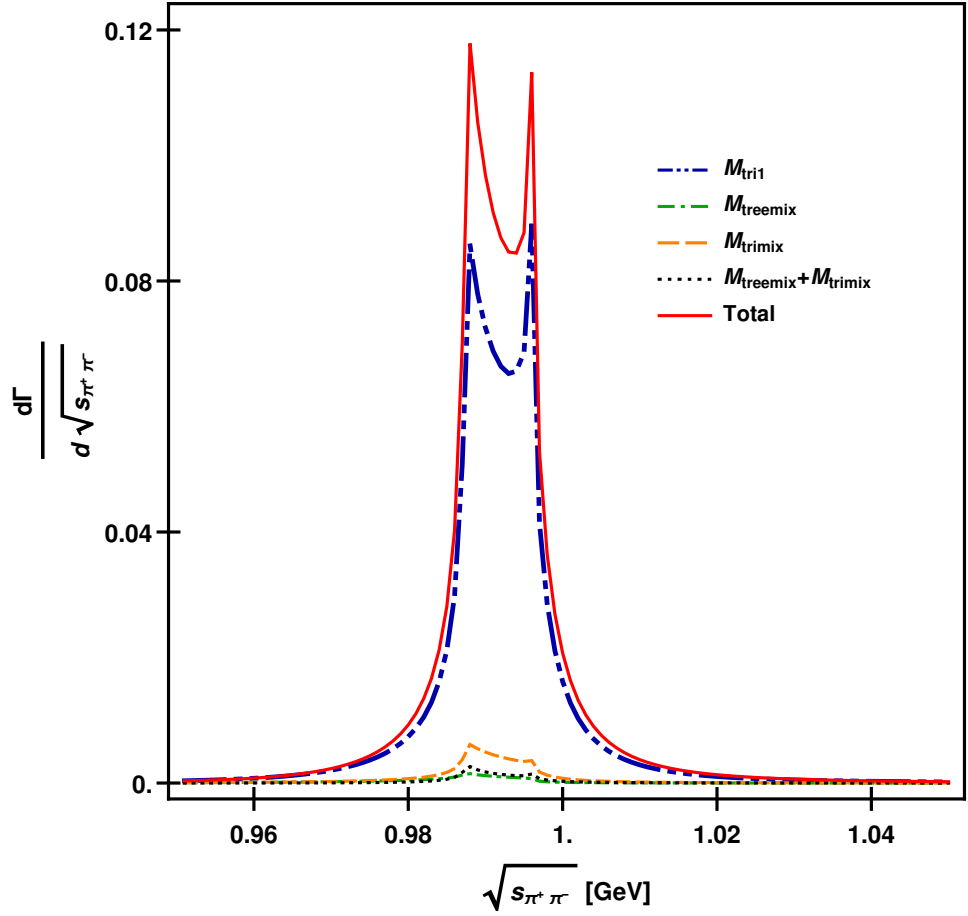}
  \caption{The $\pi^+\pi^-$ invariant mass spectrum at $\sqrt{s}=1.42$ GeV.  The full calculation is denoted by the red solid line. The blue dot-dot-dashed line is the spectrum with only triangle diagram of the $f_0\pi$ channel (Fig.~\ref{fig07} (a)). The green dot-dashed line is the contribution from the $a_0-f_0$ mixing with the bare $g_{\eta'' a_0\pi}$ coupling (Fig.~\ref{fig07} (c)), and the orange dashed line is the contribution from the $a_0-f_0$ mixing via the TS production of the $a_0(980)$ (Fig.~\ref{fig07} (b)). The black dotted line is the total contribution from the $a_0-f_0$ mixing.}\label{fig12}
\end{figure}

In the decay channel of $\eta''\to 3\pi$, the $\pi^+\pi^-$ invariant mass spectrum at $\sqrt{s}=1.42$ GeV is depicted in Fig.~\ref{fig12}. The typical narrow structure located between the charged and neutral $K\bar{K}$ thresholds is expected by the isospin violation mechanisms. In Fig.~\ref{fig12} the exclusive contributions from the isospin-violating TS diagram (Fig.~\ref{fig07} (a)), and $a_0-f_0$ mixing diagrams (Fig.~\ref{fig07} (b) and (c)) are plotted. Note that the $a_0-f_0$ mixing after the triangle rescatterings appears to be more important than the $a_0-f_0$ mixing through the tree diagram.

A unique feature with this channel is that the strength of the transition amplitude is quite insensitive to the cut-off energy. As discussed in Refs.~\cite{Wu:2011yx,Wu:2012pg}, the dispersive parts of the charged and neutral kaon loop transition amplitudes largely cancel out, and the main contributions to the isospin violations are from the absorptive parts which are in the vicinity of the on-shell kinematic region. This explains that the width effects from the intermediate $K^*$ do not change the lineshape of the narrow structure of the $\pi\pi$ spectrum.

\begin{figure}[H]
  \centering
  \includegraphics[width=5in]{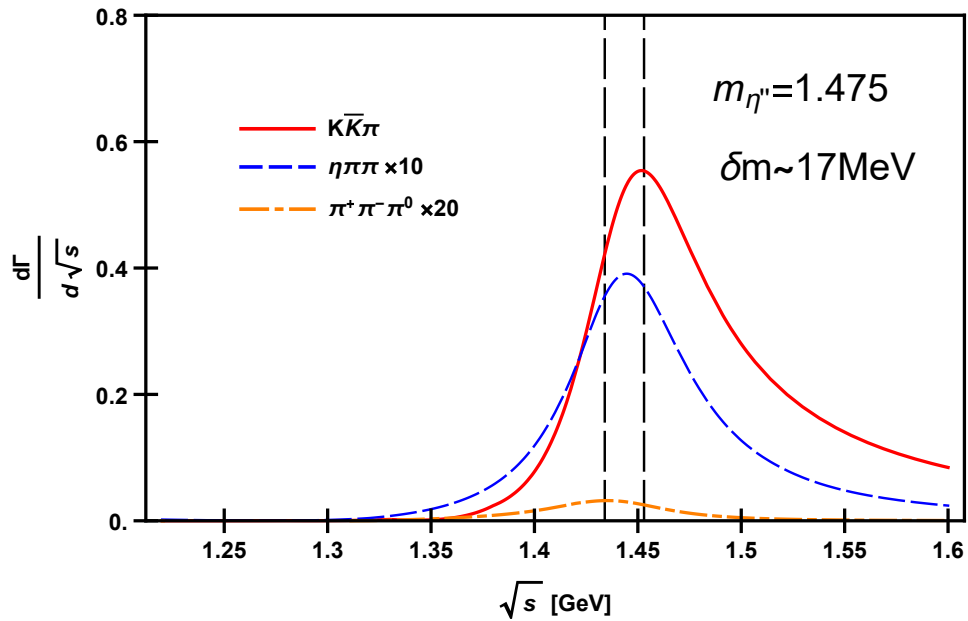}
  \caption{Invariant mass spectra of the $K\bar{K}\pi$ (red solid), $\eta\pi\pi$ (blue dashed and multiplied by 10) and $\pi^+\pi^-\pi^0$ (orange dot-dashed and multiplied by 20). The physical mass of the $\eta''$ applied in the propagator is 1.475 GeV.}\label{fig13}
\end{figure}

In Fig.~\ref{fig13} we show the spectra of $\eta''\to K\bar{K}\pi$, $\eta\pi\pi$ and $3\pi$ as a comparison of the different lineshapes for the same state in different channels. It confirms the main conclusion of Refs.~\cite{Wu:2011yx,Wu:2012pg}. Namely, due to the contributions from the TS mechanism the peak positions of the same state $\eta''$ are located in different masses in different channels. In particular, the spectrum of $\eta''$ in the $K\bar{K}\pi$ channel appears to be asymmetric and apparently deviated from a Breit-Wigner form. This is the manifestation of the $K^*\bar{K}$ dominance in $\eta''$ decays. As shown in Fig.~\ref{fig13} the shift of the peak positions due to the interferences from the TS mechanism can amount to almost 17 MeV.

It should be noted that the TS mechanism can also result in an energy-dependent description of the $\eta'' K^*\bar{K}+c.c.$ coupling. This will be reported in a followed-up work. For the purpose of understanding the impact of the TS mechanism on the width effects from the intermediate $K^*$ and the relative strengths of exclusive decay channels for $\eta''$, we actually fix the $\eta'' K^*\bar{K}+c.c.$ coupling as a constant in the present calculations.


\section{Conclusion}

In this work we revisit the topical problem of $\eta(1405/1475)$ in $J/\psi$ radiative decays into $K\bar{K}\pi$, $\eta\pi\pi$ and $3\pi$, where $\eta(1405)$ and $\eta(1475)$ are treated as being originated from the same state $\eta''$ but strongly affected by the triangle singularity mechanism. Different from the previous studies, we show that the non-zero width effects can be well-understood based on the analytical properties of triangle integral. We adopt the formulation by 't Hooft and Veltman~\cite{tHooft:1978jhc} to introduce a complex mass for the internal state. The triangle loop integral can be expressed in terms of the Spence functions for which the analytical properties can be investigated. We then show that the triangle singularity conditions can be described by motions of a set of kinematic functions in the complex plane.

In the case where the internal states have a small width, the absorptive part of the scalar loop integral can be expressed in a very compact form, which is exactly the same as that obtained from the cutting rule. The simplification of the analytic absorptive part is difficult. But we make the conjecture that these two expressions are identical based on the numerical checks. To demonstrate this, we isolate the width dependence and prove that there is an typical energy scale $\beta$ (determined by kinematic conditions) for the scalar loop integral, referring to which we can see the explicit influence of the finite width on the absorptive part.

Despite of the fact that the width of $K^*$ meson can be regarded as a small value compared with its mass, it is a big value compared with $\beta$.  Thus, the large resulting suppression can be understood. We also suggest a possible way to get the accurate analytic absorptive part of scalar loop integral, which may be useful in the case where the width of internal masses is no longer a small value.

For the decays of $\eta''$ we obtain an overall description of $\eta''\to K\bar{K}\pi$, $\eta\pi\pi$ and $3\pi$ taking into account the $K^*$ width effects. It shows that the $K\bar{K}\pi$ channel is the dominant one in its decays, and the tree diagram of $\eta''\to K\bar{K}^*+c.c.$ is the main contribution. In this channel the TS contribution is relatively small in comparison with the tree process. For the decay channel of $\eta''\to \eta\pi\pi$ its small branching ratio can be understood by the destructive interference between tree and triangle diagrams. Note that the $^3P_0$ quark model prediction for the branching ratios of $\eta'(2S)\to a_0(980)\pi$ and $K\bar{K}^*+c.c.$ are compatible~\cite{Yu:2011ta}. Without the contributions from the TS mechanism it would be very difficult to understand the much weakened decay of $\eta''\to a_0(980)\pi$ as the first radial excited state of $\eta'(958)$. An improved measurement of the branching ratio $J/\psi\to\gamma\eta''\to\gamma a_0\pi\to\gamma\eta\pi\pi$ would be useful for further quantifying the TS contributions. For the decay of $\eta''\to 3\pi$ we identify that the TS enhanced $a_0-f_0$ mixing should also contribute to the isospin violating decay of $\eta''\to 3\pi$. Taking into account the width effects from the internal $K^*$ we obtain a consistent picture for understanding the role played by the TS concerning the $\eta(1405)/\eta(1475)$ puzzle.

At this moment, there are still large experimental uncertainties with the measurement of the two-body decay widths. Therefore, partial wave analysis of these three channels combining the TS mechanism is strongly required. We would expect that experimental results from BESIII in the near future will be able to set all these controversial issues about $\eta(1405)/\eta(1475)$.

The $\eta(1405)/\eta(1475)$ puzzle has been a key issue for a better understanding of the pseudoscalar spectrum. For the purpose of searching for the pseudoscalar glueball state the settlement of this puzzle would provide a useful guidance for future efforts in both experiment and theory. Namely, our attention should be paid to the higher mass regime~\cite{Wu:2011yx} and criteria for pseudoscalar glueball candidates should be further studied in detail.

\section*{Acknowledgement}
This work is supported, in part, by the National Natural Science Foundation of China (NSFC) under Grant Nos. 11425525 and 11521505, by DFG and NSFC through funds provided to the Sino-German CRC 110 ``Symmetries and the Emergence of Structure in QCD'' (NSFC Grant No. 11621131001), and by the National Key Basic Research Program of China under Contract No.~2015CB856700.

\section*{Appendix}

We provide the decomposition of the loop amplitude into scalar loop functions. For our case, we have $m_2=m_3=m_K$, where $m_K$ is the mass of $K$ meson and $m_1=m_{K^*}$. Assuming so, the amplitude is
\begin{eqnarray}
M&=&-i\int\frac{d^4q}{(2\pi)^4}\frac{(2p_1-q)_{\mu}(-g^{\mu\nu}+\frac{q^{\mu}q^{\nu}}{q^2})(q-2p_2)_{\nu}}{D_1D_2D_3}\nonumber\\
&=&-i(s_1-m_1^2+im_1\Gamma_1+s_2-2s_3+2m_K^2-\frac{(s_1-m_K^2)(s_2-m_K^2)}{m_1^2-im_1\Gamma_1})\int\frac{d^4q}{(2\pi)^4}\frac{1}{D_1D_2D_3}\nonumber\\
&-&i\frac{(s_1-m_K^2)(s_2-m_K^2)}{m_1^2-im_1\Gamma_1})\int\frac{d^4q}{(2\pi)^4}\frac{1}{q^2D_2D_3}-i(1+\frac{s_1-m_K^2}{m_1^2-im_1\Gamma_1})\int\frac{d^4q}{(2\pi)^4}\frac{1}{D_1D_3}\nonumber\\
&-&i(1+\frac{s_2-m_K^2}{m_1^2-im_1\Gamma_1})\int\frac{d^4q}{(2\pi)^4}\frac{1}{D_1D_2}+i\int\frac{d^4q}{(2\pi)^4}\frac{1}{q^2D_1} -i\frac{m_K^2-s_1}{m_1^2-im_1\Gamma_1}\int\frac{d^4q}{(2\pi)^4}\frac{1}{q^2D_3} \nonumber\\
&-&i\frac{m_K^2-s_2}{m_1^2-im_1\Gamma_1}\int\frac{d^4q}{(2\pi)^4}\frac{1}{q^2D_2}+i\int\frac{d^4q}{(2\pi)^4}\frac{1}{D_2D_3}\nonumber
\end{eqnarray}
where $D_1=q^2-m_1^2+im_1\Gamma_1,\ D_2=(q-p_2)^2-m_2^2,\ D_3=(p_1-q)^2-m_3^2$ and $s_2=p_2^2=m_{\pi}^2$.

\bibliographystyle{unsrt}

\end{document}